\documentclass[12pt]{article}

\usepackage{amsmath,amssymb,amsthm}
\usepackage{times}
\usepackage{bm}
\usepackage{natbib}
\usepackage{graphicx}
\graphicspath{{./}}

\usepackage[ruled,noend]{algorithm2e}

\makeatletter
\renewcommand{\algocf@captiontext}[2]{#1\algocf@typo. \AlCapFnt{}#2}

\def\@algocf@capt@plain{top}
\renewcommand{\algocf@makecaption}[2]{%
  \sbox\@tempboxa{\algocf@captiontext{#1}{#2}}%
  \ifdim\wd\@tempboxa >\hsize%
    \parbox[t]{\hsize}{\algocf@captiontext{#1}{#2}}%
  \else%
    \global\@minipagefalse%
    \hbox to\hsize{\box\@tempboxa\hfill}%
  \fi%
}
\makeatother
\usepackage{CJKutf8}
\usepackage{hhline,epsfig,array,color,setspace,enumitem}
\usepackage[T1]{fontenc}
\usepackage[utf8]{inputenc}
\usepackage{mathrsfs}
\usepackage{subfigure}
\usepackage{color,colortbl}
\usepackage[normalem]{ulem}
\usepackage{mathtools}
\usepackage{multirow}
\usepackage{multicol}
\usepackage{xr-hyper}
\usepackage[hidelinks]{hyperref}

\theoremstyle{plain}
\newtheorem{theorem}{Theorem}[section]

\theoremstyle{definition}

\makeatletter
\providecommand{\@LN}[2]{}
\makeatother

\newcommand{\anon}{1}

\begin{document}
\date{}

\def\spacingset#1{\renewcommand{\baselinestretch}{#1}\small\normalsize}
\spacingset{1}

\if1\anon
{
  \title{\bf Testing Cumulative Treatment Effects in Continuous Time with ODE Dynamics and Noisy Observations}
  \author{J. Zhang\thanks{Department of Biostatistics, University of Michigan, 1415 Washington Heights, Ann Arbor, Michigan, 48109, U.S.A. Email: \texttt{jiuchen@umich.edu}.} \and A. Qu\thanks{Department of Statistics and Applied Probability, University of California, Santa Barbara, South Hall 5607A, Santa Barbara, California, 93106, U.S.A. Email: \texttt{aqu2@ucsb.edu}.}}
  \maketitle
}
\fi

\if0\anon
{
  \bigskip
  \bigskip
  \bigskip
  \begin{center}
    {\LARGE\bf Testing Cumulative Treatment Effects in Continuous Time with ODE Dynamics and Noisy Observations}
  \end{center}
  \medskip
}
\fi

\bigskip
\begin{abstract}
We develop a semiparametric continuous time inferential framework for testing cumulative treatment effects, including both immediate and carryover effects. The latent state process is modeled by an ordinary differential equation (ODE) whose drift depends on state and treatment, while the observed data are noisy measurements of this latent state at irregular, multi resolution time points. We define the average treatment effect (ATE) as the difference of expected cumulative outcomes under the always treat and always control policies and establish its identifiability from observational data. More generally, the framework accommodates contrasts between arbitrary deterministic stationary treatment policies, including clinically motivated threshold policies, with the always treat versus never treat comparison as the leading special case. A two step estimation procedure first recovers the latent state trajectories and their derivatives via spline smoothing, then solves a continuous time estimating equation derived from the infinitesimal generator of the ODE. We prove asymptotic normality of the resulting test statistic under explicit rate conditions on the spline estimator. The resulting test admits asymptotically valid inference under noisy, irregularly sampled observations, providing a studentized statistic with a standard $N(0,1)$ limit. Simulations show that the proposed method demonstrates improved stability and power over discrete time alternatives in noisy and multi resolution settings, with comparable performance under clean regular grids. We apply the method to the OhioT1DM diabetes dataset to evaluate cumulative effects of insulin treatments on glucose levels.
\end{abstract}

\vfill

\newpage
\spacingset{1.8}
\section{Introduction}
\label{sec:introduction}

Continuous time treatment processes arise naturally in many biomedical and mobile health applications, where interventions and physiological responses evolve continuously over time while observations are collected irregularly and often at heterogeneous resolutions. Examples include insulin administration and glucose regulation in diabetes management \citep{marling2020ohiot1dm}, physical activity interventions in wearable health studies, and adaptive behavioral interventions delivered through mobile devices \citep{zhang2024individualized}. In these settings, treatment effects frequently accumulate gradually through latent physiological dynamics rather than manifesting instantaneously at discrete observation times. However, most existing dynamic treatment learning and reinforcement learning methodologies are developed under discrete time formulations that assume regularly spaced observations and stagewise transitions. Such assumptions are problematic in longitudinal studies with sparse, noisy, and asynchronous measurements.

Discrete time formulations distort cumulative treatment effects under irregular observation schedules because the implied transition operator and effective discounting structure become observation dependent. When observation intervals vary across subjects and over time, the effective discount factor scales as $\gamma^{\Delta t}$ and the transition kernel depends on $\Delta t$, so the discrete time estimator carries a discretization bias that does not vanish under sparse or heterogeneous schedules and that fails to capture latent carryover dynamics evolving continuously between observation times \citep{bojinov2019time,shi2022dynamic,shi2022statistical}. The proposed framework targets cumulative treatment effects in continuous time, avoiding aggregation artifacts while accommodating noisy multi resolution observations.

These challenges are particularly pronounced in wearable and mobile health studies, where data are often collected at highly heterogeneous frequencies with substantial measurement noise. For example, continuous glucose monitor data, wearable sensor streams, and ecological momentary assessments are observed asynchronously and intermittently, while the underlying latent health states evolve continuously. Existing continuous time reinforcement learning literature has focused on policy optimization or control under fully observed stochastic systems \citep{doya2000reinforcement,jia2022policy}, whereas less attention has been devoted to statistical inference for cumulative treatment effects under noisy irregular observations \citep{kallus2022efficiently,lewis2020double}. Formal inferential procedures that explicitly account for latent state recovery error and irregular temporal sampling remain underdeveloped. Related causal inference work for continuous time interventions \citep{schomaker2024causal} typically assumes regularly spaced observations and does not target cumulative carryover effects.

Our approach also connects to the broad literature that performs causal inference through estimating equations and moment conditions. In the cross sectional and longitudinal settings, treatment effects are commonly identified and estimated through score or moment equations, including doubly robust and Neyman orthogonal constructions that render the estimating equation insensitive to first order nuisance error \citep{chernozhukov2017double,zhang2011causal}. In sequential decision making, reinforcement learning inference solves value equation moment conditions, either Bellman residual equations or efficient/doubly robust value estimators \citep{munos2006policy,shi2022statistical,shi2022dynamic,kallus2022efficiently,lewis2020double}. Our estimating equation~\eqref{eq:th1} is the continuous time counterpart of these value equation moments: it replaces the discrete Bellman residual by a generator based residual $Y+\log\gamma\,V_a+\mathcal L_a V_a$ and identifies a cumulative causal contrast from latent, noisy, irregularly sampled trajectories. Relative to discrete time value equation inference, the novelty is that the moment involves the latent state and its derivative, recovered nonparametrically in a first stage, with the spline recovery error entering at first order but driven below the $N_Y^{-1/2}$ scale by a multi resolution rate condition, and the value function and representer approximation errors entering only through a second order product; relative to the semiparametric causal estimating equation literature, the novelty is the continuous time generator identification and the multi resolution sampling design under which the first stage is negligible.

To address these challenges, we develop a continuous time inferential framework for testing cumulative treatment effects under latent ordinary differential equation (ODE) dynamics with noisy and irregular observations. We model the latent state trajectory through a continuous time dynamical system and formulate cumulative treatment effects through a value function characterization involving the infinitesimal generator of the underlying process. This formulation yields a continuous time estimating equation that extends value equation inference to irregular longitudinal settings. Unlike standard discrete time Bellman formulations, the proposed framework explicitly accommodates heterogeneous observation intervals and continuous carryover dynamics without requiring temporal aggregation onto fixed grids.

Our framework combines spline based latent trajectory recovery \citep{brunel2008parameter,huang1998projection} with estimating equation inference for continuous time value function contrasts. We establish asymptotic normality of the estimator under explicit multi resolution infill, smoothness, and undersmoothing conditions and derive a studentized test for cumulative treatment effect inference. 

Through extensive simulations, including nonlinear and irregular observation settings, we demonstrate improved estimation accuracy and more stable inference compared with discrete time alternatives, particularly under sparse and heterogeneous sampling schedules. We illustrate the practical utility of the framework using OhioT1DM wearable glucose monitoring data \citep{marling2020ohiot1dm}, where irregular continuous glucose trajectories and delayed insulin effects motivate continuous time treatment effect modeling. 

The primary contribution of this paper is a semiparametric continuous time inferential framework for cumulative treatment effects under noisy, irregular, multi resolution observations. The framework combines latent trajectory recovery via spline smoothing with continuous time estimating equation inference based on the infinitesimal generator of the underlying state process, yielding an asymptotically valid hypothesis test for cumulative carryover effects without discretizing the dynamics onto fixed temporal grids, for contrasts between arbitrary stationary treatment policies with the always treat versus never treat comparison as the leading special case.

\section{Background and Notations}
\label{sec:background}

This section provides background and notation for our continuous time inference framework for cumulative treatment effects and outlines the data generating process, observation model, and potential outcome model under this framework.

For participant \(i\) ($i = 1, \ldots, I$), the (unobserved) latent state \((\boldsymbol{S}_{it})_{t \geq 0}\) is a mapping from the infinite time horizon \([0, \infty)\) to the state space \(\mathcal{S}\), a compact subset of $\mathbb{R}^d$ representing \(d\)-dimensional time varying covariates, describing the trajectory of participant \(i\)'s state over time. The observed treatment history is denoted as \((\bar{A}_{it})_{t \geq 0}\), representing the sequence of actions chosen for participant \(i\). In this paper, the action space is set as binary, with \(\mathcal{A} = \{0, 1\}\), which represents a binary decision: either take or not take treatment. The observed outcome for participant \(i\) is denoted as \((Y_{it})_{t \geq 0}\) to provide the observed immediate reward at time \(t\).

We assume that the latent state process \((\boldsymbol{S}_{it})_{t \geq 0}\) for participant $i$ evolves according to an ordinary differential equation (ODE):
\begin{equation}
\label{eq:ode}
d\boldsymbol{S}_{it} = b(\boldsymbol{S}_{it}, A_{it})\, dt,
\end{equation}
where \(b: \mathcal{S} \times \mathcal{A} \to \mathbb{R}^d\) is the drift function describing how the state evolves as a function of its current value and the treatment. This means that conditional on the initial state and treatment history, the state trajectory is deterministic.

We focus on ODE driven latent dynamics as a first order continuous time working model for noisy mobile health trajectories. This formulation is appropriate when the dominant stochasticity arises from the treatment assignment mechanism that generated the observed data (formalized in Assumption~\ref{asp:mixing} of the Supplementary Material) and measurement noise, while the underlying physiological state evolves smoothly over short time windows.  Under latent stochastic dynamics, the proposed estimating equation targets the cumulative value function associated with the drift component of the conditional mean trajectory rather than the full stochastic process law. Under moderate latent stochasticity, this drift driven target remains the dominant signal recovered by spline smoothing, and the resulting test retains valid Type-I control and substantial power; this is verified in the sensitivity analysis of Section~\ref{sec:simulation} and Supplement~\ref{sec:sensitivity-sde}.

The latent state $\boldsymbol{S}_{it}$ is not observed directly. Instead, we observe noisy measurements
\begin{equation}
\label{eq:obs}
\boldsymbol{X}_{it} = \boldsymbol{S}_{it} + \boldsymbol{\epsilon}_{it},
\end{equation}
where $\boldsymbol{\epsilon}_{it}$ are independent, zero mean random variables representing measurement error. The observations $\boldsymbol{X}_{it}$ and the outcomes $Y_{it}$ may be recorded at different frequencies and at irregular time points, a multi resolution structure common in wearable device data.

A stationary policy $\pi$ is a time homogeneous map $\pi:\mathcal{S}\to\mathcal{A}$ that assigns to each state $\boldsymbol{s}$ the treatment $a=\pi(\boldsymbol{s})$. Under the ODE dynamics \eqref{eq:ode}, the state process $(\boldsymbol{S}_{it})_{t \ge 0}$ under policy $\pi$ is Markov with a transition semigroup $(P^{\pi}_t)_{t \ge 0}$.  Because the drift $b(\boldsymbol{s},a)$ depends on the action, the generator is policy indexed: the policy $\pi$ induces the infinitesimal generator $\mathcal{L}_\pi$ of the ODE driven process under $\pi$, acting on a differentiable function $V:\mathcal{S} \to \mathbb{R}$ as
\begin{equation}
\label{eq:generator}
\mathcal{L}_\pi V(\boldsymbol{s}) = \left\langle \nabla V(\boldsymbol{s}),\, b(\boldsymbol{s}, \pi(\boldsymbol{s}))\right\rangle,
\end{equation}
where $\langle \cdot, \cdot \rangle$ denotes the inner product. This first order form, with no second order diffusion terms, is a direct consequence of the deterministic ODE dynamics. The constant policies $\pi\equiv a$ recover the action indexed generator $\mathcal{L}_a V(\boldsymbol{s})=\left\langle\nabla V(\boldsymbol{s}),\,b(\boldsymbol{s},a)\right\rangle$ of the always-$a$ process, the special case that underlies the average treatment effect. Intuitively, $\mathcal{L}_\pi V$ is the instantaneous expected rate of change of $V$ along the latent trajectory under $\pi$, i.e., the continuous time analogue of the Bellman operator.   This is the object that governs the value function of a general policy and underlies the policy contrast estimand introduced in Section~\ref{sec:methodology}. We use \(S_{it}^{\star}(\pi)\) and \(Y_{it}^{\star}(\pi)\) to denote the associated potential state and outcome that would occur at time \(t\) if participant $i$ follows \(\pi\).

The expected cumulative outcome \(V^\pi: \mathcal{S} \rightarrow \mathbb{R}\) records the time discounted total outcome accrued from initial state \(\boldsymbol{s}\) when policy \(\pi\) is applied: $V^\pi(\boldsymbol{s}) \triangleq \int_0^{\infty} \gamma^t \mathbb{E}[Y_{it}^{\star}(\pi) \mid S_{i0}=\boldsymbol{s}] \, dt$, where \(S_{i0}\) is the initial state at time 0 for participant \(i\) and \(0<\gamma<1\) is a discount factor that balances immediate and future contributions to the outcome. We use \(V^\pi\) as the inferential target throughout: it characterizes the cumulative effect of a stationary policy through potential outcomes \citep{shi2022statistical}, without committing to a randomized behavior mechanism.

\section{Methodology}
\label{sec:methodology}
In this section we define the policy contrast estimand and present the estimation and testing procedure. The framework is stated once, for a contrast between two arbitrary stationary policies $\pi_1,\pi_2:\mathcal{S}\to\mathcal{A}$, so the inferential target is the cumulative value difference $V^{\pi_1}-V^{\pi_2}$ averaged over the initial state. The always treat versus never treat comparison, $\pi_1\equiv1$ and $\pi_2\equiv0$, is the leading special case, which defines the average treatment effect (ATE) and is the estimand of our first simulation studies; clinically motivated threshold policies such as ``treat when a monitored marker exceeds a cutoff,'' $\pi(\boldsymbol{s})=\mathbb{I}\{g(\boldsymbol{s})>c\}$, are covered by the same statements.

\subsection{Identification of stationary policy contrasts}

Our framework adapts the continuous time value function representation to a noisy, irregularly sampled observational setting, generalizing discrete time results from \citet{shi2022statistical}. The key statistical challenge is that the latent state is observed indirectly and at irregular times, so neither the state trajectory nor its derivative is directly available. Our solution is a two step procedure: first recover the latent state and its derivative nonparametrically, then solve the continuous time estimating equation for the value function. We show that the first step error is asymptotically negligible under an explicit rate condition. The estimating equation has a clear semiparametric architecture: a nonparametric first stage that recovers the latent state and its derivative via cubic B-spline smoothing of the noisy observations, and a parametric sieve second stage that solves a moment equation for the value function coefficients $\beta^*$.

Let $\mathbb{G}$ be a reference distribution with bounded density representing the distribution of the initial state, and let $V^{\pi}$ be the cumulative value function of Section~\ref{sec:background}. For any two stationary policies $\pi_1,\pi_2$ we define the \emph{policy contrast estimand}
\begin{equation}
\label{eq:ate_policy}
\tau(\pi_1,\pi_2)=\int_{\mathcal{S}}\{V^{\pi_1}(s)-V^{\pi_2}(s)\}\,\mathbb{G}(ds),
\end{equation}
the gain in expected cumulative outcome from following $\pi_1$ instead of $\pi_2$, averaged over the initial state. For the constant policies we abbreviate $V_a:=V^{\pi\equiv a}$, and the average treatment effect (ATE) is the special case $\tau:=\tau(\pi_1{\equiv}1,\,\pi_2{\equiv}0)$.

Under suitable regularity conditions (see Section~\ref{sec:theory}; specifically the consistency, sequential randomization, Markov, and conditional mean independence Assumptions~\ref{asp:CA}--\ref{asp:CMIA}, the positivity Assumption~\ref{asp:positivity}, and the generator regularity Assumption~\ref{asp:generator_reg}), the value function of any stationary policy $\pi$ satisfies the continuous time estimating equation
\begin{equation}
\label{eq:th1}
\mathbb{E}\left[\left\{Y_{it}+\log\gamma\cdot V^\pi(S_{it})+\mathcal{L}_\pi V^\pi(S_{it})\right\}\phi(S_{it})\,\mathbb{I}\{A_{it}=\pi(S_{it})\}\right]=0
\end{equation}
for any bounded measurable $\phi:\mathcal{S}\to\mathbb{R}$, where $\mathcal{L}_\pi$ is the policy generator of Section~\ref{sec:background} and $\mathbb{I}\{A_{it}=\pi(S_{it})\}$ is the \emph{policy agreement indicator}, the weight that restricts the moment to the observation times at which the recorded action agrees with the target policy. For a constant policy $\pi\equiv a$ the agreement indicator reduces to the action indicator $\mathbb{I}\{A_{it}=a\}$ and $\mathcal{L}_\pi$ to $\mathcal{L}_a$, so~\eqref{eq:th1} identifies $V_0$ and $V_1$ separately through the two action subsamples. Crucially, on the agreement event $\{A_{it}=\pi(S_{it})\}$ the observed state derivative satisfies $dS_{it}/dt=b(S_{it},A_{it})=b(S_{it},\pi(S_{it}))$, so the generator term $\mathcal{L}_\pi V^\pi(S_{it})=\langle\nabla V^\pi(S_{it}),\,dS_{it}/dt\rangle$ is computed from the spline derivative by the same formula for every policy; only the subset of observation times entering the equation changes, which is why identification requires the overlap condition of Assumption~\ref{asp:positivity}. Equation~\eqref{eq:th1} is a continuous time analogue of the value equation moment conditions in \citet{doya2000reinforcement} and \citet{jia2022policy}.

Equation \eqref{eq:th1} identifies $V^{\pi}(s)$ through a population moment involving the \emph{latent} state and its derivative; because the state is observed only with noise, $V^\pi$ is feasibly estimated from the observed data by substituting the spline recovered trajectory and derivative, under the first stage recovery conditions of Section~\ref{sec:theory}. Since $\tau(\pi_1,\pi_2)$ is a functional of $V^{\pi_1}$ and $V^{\pi_2}$, the policy contrast, and in particular the average treatment effect, is also identifiable under these conditions. The formal assumptions and a detailed identification analysis are provided in Section~\ref{sec:theory}.

To compute $\mathcal{L}_\pi V^\pi(S_{it})$ using equation~\eqref{eq:generator}, we need the drift $b(S_{it}, A_{it})$, which equals $dS_{it}/dt$ under the ODE assumption~\eqref{eq:ode}. Since the latent state is observed with noise (equation~\eqref{eq:obs}), we recover the state trajectory and its derivative via spline smoothing, a nonparametric regression approach \citep{brunel2008parameter}.

Specifically, let $\left\{\chi^{\top} w: w \in \mathbb{R}^{M_s\times d}\right\}$ be a large linear approximation space for $S_{it}$, where $\chi(\cdot)$ is a vector containing cubic B-spline basis functions $\chi_m(\cdot)$ for $m = 1,\cdots,M_s$. The spline coefficients $\hat{w}_i$ for participant $i$ are obtained by fitting the spline to the noisy observations $X_{it}$, so that $\hat{S}_{it} =\chi^{\top} \hat{w}_i$ and the derivative $d \hat{S}_{it}/d t$ can be estimated as $\nabla\chi^{\top} \hat{w}_i$, where $\nabla\chi$ is the derivative of the basis functions $\chi$.  Substituting the spline derivative $\nabla\chi^{\top}\hat{w}_i$ for the drift $b(S_{it},A_{it})$ in equation~\eqref{eq:generator} yields $\mathcal{L}_\pi V^\pi(S_{it})$ on the agreement event $\{A_{it}=\pi(S_{it})\}$ from the observed data.

\subsection{Test procedure}

We now describe the estimation and testing procedure based on the estimating equation~\eqref{eq:th1}. The procedure first recovers the latent state and its derivative via spline smoothing, then solves the continuous time estimating equation for the value function coefficients.

Let $\mathcal{V}=\left\{\Psi^{\top}(s) \beta: \beta \in \mathbb{R}^M\right\}$ be a large linear approximation space for the value functions $V^{\pi_k}(s)$, $k=1,2$, where $\Psi(\cdot)$ is a vector containing basis functions $\Psi_m(\cdot)$ for $m = 1,\cdots,M$ on $\mathcal{S}$. In our numerical studies, we use polynomial bases (degree $3$ in the univariate designs and degree $2$ in the multivariate and real data designs); cubic B-spline bases with $K$ interior knots are an alternative for higher dimensional state spaces.

Fix the target pair $(\pi_1,\pi_2)$ and, for $k\in\{1,2\}$, write $\mathbb{I}_k(it):=\mathbb{I}\{A_{it}=\pi_k(S_{it})\}$ for the policy agreement indicator; for the constant policies $\pi_1\equiv1,\pi_2\equiv0$ these are the action indicators $\mathbb{I}(A_{it}=1)$ and $\mathbb{I}(A_{it}=0)$. Rather than assume the value functions lie exactly in the sieve $\mathcal{V}$, we define $\boldsymbol{\beta}^*=(\beta_{\pi_2}^{* \top}, \beta_{\pi_1}^{* \top})^{\top}$ as the solution of the population estimating equation obtained by setting $\phi=\Psi_m$, $m=1,\dots,M$, in Theorem \ref{thm:bellman}:
$$
\begin{aligned}
\mathbb{E}\left[\left\{Y_{it}  + \log \gamma \cdot \Psi^{\top}(S_{it}) \beta^*_{\pi_k} + \left\langle\nabla \Psi^{\top}(S_{it}) \beta^*_{\pi_k}, D\right\rangle \right\} \Psi\left(S_{it}\right) \mathbb{I}_k(it) \right]=0,
\quad k \in\{1,2\},
\end{aligned}
$$
where $\mathbb{I}(\cdot)$ denotes the indicator function, $D=D_{it}:=\dot S_{it} \in \mathbb{R}^{d}$ is the latent state derivative (equal to $b(S_{it},A_{it})$ under the ODE~\eqref{eq:ode}), $\nabla{\chi} = \left ( \nabla \chi_1 , \cdots, \nabla \chi_{M_s}  \right) \in \mathbb{R}^{M_s}$, and  $\nabla{\Psi}(s) = \left ( \nabla \Psi_1 (s), \cdots, \nabla \Psi_M (s) \right) \in \mathbb{R}^{d \times M}$. In the empirical estimating equations below, $\widehat{D}_{it} = \hat w_i^{\top}\nabla\chi(t)$, the derivative of the fitted spline, replaces $D_{it}$ at each observation time. When $V^{\pi_k}\in\mathcal{V}$ this $\boldsymbol{\beta}^*$ reproduces the value functions exactly; otherwise it is the sieve pseudo-true parameter, and the approximation error it induces in $\tau(\pi_1,\pi_2)$ is shown to be $o\{(N_Y)^{-1/2}\}$, hence asymptotically negligible, in Supplement~\ref{subsec:sieve_bias}. The above equations can be rewritten as $\mathbb{E}\left(\Sigma_{it} \beta^*\right)=\mathbb{E} \eta_t$, where $\Sigma_{it}$ is a block diagonal matrix given with $\Psi\left(S_{it}\right) \mathbb{I}_2(it) \cdot \left\{-\log \gamma \Psi^{\top}\left(S_{it}\right)  - D ^{\top} \nabla{\Psi}(S_{it}) \right\}$ and $\Psi\left(S_{it}\right) \mathbb{I}_1(it) \cdot \left\{-\log \gamma \Psi^{\top}\left(S_{it}\right)  - D^{\top} \nabla{\Psi}(S_{it})  \right\}$, $\eta_{it}=\left\{\Psi\left(S_{it}\right)^{\top} \mathbb{I}_2(it) Y_{it}, \Psi\left(S_{it}\right)^{\top}\mathbb{I}_1(it) Y_{it}\right\}^{\top}$.

Let $\widehat{\Sigma}=N_Y^{-1}\sum_{1\le i \le I} \sum_{t \in \mathbb{T}_i^Y} \boldsymbol{\Sigma}_{it}$ and $\widehat{\eta}=N_Y^{-1}\sum_{1\le i \le I} \sum_{t \in \mathbb{T}_i^Y}  \eta_{it}$ where $I$ is the number of subjects and $N_Y := \sum_{i=1}^I |\mathbb{T}_i^Y|$ is the total number of outcome observation times across subjects. Under Assumption~\ref{asp:sampling} the \emph{outcome} grid has spacing bounded below and above, so $N_Y \asymp \sum_i T_i$ and $N_Y$ may be interpreted equivalently as the total follow up time; the dense (infill) sampling that drives the first stage spline accuracy is required of the \emph{state} grid only, and no equivalence between the outcome count and the state observation count is used anywhere. It follows that $\mathbb{E}\left\{\widehat{\Sigma} \beta^*\right\}=\mathbb{E}\{\widehat{\eta}\}$. Consequently, $\beta^*$ is estimated by
$\widehat{\boldsymbol{\beta}}=\{\widehat{\beta}_{\pi_2}^{\top}, \widehat{\beta}_{\pi_1}^{\top}\}^{\top}=\widehat{\boldsymbol{\Sigma}}^{-1} \widehat{\eta}$. Thus, the policy contrast can be estimated by the plug in estimator $\widehat{\tau}=\int_s \Psi^{\top}(s)\{\widehat{\beta}_{\pi_1}-\widehat{\beta}_{\pi_2}\}\, \mathbb{G}(d s)$, which for the constant policies is the estimator of the average treatment effect. Because on each agreement subsample the spline derivative already equals the target policy drift, the same first step recovery serves every policy; only the agreement weights change across targets.

One caveat on smoothness deserves note for threshold policies: the value function $V^{\pi_k}$ must inherit the regularity required by the sieve, and for a hard threshold policy $V^{\pi_k}$ may have a derivative discontinuity at the switching surface $\{g(s)=c\}$. Because the threshold is specified by the analyst, this is handled by adapting the sieve to the known switching surface rather than by modifying the policy: in one dimension we place a knot of multiplicity one less than the spline order exactly at the threshold $c$ (a triple knot for a cubic spline, retaining continuity while freeing the derivative), and more generally we augment the basis with functions of the hinge $(g(s)-c)_+$, so that piecewise H\"older smoothness of $V^{\pi_k}$ on either side of the surface restores the approximation rate required by Assumption~\ref{asp:basisfunction}, whose $C^1$ display is then understood piecewise away from the switching surface, with the surface neighborhood controlled by the margin condition (Assumption~\ref{asp:margin} in the Supplementary Material). The theory covers hard threshold policies when the sieve is adapted to the surface in this way; the numerical studies and the data analysis use polynomial bases without hinge terms (Sections~\ref{sec:simulation}--\ref{sec:realdata}), so their threshold results are exploratory relative to this guarantee. The block notation convention under which the supplementary assumptions and proofs are stated for general policy contrasts, together with the four policy specific conditions, is recorded in Section~\ref{subsec:assumptions_normality} of the Supplementary Material. A separate first stage issue arises for hard threshold policies because Algorithm~\ref{alg:test} evaluates the agreement indicator at the estimated state, $\mathbb I\{A_{it}=\pi(\widehat S_{it})\}$, which can flip near the switching surface under state estimation error. Under a margin condition (the stationary density of $g(S)$ bounded near the threshold), Lemma~\ref{lem:indicator} of the Supplementary Material shows that the flipped indicators are $o_p\{(N_Y)^{-1/2}\}$-negligible under the first stage rate condition~\eqref{eq:spline_rate}, and Remark~\ref{rem:switching} there records existence and uniqueness of the policy induced ODE at the discontinuous switching surface in the Filippov sense.

A key advantage of the proposed estimator is its ability to accommodate multi resolution and irregularly observed data \citep{zhang2024individualized}. When observations are multi resolution but on a regular grid, we evaluate the estimating equation at the time points where both $S_{it}$ and $Y_{it}$ are observed and use the remaining state observations to refine the numerical derivative $D$. When the grid is irregular, the spline path $\chi^{\top}\hat{w}_i$ supplies predicted state values at any outcome observation time. Specifically, we let $\hat S_{it} = \chi^{\top} \hat w_i$ for $t \in \mathbb{T}^{Y}$, where $\mathbb{T}^{Y}$ is the index set for the outcome observations $Y_{it}$.

Furthermore, we utilize $\widehat{\tau}$ to construct our test statistics. Let $u_{\mathbb{G}}=\int_{s \in \mathbb{S}} \Psi(s)\, \mathbb{G}(d s)$ and $\boldsymbol{U}=\{-u_{\mathbb{G}}^{\top},\ u_{\mathbb{G}}^{\top}\}^{\top}$, indicating that $\widehat{\tau}=\boldsymbol{U} \widehat{\boldsymbol{\beta}}$. We can show the asymptotic normality of $\sqrt{N_Y}\left\{\widehat{\tau}-\tau_0\right\}$  in Section \ref{sec:theory}. Additionally, the variance of $\sqrt{N_Y}\widehat{\tau}$ can be consistently estimated  by $\widehat{\sigma}^2=\boldsymbol{U}^{\top} \widehat{\boldsymbol{\Sigma}}^{-1} \widehat{\boldsymbol{\Omega}}\{\widehat{\boldsymbol{\Sigma}}^{-1}\}^{\top} \boldsymbol{U}$, as $N_Y$ grows to infinity, where $\widehat{\boldsymbol{\Sigma}}^{-1} \widehat{\boldsymbol{\Omega}}\left\{\widehat{\boldsymbol{\Sigma}}^{-1}\right\}^{\top}$ is the sandwich estimator for the variance of $\sqrt{N_Y}\left\{\widehat{\boldsymbol{\beta}}-\boldsymbol{\beta}^*\right\}$. Here $\widehat{\boldsymbol{\Omega}}$ denotes a consistent estimator of the long run covariance matrix of the retained score process based on the outcome observation times $t\in\mathbb{T}_i^Y$. When the retained score summands are effectively uncorrelated, the simple empirical outer product estimator is sufficient; under serial dependence, a HAC-type long run covariance estimator should be used instead. The continuous time value equation error for block $k$ is $\widehat{\varepsilon}_{{it}, k}= Y_{it} + \log \gamma \cdot \Psi^{\top}(\widehat{S}_{it}) \widehat{\beta}_{\pi_k} +  \left\langle\nabla \Psi^{\top}(\widehat{S}_{it}) \widehat{\beta}_{\pi_k}, \widehat{D}_{it}\right\rangle $, where $\widehat{S}_{it}$ and $\widehat{D}_{it}$ are the spline smoothed state and its derivative at $t$.  The conditional expectation of the value equation error given the agreement event and the state is zero asymptotically by Theorem~\ref{thm:bellman}. This yields our test statistic $\sqrt{N_Y} \widehat{\tau} / \widehat{\sigma}$.

\begin{algorithm}[!h]
\caption{Continuous time cumulative treatment effect test}
\label{alg:test}

\begin{enumerate}[]
\item \textit{(Input and tuning parameters)} Take noisy state observations $\{X_{it}: t \in \mathbb{T}_i^S\}$, outcomes $\{Y_{it}: t \in \mathbb{T}_i^Y\}$, and treatments $\{A_{it}\}$ for $i = 1, \ldots, I$, and specify the target policy pair $(\pi_1,\pi_2)$ (e.g., always treat versus never treat). Set the discount factor $\gamma$, spline basis dimension $M_s$, smoothing parameter $\lambda$, and value function basis dimension $M$.

\item \textit{(State trajectory recovery)} For each subject $i = 1, \ldots, I$:
(i) Fit the cubic B-spline $\hat{S}_{it} = \chi^\top \hat{w}_i$ to $\{X_{it}: t \in \mathbb{T}_i^S\}$ with smoothing parameter $\lambda$.
(ii) Compute the derivative estimate $\hat{D}_{it} = d\hat{S}_{it}/dt = \nabla\chi^\top \hat{w}_i$.

\item \textit{(Continuous time estimating equation)} Choose the value function basis $\Psi(\cdot) = (\Psi_1(\cdot), \ldots, \Psi_M(\cdot))^\top$. Form the agreement indicators $\mathbb{I}_k(it)=\mathbb{I}\{A_{it}=\pi_k(\hat S_{it})\}$, compute $\widehat{\Sigma}$ and $\widehat{\eta}$ using $\{(\hat{S}_{it}, \hat{D}_{it}, Y_{it}, \mathbb{I}_k(it)): t \in \mathbb{T}_i^Y\}$, and solve $\widehat{\boldsymbol{\beta}} = \widehat{\Sigma}^{-1}\widehat{\eta}$ to obtain $\widehat{V}^{\pi_2}$ and $\widehat{V}^{\pi_1}$.

\item \textit{(Test statistic and decision)} Compute $\widehat{\tau} = \boldsymbol{U}^\top \widehat{\boldsymbol{\beta}}$ and the sandwich variance $\widehat{\sigma}^2$. Form $Z = \sqrt{N_Y}\,\widehat{\tau}/\widehat{\sigma}$ and reject $H_0: \tau(\pi_1,\pi_2) \le 0$ if $Z > z_\alpha$.
\end{enumerate}

\end{algorithm}

The complete procedure, summarized in Algorithm~\ref{alg:test}, involves three groups of tuning parameters. (i)~\emph{Spline smoothing:} The spline basis dimension $M_s$ and smoothing parameter $\lambda$ control the bias variance tradeoff in recovering the latent state. We recommend choosing $\lambda$ via generalized cross-validation; the spline rate condition \eqref{eq:spline_rate} provides the theoretical requirement. (ii)~\emph{Value function basis:} For low dimensional state spaces ($d \le 3$), polynomial bases of moderate degree are effective; for higher dimensions, cubic B-splines with $K$ interior knots offer flexibility while controlling complexity. The sieve approximation error must satisfy $o\{(N_Y)^{-1/4}\}$ (Assumption~\ref{asp:basisfunction}). (iii)~\emph{Discount factor:} The discount factor $\gamma \in (0,1)$ controls the effective time horizon $1/|\log \gamma|$ over which treatment effects are accumulated. Smaller $\gamma$ emphasizes short term effects; values nearer~1 capture longer term carryover. In our numerical studies, $\gamma=0.8$ performed well across the settings considered, and the supplementary sweeps also examine $\gamma\in\{0.5,0.7\}$.  Practitioners can choose $\gamma$ to match the clinically relevant carryover window of the intervention, for example the agent's half life (for insulin, 3 to 6 hours), the persistence horizon of a behavioral intervention, or a prespecified endpoint horizon. 

\section{Theory}
\label{sec:theory}

This section establishes identification of the average treatment effect and derives the asymptotic normality of the proposed test statistic. Section~\ref{sec:theory}.1 gives identification under continuous time analogues of the standard causal assumptions. Section~\ref{sec:theory}.2 states the rate condition controlling the first step spline error and the corresponding asymptotic distribution of the studentized statistic. Section~\ref{sec:theory}.3 quantifies the discretization and noise amplification bias of discrete time value estimators, explaining why they lose power as the observation grid refines under measurement noise while the proposed estimator does not.

\subsection{Identification}
We first provide sufficient conditions under which the value functions $\{V_a: a\in\{0,1\}\}$ are identifiable from the observable process. Because the average treatment effect is the constant policy case of \eqref{eq:ate_policy}, a functional of $V_0$ and $V_1$, identification of the value functions implies identification of the average treatment effect.

To keep the main text compact, the formal causal conditions are stated and discussed in Section~\ref{subsec:causal_assumptions} of the Supplementary Material: consistency (Assumption~\ref{asp:CA}), continuous time sequential randomization (Assumption~\ref{asp:SR}), a Markov condition (Assumption~\ref{asp:MA}), conditional mean independence (Assumption~\ref{asp:CMIA}), and policy overlap/positivity (Assumption~\ref{asp:positivity}), together with the potential outcome notation in which they are expressed, their plausibility in mobile health applications, and practical diagnostics. In brief: the recorded state must capture the information used to assign treatment, the dynamics must be Markov in the recorded state, the mean outcome must depend on the past only through the current state and action, and the observed data must contain, across the relevant state range, decisions agreeing with each target policy. The generator regularity conditions are collected in Assumption~\ref{asp:generator_reg}.

\begin{theorem}
\label{thm:bellman}
Under the ODE assumption \eqref{eq:ode}, Assumptions \ref{asp:CA} to \ref{asp:generator_reg}, for any $t \geq 0$, any stationary policy $\pi:\mathcal{S}\to\mathcal{A}$, and any bounded measurable $\phi:\mathcal{S}\to\mathbb{R}$,
\begin{equation}
\label{eq:bellman_policy}
\mathbb{E}\left[\mathbb{I}\{A_{it}=\pi(S_{it})\}\left\{Y_{it}+\log\gamma\cdot V^\pi(S_{it})+\mathcal{L}_\pi V^\pi(S_{it})\right\}\phi(S_{it})\right]=0.
\end{equation}
For the constant policies $\pi\equiv a$, the agreement indicator is the action indicator $\mathbb{I}\{A_{it}=a\}$, the generator is $\mathcal{L}_a$, and the overlap condition reduces to $\Pr\{A_{it}=a\mid S_{it}=s\}>0$.
\end{theorem}

Theorem~\ref{thm:bellman} shows that the value function of any stationary policy can be identified from the observed data via the estimating equation~\eqref{eq:th1}.  A key feature is that the generator under the ODE assumption takes the first order form \eqref{eq:generator}, so the estimating equation involves only the gradient of $V^\pi$ and the drift $b$, with no second order term, eliminating diffusion coefficient estimation.

The discrete time Bellman equation used in \citet{shi2022statistical, shi2022dynamic} can be viewed as a special case where $\mathcal{L}_a V_a(S_{it})$ is approximated by a finite difference quotient $\{V_a(S_{i,t+1}) - V_a(S_{it})\}/\Delta t$. Our continuous time formulation avoids the discretization error inherent in this approximation.

\subsection{Asymptotic properties}

To study the asymptotic properties of our procedure, we work under Assumptions~\ref{asp:sampling} to~\ref{asp:moments}, stated in the Supplementary Material, which tailor standard regularity conditions from sequential inference to our continuous time, irregularly sampled setting.

In particular, Assumption~\ref{asp:sampling}  specifies the two grid sampling design on which the asymptotic analysis rests: a densifying state grid that drives the first stage spline accuracy and a bounded spacing outcome grid over which the $\sqrt{N_Y}$-normalized estimating equation is summed, so refining the state resolution never degrades the dependence structure of the central limit sum. Assumption~\ref{asp:mixing} imposes primitive conditions on the treatment mechanism and dynamics under which the joint state--treatment process is proved to be strictly stationary and exponentially $\beta$-mixing (Proposition~\ref{prop:behavior} in the Supplementary Material), which also yields the positivity of Assumption~\ref{asp:positivity} at treatment decision times; positivity at general observation times remains a separate assumption.  Assumptions~\ref{asp:basisfunction} to~\ref{asp:moments} impose standard sieve approximation accuracy, value equation error regularity, and moment conditions that yield well conditioned normal equations and a central limit theorem for the score under irregular sampling. Because the per subject scores are serially dependent, the limit theorem invoked in the proof of Theorem~\ref{thm:normal} is a triangular array central limit theorem for $\beta$-mixing sequences (with independence across subjects and exponential mixing within subject, Assumptions~\ref{asp:sampling}--\ref{asp:mixing}), and the serial dependence is absorbed into a long run (HAC type) covariance rather than treated as a martingale difference sequence.

Crucially, these assumptions implicitly require that the first step spline estimation error is negligible. Specifically, the asymptotic normality result holds provided the spline estimator satisfies
\begin{equation}
\label{eq:spline_rate}
\sqrt{N_Y}\left(\|\hat{S} - S\|_{\infty}+\|\hat{D}-D\|_{\infty}\right)=o_p(1),
\end{equation}
where $N_Y$ is the total number of outcome observation times entering the second step estimating equation, as defined in Section~\ref{sec:methodology}.  Lemma~\ref{lem:spline_rate} in the Supplementary Material establishes~\eqref{eq:spline_rate} from primitive conditions: piecewise H\"older smoothness of the trajectory between treatment changes, a mildly undersmoothed spline with repeated knots at the change times, and sufficiently dense state sampling relative to the outcome count. 

\begin{theorem}
\label{thm:normal}
    Under the ODE assumption \eqref{eq:ode}, Assumptions~\ref{asp:CA}--\ref{asp:moments} together with the first stage spline rate condition~\eqref{eq:spline_rate}, which Lemma~\ref{lem:spline_rate} in the Supplementary Material establishes from the primitive conditions (P1)--(P4) stated there; when a hard threshold policy is among the targets, additionally the margin condition of Assumption~\ref{asp:margin} in the Supplementary Material, we have
$$\sqrt{N_Y} (\widehat{\tau}-\tau) / \widehat{\sigma}\xrightarrow{d} N(0,1).$$
\end{theorem}

Theorem~\ref{thm:normal} can be read as an oracle linearization plus first step negligibility. The oracle linearization is the standard sieve Z-estimator expansion that would hold if the latent state and its derivative were observed exactly; the first step negligibility, controlled by the spline rate condition \eqref{eq:spline_rate}, ensures that the error of spline recovery contributes $o_p((N_Y)^{-1/2})$ to the studentized statistic. Together they yield a $\sqrt{N_Y}$-asymptotically valid studentized test with a standard $N(0,1)$ limit, despite the latent state being observed indirectly and at irregular times. The Supplementary Material formalizes both steps and provides the long run covariance estimator used for studentization.  Using this result, we test \(H_0:\tau\le 0\) versus \(H_a:\tau>0\) (or \(H_0:\tau=0\) versus \(H_a:\tau\neq0\)) with the statistic \(Z=\sqrt{N_Y}\,\widehat{\tau}/\widehat{\sigma}\), rejecting at level \(\alpha\) if \(Z>z_\alpha\) (or \(|Z|>z_{\alpha/2}\) for the two sided test), where \(z_\alpha\) is the upper \(\alpha\)-quantile of the standard normal distribution.

\subsection{Discretization bias of discrete-time estimators}
\label{sec:discretization}

The simulations and the OhioT1DM analysis repeatedly show discrete-time value learning (SAVE) losing power as the observation grid refines under measurement noise; the theorem below explains this behavior and formalizes the advantage of the proposed estimator. Discrete-time methods replace the generator $\mathcal{L}_a V(S_t)=\langle\nabla V,b\rangle$ by a one step difference of the value function across successive grid points and evaluate it at the raw noisy observations $X_t=S_t+\epsilon_t$ rather than the latent state. Fix an action $a$ (equivalently, work on the agreement subsample of a policy $\pi$) and a grid spacing $\Delta$, and write the discrete-time population moment at a value function $V$ and weight $\phi$ as
\begin{equation}
\label{eq:discrete_moment}
m_\Delta(V;\phi)=\mathbb{E}\!\left[\Big\{Y_t+\tfrac{1}{\Delta}\big(\gamma^\Delta V(X_{t+\Delta})-V(X_t)\big)\Big\}\,\phi(X_t)\,\mathbb{I}^\Delta_a\right],
\end{equation}
which is the continuous-time moment~\eqref{eq:th1} with the generator replaced by a finite difference and the latent state replaced by its noisy measurement. Here $\mathbb{I}^\Delta_a:=\mathbb{I}\{A_t=a\}\,\mathbb{I}\{\text{no treatment decision time in }(t,t+\Delta]\}$ retains the transition pairs over which the action is held at $a$, the standard practice for discrete time value learning and the most favorable evaluation of the discrete moment: under the refresh mechanism of Assumption~\ref{asp:mixing} an intervening switch would contribute an additional $O(\Delta)$ term, and because the decision times are Poisson and independent of everything else, the no switch requirement is an independent thinning that multiplies the moment by $e^{-\lambda\Delta}=1+O(\Delta)$, leaving the displayed leading terms unchanged. Equation~\eqref{eq:discrete_moment} is the \emph{time adapted} discrete value moment: the per step discount $\gamma^\Delta$ and the $1/\Delta$ scaling calibrate it so that its population target converges to the continuous time value function in the noise free limit as $\Delta\to0$.  The theorem below therefore isolates the effect of discretization and measurement noise on the \emph{best calibrated} member of the discrete time class. A standard discrete time value learner run with a fixed per step discount $\gamma$ at every grid point (the way SAVE is implemented in Section~\ref{sec:simulation}) faces an additional, more elementary difficulty: its implied discount per unit time is $\gamma^{1/\Delta}$, so its estimand itself changes with the observation grid, in line with the design dependence discussed in Section~\ref{sec:introduction}. The implemented SAVE therefore shares the finite difference structure that drives the noise amplification term and carries the grid dependent target on top of it. Let $V^*=V_a$ be the true value function solving the continuous-time moment.

\begin{theorem}[Discretization and noise-amplification bias]
\label{thm:discretization}
Suppose $V^*\in C^4(\mathcal{S})$ and $\phi\in C^2(\mathcal{S})$ have bounded derivatives, the drift $b(\cdot,a)$ is $C^2$, the sieve represents the value function exactly, $V^*=\Psi^\top\beta^*$ for some $\beta^*$, and the measurement errors scale as $\epsilon_t=\sigma_\epsilon Z_t$, where the $Z_t$ are independent across $t$, mean zero and symmetric with $\mathrm{Cov}(Z_t)=I_d$ and uniformly bounded fourth moments (an arbitrary covariance $\Sigma_Z$ only rescales the displayed noise functional). Then, as $\Delta\to0$ and $\sigma_\epsilon\to0$,
\[
m_\Delta(V^*;\phi)=\tfrac{\Delta}{2}\,\mathbb{E}\!\big[(\log\gamma+\mathcal{L}_a)^2V^*(S_t)\,\phi(S_t)\,\mathbb{I}\{A_t{=}a\}\big]-\tfrac{\sigma_\epsilon^2}{\Delta}\,\mathbb{E}\!\big[\nabla V^*(S_t)^\top\nabla\phi(S_t)\,\mathbb{I}\{A_t{=}a\}\big]+R,
\]
with remainder $R=O(\Delta^2+\sigma_\epsilon^2+\sigma_\epsilon^4/\Delta)$. Consequently, provided the two leading functionals are not collinear across the sieve $\{\Psi_j\}$, the population solution $\bar\beta_\Delta$ of the discrete time moment satisfies $\big\|\bar\beta_\Delta-\beta^*\big\|=\Theta\!\big(\Delta+\sigma_\epsilon^2/\Delta\big)$, minimized over the grid spacing at $\Delta^\star\asymp\sigma_\epsilon$, where the bias is $\Theta(\sigma_\epsilon)$; thus, to the order of the expansion, no choice of grid spacing reduces the bias below the order of the measurement noise.
\end{theorem}

Three consequences follow. First, the bias is a genuine bias--bias tradeoff in the grid: coarsening $\Delta$ inflates the $O(\Delta)$ Euler discretization error, in which the finite difference misses the second order term $\tfrac12(\log\gamma+\mathcal{L}_a)^2V^*$; refining $\Delta$ inflates the $O(\sigma_\epsilon^2/\Delta)$ noise-amplification error, because differencing two noisy states divides the measurement noise by $\Delta$. The optimum $\Delta^\star\asymp\sigma_\epsilon$ leaves an irreducible $\Theta(\sigma_\epsilon)$ bias that no choice of grid removes; these results are consistent with the combined grid dependence and noise amplification mechanisms behind SAVE's degradation under grid refinement in Section~\ref{sec:simulation}. Second, the proposed estimator escapes the tradeoff: it applies the generator to the spline-smoothed trajectory $\widehat{S}$, whose sup-norm error is $o_p((N_Y)^{-1/2})$ in the multi-resolution regime (Lemma~\ref{lem:spline_rate}), so the noise is removed before differentiation and the bias vanishes as $N_Y\to\infty$ for any fixed $\sigma_\epsilon$. Third, the theorem characterizes the estimators that difference the \emph{raw} noisy states, the class that includes SAVE as implemented, and identifies the mechanism, noise amplification through finite differencing, that governs that class; a discrete time method that smooths the state sequence before differencing escapes the $\sigma_\epsilon^2/\Delta$ term, in effect reproducing the first stage of the proposed procedure.  The proof is given in Supplement~\ref{sec:proof-discretization}.

\section{Numerical Study}
\label{sec:simulation}

In this section, we conduct simulations to investigate the empirical performance of the continuous time cumulative treatment effect test and compare it with existing methods under different settings. Specifically, we compare the proposed method with three competing methods; namely, the t test \citep{student1908probable}, the double machine learning (DML) test \citep{chernozhukov2017double}, and the SequentiAl Value Evaluation (SAVE) method \citep{shi2022statistical}.

The t test is applied to the outcome observations $Y_{it}$ for two groups defined by $A_{it} = 1$ and $A_{it} = 0$. This test, which assumes unequal variances, is implemented by the Python package \texttt{scipy.stats}. The double machine learning method utilizes the state observations to estimate the effect of treatment, which is implemented by the Python package \texttt{doubleml}. For the SequentiAl Value Evaluation (SAVE), a method designed for a time varying setting, we apply its discrete time Bellman regression at the time points where both state and outcome are observed. Implementation details of all four methods (nuisance learners and folds, bases, discounts, variance estimators, and tuning selection rules) are collected in Supplement Section~\ref{sec:repro}.

The methods do not all target the same quantity, and the comparison should be read accordingly. The proposed test targets the continuous time cumulative value contrast $\tau$ of~\eqref{eq:ate_policy}. SAVE is a \emph{related discrete time benchmark}: it solves a discrete time value equation on the observed grid via one step finite differences of the noisy state with a fixed per step discount, so its estimand is the discrete time cumulative value on that grid and is grid dependent (its implied discount per unit time is $\gamma^{1/\Delta}$, which degenerates as $\Delta\to0$; see Section~\ref{sec:theory}.3), so its target varies with the grid and departs from $\tau$ as $\Delta\to0$; the time adapted calibration with per step discount $\gamma^\Delta$, the most favorable member of the discrete time class analyzed in Theorem~\ref{thm:discretization}, targets $\tau$ in the fine grid, noise free limit; the t test and DML instead target \emph{contemporaneous} contrasts: a marginal mean difference in $Y_{it}$ across the two action groups, and the coefficient of $A_{it}$ in a partially linear regression on the current state, respectively. The t test and DML rows therefore quantify what a contemporaneous analysis misses when the signal accrues through the latent dynamics, while SAVE is the closest available benchmark within the discrete time value equation class, and its rows quantify the combined effect of discretization, noise amplification, and the grid dependence of the discrete time target.

\subsection[Univariate simulations]{Univariate simulations}
\begin{figure}[!h]
    \centering
    \includegraphics[width=0.8\linewidth]{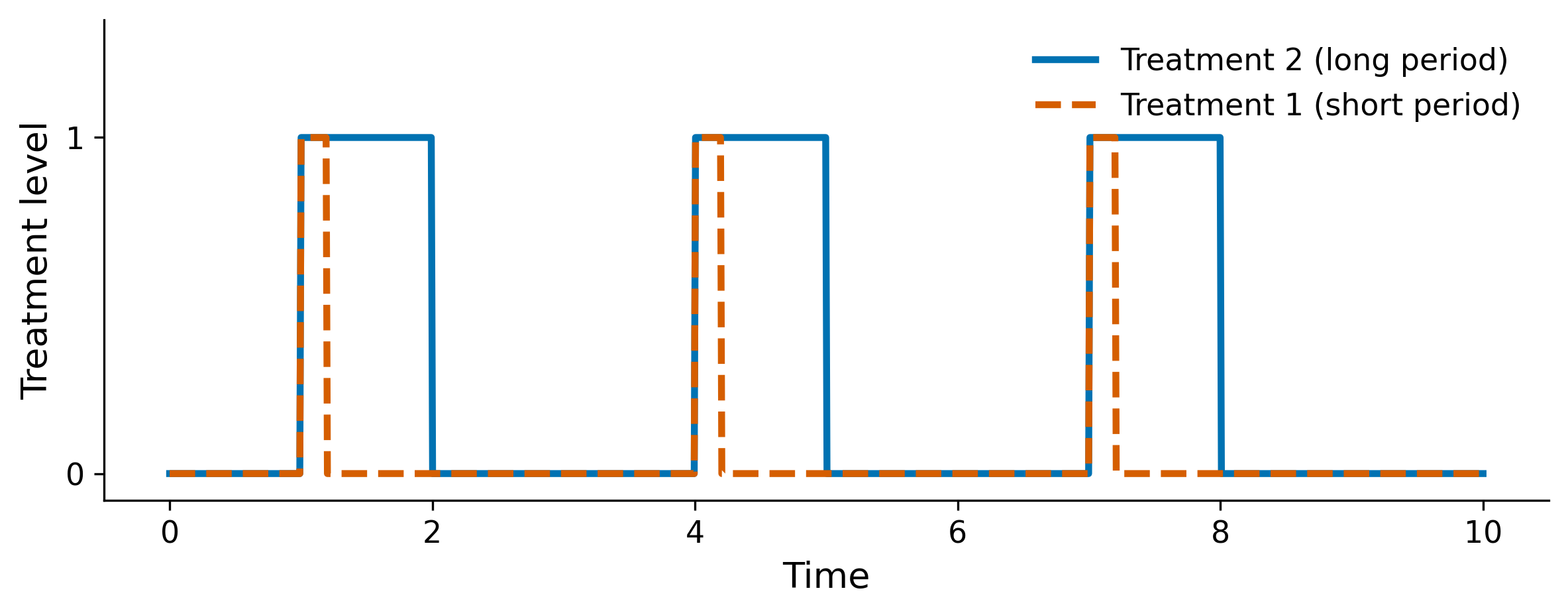}
    \caption{Treatment schedules used in the one dimensional simulation: Treatment 1 (short period; orange, dashed) and Treatment 2 (long period; blue, solid).}
    \label{fig:treat}
\end{figure}

Throughout this section, the first set of simulation studies evaluates the leading special case of the policy contrast, the always treat versus never treat comparison ($\pi_1\equiv1$, $\pi_2\equiv0$), whose estimand is the average treatment effect $\tau=\int\{V_1-V_0\}\,d\mathbb{G}$; the policy contrast simulation later in this section evaluates general threshold policies. We evaluated the effectiveness of the proposed method against the t-test, the double machine learning test, and the SequentiAl Value Evaluation (SAVE) in two simulation settings. As shown in Figure~\ref{fig:treat}, the short treatment period resembles bolus insulin therapy, where rapid changes are expected, while the long treatment period mimics basal insulin, where the treatment effect accumulates slowly over time. The data generating process uses the following deterministic ODE with additive observation noise:

\begin{equation}
    \begin{aligned}
    dS_t &= (-0.2 S_t + \delta A_t)\, dt \\
    X_t &= S_t + \varepsilon^X_t,\qquad Y_t = S_t + \varepsilon^Y_t
\end{aligned}
\label{sim0}
\end{equation}

with $\varepsilon^X_t,\varepsilon^Y_t \overset{\text{iid}}{\sim} N(0,\epsilon^2)$, initial state $S_{i,0}\sim N(0, 0.5^2)$ drawn independently across subjects, and time step $0.01$. The drift $b(S_t, A_t) = -0.2 S_t + \delta A_t$ is fully deterministic; the noise level $\epsilon$ is the observation noise scale appearing in equation~\eqref{eq:obs}.

The treatment effect $\delta$ was set to 0.3 unless otherwise specified. We sampled $t$ from 0 to 10: $Y_t$ is recorded every 0.2 and the noisy state on the full simulation mesh (step 0.01), a twenty to one state to outcome ratio for the spline stage; SAVE, the t test, and DML use the time points where both are recorded, and the supplementary sweep (Table~\ref{tb:sim2}) studies sparser state grids with $n_s\in\{50,100,200\}$. We simulate $I=50$ independent subjects per replicate. For the value function approximation, we used polynomial basis functions of degree~3 in the univariate case and degree~2 in the multivariate case. The spline smoothing for the state trajectory used cubic B-splines with a smoothing parameter set adaptively from a Rice style noise variance estimate, $\lambda = n_s\,\widehat{\sigma}^2_\varepsilon$. The discount factor was set to $\gamma = 0.8$.

The null hypothesis is that the new treatment ($A_{it} = 1$) is not better than the old treatment ($A_{it} = 0$), while the alternative hypothesis is that the new treatment is better. The performance of these tests is measured by their empirical power. We computed each entry as the rejection rate over 200 repetitions at significance level $0.05$, and replicated this across 50 Monte Carlo runs to obtain the standard deviation reported in parentheses. 

\begin{table}[!h]
\centering

\begin{tabular}{c|c|c|cccc}
\hline\hline
               & $\epsilon$ & $\delta$
               & t-test        & DML-test        & SAVE            & Proposed \\
\hline
\multirow{4}{*}{Treatment 1}
  & \multirow{2}{*}{0.1} & 0.3 % Power
      & 0.000 (0.000) & 0.046 (0.013) & 0.840 (0.026) & \textbf{0.995 (0.005)} \\
  &                      & 0.0 % Type-I
      & 0.000 (0.000) & 0.048 (0.015) & 0.030 (0.012) & 0.020 (0.010) \\ \cline{2-7}
  & \multirow{2}{*}{0.3} & 0.3
      & 0.020 (0.010) & 0.037 (0.015) & 0.030 (0.012) & \textbf{0.815 (0.027)} \\
  &                      & 0.0
      & 0.020 (0.010) & 0.035 (0.013) & 0.025 (0.011) & 0.065 (0.017) \\
\hline
\multirow{4}{*}{Treatment 2}
  & \multirow{2}{*}{0.1} & 0.3
      & 0.000 (0.000) & 0.045 (0.014) & \textbf{1.000 (0.000)} & 0.995 (0.005) \\
  &                      & 0.0
      & 0.000 (0.000) & 0.048 (0.016) & 0.000 (0.000) & 0.025 (0.011) \\ \cline{2-7}
  & \multirow{2}{*}{0.3} & 0.3
      & 0.045 (0.015) & 0.035 (0.014) & 0.475 (0.035) & \textbf{0.895 (0.022)} \\
  &                      & 0.0
      & 0.020 (0.010) & 0.034 (0.012) & 0.005 (0.005) & 0.030 (0.012) \\
\hline\hline
\end{tabular}
\caption{Test power and Type-I error of competing methods under different treatment assignments and observation noise levels $\epsilon$.}
\label{tb:sim1}
\end{table}

Table~\ref{tb:sim1} compares the power of the methods across two treatment durations, short‑term (Treatment~1) and long‑term (Treatment~2), under low and high observation noise levels.  The proposed test leads or is essentially tied with SAVE in every alternative cell, yet it still controls Type‑I error at or near the nominal $0.05$ level.

The \textit{t}-test's assumption of independent observations prevents it from exploiting temporal dependence, leaving it essentially powerless against the cumulative carryover signal. The double machine learning test, although it incorporates flexible random‑forest learners through the partial linear regression of \citet{chernozhukov2017double}, also struggles to capture the continuous‑time temporal dependence under sparse outcome observations: across every $(\delta,\epsilon)$ combination its rejection rate remains within Monte Carlo noise of the nominal $0.05$ level (between $0.035$ and $0.046$ when $\delta=0.3$), barely separating power from Type-I error.  The SequentiAl Value Evaluation (SAVE) fares better, particularly for long treatment periods with low noise. However, its power declines sharply when noise increases or the treatment duration shortens, reflecting its limited ability to model fast‑changing dynamics observed under measurement error. The mechanism is direct: SAVE's discrete time Bellman residual is built from one step finite differences of the observed state, and finite differences amplify additive observation noise; under $\epsilon_{obs}=0.3$ this amplified residual produces variance inflation that worsens as the observation grid refines, since finer differencing amplifies the variance of the discrete Bellman residual, precisely the finite difference noise amplification formalized in Theorem~\ref{thm:discretization}. Spline smoothing avoids this amplification by recovering the smooth latent trajectory before the estimating equation is evaluated, which is why the proposed method retains power where SAVE collapses.

In contrast, the proposed method maintains strong performance across all scenarios.   In the most challenging short‑term, high‑noise setting the proposed test retains power $0.815$, roughly $27$ times the SAVE power of $0.030$.  When the long treatment duration is combined with high observation noise, the proposed method delivers power $0.895$ versus SAVE's $0.475$.  These observations exhibit a mild Type-I inflation, reaching $0.065$ in the short term high noise configuration of Table~\ref{tb:sim1} and comparable levels in Tables~\ref{tb:sim2} and~\ref{tb:sim3}. This inflation reflects finite sample bias of the sandwich variance estimator when the spline smoothed derivative is itself noisy: the empirical score covariance underestimates the true long run variance, leading to slightly anti conservative studentization. A consistent HAC-type long run covariance estimator in Algorithm~\ref{alg:test} is theoretically appropriate under stronger serial dependence and may improve finite sample calibration.

\subsection[Multivariate simulations]{Multivariate simulations}
To evaluate the proposed approach beyond the one dimensional case, we run three two dimensional latent state scenarios with progressively richer structure: (Sim1) independent two dimensional drift, (Sim2) two dimensional drift with cross state coupling, and (Sim3) two dimensional drift with treatment dependent dynamics that can reverse the state trend under intervention. Multi dimensional state spaces are common in real world applications, such as monitoring multiple physiological markers or financial assets, where interactions between variables must be accurately captured. The latent state \( S_t \) was generated by the deterministic drifts specified in~(\ref{sim1})--(\ref{sim3}), with observations $X_{j,t}=S_{j,t}+\varepsilon^X_{j,t}$ and $Y_t=(S_{1,t}+S_{2,t})/2 + \varepsilon^Y_t$ contaminated by additive noise in all three scenarios. The sampling design is unchanged ($Y_t$ every $0.2$, the noisy state on the $0.01$ mesh, $\epsilon_{obs}=0.1$); Table~\ref{tb:sim2} reports sparser state grid versions of Sim1, where the proposed method's power is accordingly lower.

\texttt{Sim1}:
\begin{equation}
dS_{1,t} = (-0.1 S_{1,t} + \delta A_t)\, dt, \qquad
dS_{2,t} = (-0.3 S_{2,t} + 0.5\delta A_t)\, dt
\label{sim1}
\end{equation}

Sim2 (\ref{sim2}) introduces cross state coupling, and Sim3 (\ref{sim3}) adds treatment dependent dynamics on top of that coupling. In Sim2, the two state processes are positively correlated, reflecting scenarios such as blood pressure and heart rate, where physiological measures are interconnected.  In Sim3, the treatment effect reverses the trend of the state processes, simulating cases where treatment interventions not only stabilize but actively reverse the progression of a condition. 

\texttt{Sim2}:
\begin{equation}
    \begin{aligned}
    dS_{1,t} &= (-0.1 S_{1,t} +0.2 S_{2,t} + \delta A_t)\, dt \\
    dS_{2,t} &= (0.1 S_{1,t}- 0.3 S_{2,t} + 0.5\delta A_t)\, dt
\end{aligned}
\label{sim2}
\end{equation}

\texttt{Sim3}:
\begin{equation}
    \begin{aligned}
    dS_{1,t} &= \{(2A_t - 1)(-0.1 S_{1,t} + 0.2 S_{2,t}) + \delta A_t\}\, dt \\
    dS_{2,t} &= \{(2A_t - 1)(0.1 S_{1,t}- 0.3 S_{2,t}) + 0.5\delta A_t\}\, dt
\end{aligned}
\label{sim3}
\end{equation}

Table \ref{tb:sim4} shows that the proposed method has a substantially higher rejection probability than the competitors for the cumulative alternative it is designed to detect, in every scenario; as discussed above, the t test and DML target contemporaneous null hypotheses that are nearly true in these designs, so their low rejection rates reflect a different inferential target rather than a defect of those methods. In Sim1, the proposed method achieves a power of $0.994$, far above the SequentiAl Value Evaluation (SAVE)'s $0.212$. In Sim2, where the two state processes are correlated, the proposed method again attains $0.995$ versus SAVE's $0.224$. In Sim3, despite the trend reversing dynamics, the proposed method retains a power of $0.998$, compared with SAVE's $0.104$. This advantage stems from the multi resolution, continuous time formulation, which jointly models temporal dependencies and cross state interactions and thus captures even subtle treatment effects in multi dimensional state processes. 

\begin{table}[!h]
\centering

\begin{tabular}{c|cccc}
\hline
\hline
   & t-test & DML-test & SAVE  & Proposed  \\
\hline
Sim1 & 0.001 (0.002) & 0.039 (0.014)   &   0.212 (0.026)  &  \textbf{0.994 (0.005)} \\
Sim2 & 0.000 (0.000) & 0.042 (0.015)   &   0.224 (0.027)  &  \textbf{0.995 (0.005)} \\
Sim3 & 0.000 (0.000) & 0.019 (0.009)   &   0.104 (0.023)  &  \textbf{0.998 (0.003)} \\
\hline
\hline
\end{tabular}
 \caption[Power of competing methods across multivariate settings at treatment effect $\delta=0.05$.]{Power of competing methods across multivariate settings at treatment effect $\delta=0.05$.}
 \label{tb:sim4}
\end{table}

Additional simulations in the Supplementary Material (Tables~\ref{tb:sim2} and~\ref{tb:sim3}) report the proposed method's power on Sim1 as the sample size and the treatment effect vary.  The proposed method shows higher power than the SequentiAl Value Evaluation (SAVE) for all but the smallest sample size and across the full $\delta$ range tested, while the t-test and DML-test remain near the nominal $0.05$ level throughout.  A targeted regular grid comparison (Table~\ref{tb:regular_grid}) confirms that the proposed method matches SAVE at near perfect power under clean observations ($\epsilon_{obs}=0.1$) and shows a three to six fold power advantage under realistic noise ($\epsilon_{obs}=0.3$), while DML remains within Monte Carlo noise of the nominal $0.05$ level.

\subsection[Policy contrasts and robustness]{Policy contrasts and robustness}
To confirm that inference for the general policy contrast $\tau(\pi_1,\pi_2)$ of Section~\ref{sec:methodology} is as valid as for the binary special case, we simulate the 1D ODE of~\eqref{sim0} under a piecewise constant assignment mechanism assigning each action with probability $1/2$ over blocks of one time unit, so that overlap (Assumption~\ref{asp:positivity}) holds for every threshold policy. We test, by the proposed method with the action indicator replaced by the policy agreement indicator $\mathbb{I}\{A_{it}=\pi_k(S_{it})\}$, four contrasts: always treat versus never treat ($\pi_1\equiv1$ vs $\pi_2\equiv0$), two threshold policies $\pi(s)=\mathbb{I}\{s>c\}$ versus never treat ($c\in\{0,0.8\}$ on the standardized state), and the two threshold policies against each other ($c=0$ vs $c=0.8$). The sieve is the degree $3$ polynomial basis without hinge terms, so this validation is likewise exploratory relative to the hinge adapted guarantee. Table~\ref{tb:policy_sim} reports rejection rates over $30$ Monte Carlo runs of $100$ replications each at $\delta=0.3$ (power) and $\delta=0$ (Type-I), for observation noise $\epsilon_{obs}\in\{0.1,0.3\}$.

\begin{table}[!h]
\centering
\begin{tabular}{l|cc|cc}
\hline\hline
 & \multicolumn{2}{c|}{Power ($\delta=0.3$)} & \multicolumn{2}{c}{Type-I ($\delta=0$)} \\
Policy contrast & $\epsilon_{obs}{=}0.1$ & $0.3$ & $\epsilon_{obs}{=}0.1$ & $0.3$ \\
\hline
always vs.\ never                 & $0.994\ (.008)$ & $0.959\ (.019)$ & $0.023\ (.015)$ & $0.021\ (.013)$ \\
$\pi_{c=0}$ vs.\ never            & $0.992\ (.008)$ & $0.950\ (.026)$ & $0.029\ (.015)$ & $0.023\ (.015)$ \\
$\pi_{c=0.8}$ vs.\ never          & $0.947\ (.024)$ & $0.790\ (.045)$ & $0.037\ (.019)$ & $0.043\ (.016)$ \\
$\pi_{c=0}$ vs.\ $\pi_{c=0.8}$    & $0.981\ (.015)$ & $0.851\ (.031)$ & $0.019\ (.013)$ & $0.017\ (.013)$ \\
\hline\hline
\end{tabular}
\caption[Policy contrast estimand: rejection rates of the proposed test.]{Policy contrast estimand: rejection rates of the proposed test (standard deviation across Monte Carlo runs in parentheses). $\delta=0.3$ columns are power; $\delta=0$ columns are Type-I error. The always treat versus never treat row is the binary special case.}
\label{tb:policy_sim}
\end{table}

The proposed test attains high power for every policy contrast and controls Type-I error at or below the nominal $0.05$ level throughout, confirming that the policy contrast estimand is estimated and tested with the same validity as the binary contrast.  This validates the substitution described in Section~\ref{sec:methodology}.2, of which the always treat versus never treat test is the special case.

Finally, to assess robustness to violations of the deterministic ODE assumption, we run a sensitivity simulation in which the latent state evolves according to a stochastic differential equation $dS_t = b(S_t, A_t)\,dt + \sigma\,dW_t$ with diffusion levels $\sigma \in \{0.1, 0.3\}$, while the proposed ODE method is applied as is (Supplement~\ref{sec:sensitivity-sde}, Table~\ref{tb:sim_sde}). The setting reuses the short period and long period treatment schedules and the observation grid of the main 1D simulation.  The proposed method retains substantially higher power than SAVE at both diffusion levels, and empirical rejection under the null is near nominal in all eight scenarios examined (Table~\ref{tb:sim_sde}). These results indicate that spline smoothing recovers the drift driven component of the latent trajectory even when a nontrivial Brownian diffusion is present, demonstrating finite sample robustness in the simulated settings and supporting the ODE specification as a working model rather than a strictly deterministic commitment. We discuss SDE extensions as future work in Section~\ref{sec:discussion}.

\section{Real Data Application}
\label{sec:realdata}
\begin{figure}[!h]
    \centering
    \makebox[\linewidth][c]{\includegraphics[width=0.53\linewidth]{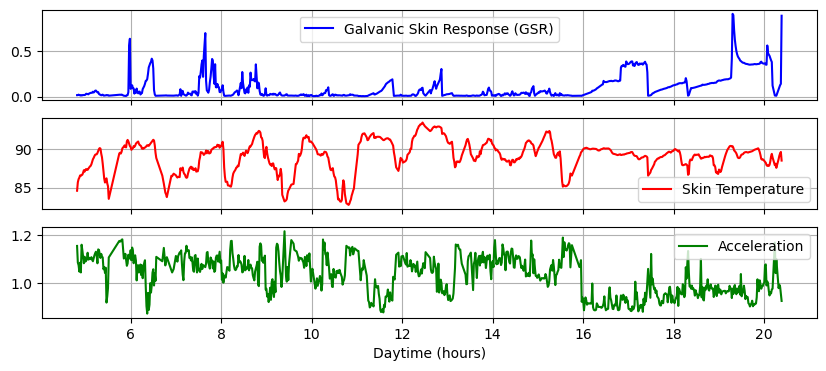}\hspace{0.005\linewidth}\includegraphics[width=0.53\linewidth]{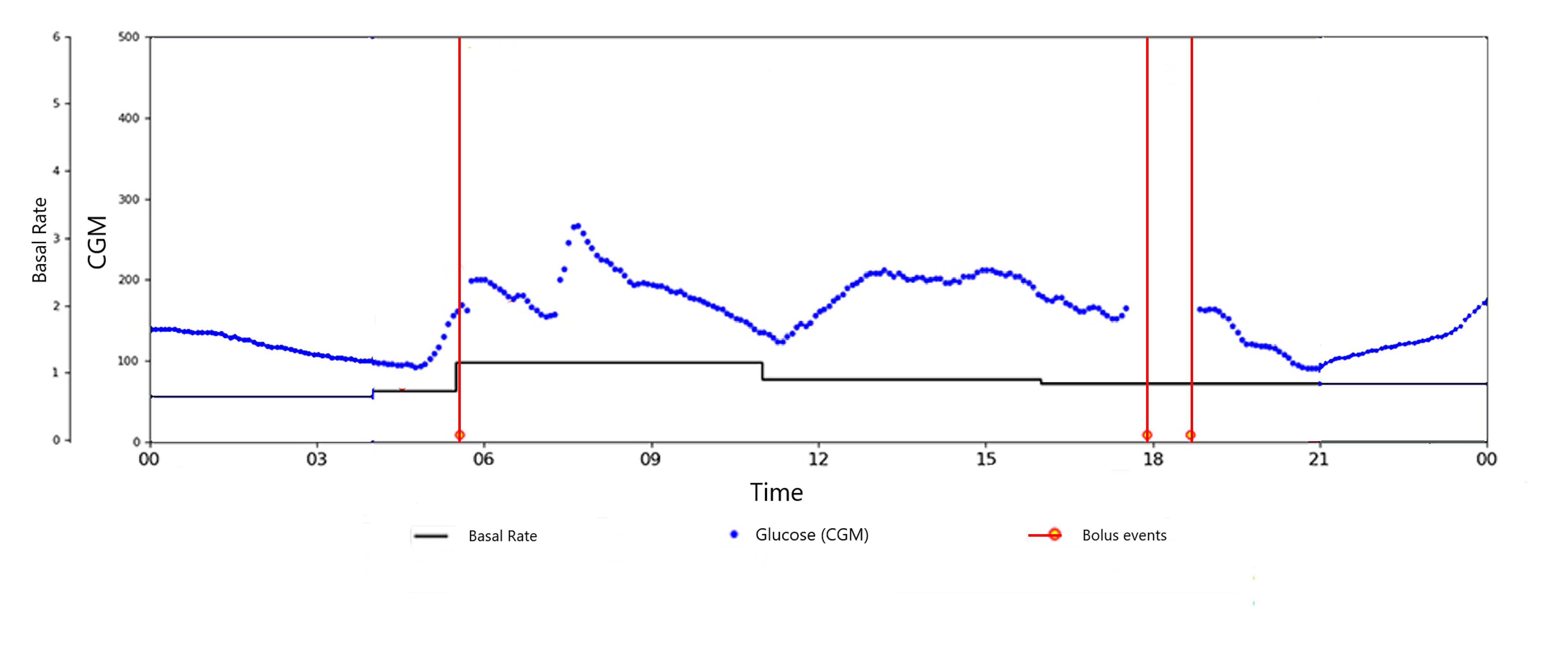}}
    \caption{OhioT1DM multi resolution structure for one patient over 24 hours. Left: state channels, with physiological signals at 1 minute resolution and CGM glucose every 5 minutes. Right: CGM outcome and irregularly timed basal and bolus insulin events.}
    \label{fig:state}
\end{figure}

We apply the proposed method to the OhioT1DM dataset \citep{marling2020ohiot1dm}, a publicly available dataset designed to closely monitor blood glucose. It provides eight weeks of data for twelve patients with type 1 diabetes on insulin pump therapy, in two cohorts released in 2018 and 2020, including continuous glucose monitoring (CGM) every 5 minutes, basal and bolus insulin doses, self reported life events, and physiological signals, namely galvanic skin response (GSR), skin temperature, and acceleration, at 1 minute resolution for the 2020 cohort. We focus on the 2020 cohort, whose fine grained physiological data combined with the 5 minute CGM produce the multi resolution structure the method is designed for; Figure~\ref{fig:state} shows a representative 24 hour window of the state channels (left) together with the outcome process and insulin events (right). The scientific question is whether glucose exhibits a cumulative response to insulin dosing, beyond what contemporaneous comparisons reveal: insulin acts over a multi hour window, so its effect accrues continuously between sparse dosing events. The analysis yields three findings: a direction stable cumulative association between bolus dosing and glucose, whose positive sign reflects the timing of dosing rather than a glucose raising effect of insulin; conclusions governed by the level of uncertainty quantification rather than the choice of method; and inconclusive basal threshold contrasts for lack of overlap.

All analyses share one pipeline. The state is the four dimensional augmented state $(\mathrm{glucose},\mathrm{GSR},\mathrm{skin\ temperature},\mathrm{acceleration})$, spline recovered from the native grids and evaluated at the CGM times; CGM glucose is the outcome, appearing also as a state coordinate as in Section~\ref{sec:simulation}. Excluding days with inadequate coverage leaves $188$ patient days from the six patients of the 2020 cohort. Days from the same patient are serially dependent, so the patient is the independent replication unit: the long run covariance is a patient clustered sum of scores, all $p$ values and $95\%$ confidence intervals use the $t_{5}$ reference, every estimate is reported with its leave one patient out (LOPO) range, and with six patients the application is exploratory. Time is in hours with $\gamma=0.8$ (effective horizon $1/|\log\gamma|\approx4.5$ hours), so the estimates are discounted cumulative quantities on the standardized glucose $\times$ hours scale, not instantaneous effect sizes. The remaining implementation choices are fixed across analyses and collected in Supplement Section~\ref{sec:realdata-implementation}. SAVE is excluded: with insulin events at a few time points per day, its discrete time regression is dominated by untreated epochs.

\subsection{Bolus insulin: a descriptive cumulative timing association}
\label{sec:realdata-bolus}

For bolus insulin we set $A_{it}=1$ when a bolus dose was administered since the preceding CGM reading. Boluses accompany meals and hyperglycemia \citep{marling2020ohiot1dm}, and meals are substantially under reported, so the regime depends on triggers the recorded state only partially captures; we therefore evaluate it descriptively, as a cumulative timing association with never dosing (design diagnostics in Supplement Section~\ref{sec:realdata-implementation}). The comparison adds a patient clustered OLS baseline to the t test and DML of Section~\ref{sec:simulation}: the same contemporaneous mean difference as the t test, studentized with patient clustered standard errors. The competitors target contemporaneous estimands while the proposed statistic targets the cumulative value contrast, so the rows of Table~\ref{tb:realdata} are complementary, not competing, summaries.

\begin{table}[!h]
\centering
\begin{tabular}{l|c|c}
\hline\hline
Method & Estimate [$95\%$ CI] & LOPO range \\
\hline
t test (iid variance)        & $+0.21$ [$+0.14,+0.28$] & $[+0.12,+0.29]$ \\
Patient clustered OLS         & $+0.21$ [$-0.11,+0.53$] & $[+0.12,+0.29]$ \\
DML-test                      & $+0.02$ [$-0.01,+0.05$] & $[+0.01,+0.03]$ \\
Proposed                      & $+0.17$ [$-0.05,+0.38$] & $[+0.10,+0.22]$ \\
\hline\hline
\end{tabular}
\caption{Bolus insulin: observed dosing regime versus never dosing (descriptive contrast). Each method is evaluated once on the full data, standardized glucose scale. The t test and clustered OLS share the same contemporaneous estimate and differ only in the variance (iid versus patient clustered $t_{5}$); the proposed cumulative contrast uses the patient clustered variance.}
\label{tb:realdata}
\end{table}

Table~\ref{tb:realdata} evaluates every method once on the full data and carries the first two findings. First, the cumulative contrast between the observed dosing regime and never dosing is positive and stable in direction, and its sign is informative about timing rather than pharmacology. The proposed estimate is $\widehat\tau=+0.17$ with patient clustered $95\%$ confidence interval $[-0.05,+0.38]$, positive for every held out patient (LOPO range $[+0.10,+0.22]$) though not significant with six patients. Because dosing engages when meals and hyperglycemia perturb glucose, the regime accrues more discounted glucose than the calm undosed periods from which the never dose value is extrapolated: the positive sign reflects this timing, not a glucose raising effect of insulin. The contemporaneous methods sharpen this reading: the t test and the clustered OLS find the same positive level difference at dosing times, DML shows that difference is absorbed by conditioning on current glucose ($+0.02$ $[-0.01,+0.05]$), and the cumulative statistic differs in what it measures, the multi hour carryover the value function integrates, rather than in whether it finds an association. Second, the level of uncertainty quantification, not the choice of method, governs the conclusions. The t test and the clustered OLS share the identical contemporaneous estimate $+0.21$, yet the iid interval $[+0.14,+0.28]$ is less than a quarter the width of the patient clustered interval $[-0.11,+0.53]$, a gap that reflects the pseudoreplication of treating serially dependent five minute readings as independent; a day clustered variance would likewise inflate the proposed $z$ from $2.0$ to $6.4$. Every method's LOPO range keeps its sign, and no patient clustered method reaches significance; a hierarchical patient bootstrap interval is wider and likewise covers zero (Supplement Section~\ref{sec:realdata-implementation}). The bolus result is a directionally positive but statistically non decisive screening contrast, not a causal effect size or evidence that insulin raises glucose.

\subsection{Basal insulin: comparing two threshold dosing policies}
\label{sec:realdata-policy}

For basal insulin the clinically relevant question is how to adjust the dose rather than whether to dose, so we compare two \emph{exploratory binary relative elevation strategies}: threshold rules calling for basal delivery above the patient's own average when CGM glucose exceeds $c$, for $c\in\{150,180\}$ mg/dL (tight correction target versus standard hyperglycemia threshold). These are not fully executable dosing policies: the action records elevation above the personal reference without specifying the dose increment. The policy is a coordinate threshold on the augmented state, $\pi_c(\boldsymbol s)=\mathbb{I}\{s_{\mathrm{glucose}}>c\}$, so the physiological channels continue to provide the measured state adjustment (time of day, an important determinant of scheduled dosing, is not part of the state; see the limitations at the end of this section); each policy is compared against the other and against the no elevation policy $\pi\equiv0$; the full data estimates use all $188$ days, while the descriptive subsample rejection frequencies of Table~\ref{tb:realdata_policy} resample $50$ days per replicate. Because the value function basis (Supplement Section~\ref{sec:realdata-implementation}) contains no threshold hinge terms, these threshold results are exploratory relative to the hinge adapted sieve required by the formal guarantee of Section~\ref{sec:theory}. Action $A_{it}=1$ means the basal rate exceeds the patient's own average, so the comparator $\pi\equiv0$ is no elevation above the reference basal rate, not the withdrawal of basal insulin: the estimand contrasts elevated versus non elevated basal delivery around each patient's personalized reference. The reference is computed from the patient's full record; recomputing it from only the first half of each record leaves the induced actions unchanged for all six patients. Because the reference is patient specific, $A_{it}=1$ corresponds to different physical dose rates for different patients: the drift model $b(s,a)$ with a common action label across patients is a \emph{working approximation across personalized action scales}, pooling the physical treatment versions that share the label. Including the patient specific reference rate in the state (so that actions share a physical scale) is a natural refinement the present six patient sample cannot support.

The overlap diagnostic of Assumption~\ref{asp:positivity} is reported in marginal and stratified form: the marginal agreement rate is $\widehat{\Pr}\{A_{it}=\pi_c(S_{it})\}$, the fraction of all epochs at which the recorded action matches the policy, and the treat stratum agreement is $\widehat{\Pr}\{A_{it}=1\mid s_{\mathrm{glucose}}>c\}$, the fraction of epochs where the policy prescribes dosing at which an elevation was actually recorded. The distinction matters because positivity is a conditional requirement: a high marginal agreement rate can be driven entirely by the no treatment region, and Table~\ref{tb:realdata_overlap} in the Supplementary Material shows exactly this pattern, with marginal agreement of $0.48$ to $0.85$ against treat stratum agreement of $0.15$; that the treat stratum rate is the same at both thresholds, and essentially equals the marginal treated fraction, is the signature of scheduled rather than glucose triggered elevation. The stratified rate is a coarse summary of the conditional condition $\inf_s\Pr\{A_{it}=\pi(s)\mid S_{it}=s\}>0$, so we refine it with a model based agreement propensity fitted over all four state channels: its treat stratum quantiles approach zero in the lower tail, so conditional support comes close to failing in part of the treat region, which is why the basal contrasts below are reported as inconclusive (details in Supplement Section~\ref{sec:realdata-implementation}). 

\begin{table}[!h]
\centering
\begin{tabular}{l|c|c|c}
\hline\hline
Contrast & $\widehat\tau$ [$95\%$ CI] & LOPO range & subsample rej.\ freq. \\
\hline
$\pi_{150}$ vs.\ $\pi_{180}$  & $+0.31$ [$-7.91,+8.53$] & $[-3.15,+0.43]$ & $0.002\ (0.004)$ \\
$\pi_{150}$ vs.\ $\pi\equiv0$ & $+0.22$ [$-9.07,+9.51$] & $[-1.99,-0.02]$ & $0.025\ (0.019)$ \\
$\pi_{180}$ vs.\ $\pi\equiv0$ & $-0.09$ [$-1.54,+1.36$] & $[-0.92,+1.17]$ & $0.026\ (0.015)$ \\
\hline\hline
\end{tabular}
\caption{Basal threshold policies: contrast estimates on the full data with patient clustered $t_{5}$ intervals, leave one patient out (LOPO) ranges, and descriptive subsample rejection frequencies over $50$ day resamples; all estimates use the empirical reference distribution $\mathbb G$, and the per policy overlap diagnostics are collected in Table~\ref{tb:realdata_overlap} of Supplement Section~\ref{sec:realdata-implementation}.}
\label{tb:realdata_policy}
\end{table}

Third, the basal policy contrasts are inconclusive. No statistically detectable differences were found (Table~\ref{tb:realdata_policy}), but conditional support in the treatment region is weak (treat stratum agreement $0.15$), the confidence intervals are wide accordingly, hierarchical bootstrap intervals also cover zero with widths varying substantially (Supplement Section~\ref{sec:realdata-implementation}), and the LOPO estimates fluctuate in sign. The contrasts are therefore inconclusive rather than evidence of equivalence, because weak overlap implies weak information rather than a null effect. A sensitivity grid over time of day windows and thresholds (Supplement Section~\ref{sec:realdata-robustness}) accords with this reading.

Limitations include the Markov assumption on the recorded state, self reported annotations with imprecise timing, six patient exploratory inference, and a scheduled, time of day dependent treatment mechanism that satisfies the stationarity assumptions approximately at best; augmenting the state with time of day is a natural refinement, and multi cohort checks across the 2018 and 2020 releases are a natural extension.

\section{Discussion}
\label{sec:discussion}

We have developed a continuous time inferential framework providing a hypothesis test, a studentized test statistic, and explicit asymptotic guarantees for cumulative treatment effects under ODE state dynamics with noisy observations. The method defines the average treatment effect through continuous time value functions, identifies it via a generator based estimating equation, and provides valid inference through a two step procedure combining spline smoothing with sieve estimation. Our theoretical results establish asymptotic normality under explicit conditions, including a rate condition ensuring the first step spline error is asymptotically negligible.

The ODE assumption~\eqref{eq:ode} is a deliberate modeling choice that balances tractability with applicability. It is well suited for settings where the state dynamics are governed by smooth physiological or physical processes and stochasticity enters primarily through measurement noise and the treatment assignment mechanism, as is common in mobile health applications with wearable devices. When intrinsic state variability is large relative to measurement noise, a stochastic differential equation (SDE) model with a nontrivial diffusion term may be more appropriate. 

We highlight three potential directions for future research. The first is the development of an online testing framework, which would serve as a valuable tool for real time monitoring of treatment effects in ongoing clinical trials. The second is an extension of the test procedure to SDE settings where the state process has intrinsic stochastic dynamics. This would require consistent estimation of both drift and diffusion coefficients from noisy, irregularly sampled data, building on methods such as those in \citet{florens1993estimating} and \citet{gloter2000discrete}. The third direction is to extend the framework beyond the stationary policy contrasts developed in this paper to non stationary or adaptive treatment policies, broadening applicability to adaptive treatment strategies in precision medicine. 

We close with a concise statement of limitations. The ODE specification is a working model, and inference is conditional on it being a reasonable approximation to the dominant drift driven dynamics; under strongly stochastic latent dynamics the estimating equation targets a drift component projection rather than the full SDE value function. Identification rests on continuous time sequential randomization given the recorded state, so unmeasured time varying confounders not captured by the state vector would bias the procedure. Derivative recovery is sensitive to spline tuning and degrades as the state dimension grows, which constrains practical use to low to moderate dimensional state spaces with the cubic B-spline sieve we employ. Finally, our asymptotic theory assumes a stationary treatment assignment mechanism; extensions to non stationary or adaptive treatment policies are left to future work.

\section*{Disclosure statement}

The authors report no conflicts of interest.

\section*{Data availability statement}

The OhioT1DM data used in the real data application are publicly available through the cited source \citep{marling2020ohiot1dm}. Additional simulation details, assumptions, and proofs are described in the supplementary material.

\section*{Declaration of the use of generative AI and AI assisted technologies}

During the preparation of this work the authors used ChatGPT and Claude to assist with coding, checking theoretical proofs, and polishing the writing. After using these tools the authors reviewed and edited the content and take full responsibility for the content of the publication.

\phantomsection\label{supplementary-material}
\bigskip

\begin{center}
{\large\bf SUPPLEMENTARY MATERIAL}
\end{center}

\begin{description}
\item[Supplementary material:]
The supplementary materials provide more simulations, additional assumptions, and proofs of the theorems.
\end{description}

\setlength{\bibsep}{0.5pt plus 0.3pt}
\renewcommand{\bibfont}{\small}
\bibliographystyle{agsm}
\bibliography{reflist}

@article{gloter2000discrete,
  title={Discrete sampling of an integrated diffusion process and parameter estimation of the diffusion coefficient},
  author={Gloter, Arnaud},
  journal={ESAIM: Probability and Statistics},
  volume={4},
  pages={205--227},
  year={2000},
  publisher={EDP Sciences}
}

@article{florens1993estimating,
  title={On estimating the diffusion coefficient from discrete observations},
  author={Florens-Zmirou, Danielle},
  journal={Journal of applied probability},
  volume={30},
  number={4},
  pages={790--804},
  year={1993},
  publisher={Cambridge University Press}
}

@article{student1908probable,
  title={The probable error of a mean},
  author={Student},
  journal={Biometrika},
  volume={6},
  number={1},
  pages={1--25},
  year={1908},
  publisher={JSTOR}
}

@article{chernozhukov2017double,
  title={Double/debiased/neyman machine learning of treatment effects},
  author={Chernozhukov, Victor and Chetverikov, Denis and Demirer, Mert and Duflo, Esther and Hansen, Christian and Newey, Whitney},
  journal={American Economic Review},
  volume={107},
  number={5},
  pages={261--265},
  year={2017},
  publisher={American Economic Association 2014 Broadway, Suite 305, Nashville, TN 37203}
}

@article{shi2022statistical,
  title={Statistical inference of the value function for reinforcement learning in infinite-horizon settings},
  author={Shi, Chengchun and Zhang, Sheng and Lu, Wenbin and Song, Rui},
  journal={Journal of the Royal Statistical Society Series B: Statistical Methodology},
  volume={84},
  number={3},
  pages={765--793},
  year={2022},
  publisher={Oxford University Press}
}

@inproceedings{marling2020ohiot1dm,
  title={The OhioT1DM dataset for blood glucose level prediction: Update 2020},
  author={Marling, Cindy and Bunescu, Razvan},
  booktitle={CEUR workshop proceedings},
  volume={2675},
  pages={71},
  year={2020},
  organization={NIH Public Access}
}

@article{zhang2011causal,
  title={Causal inference for continuous-time processes when covariates are observed only at discrete times},
  author={Zhang, Mingyuan and Joffe, Marshall M and Small, Dylan S},
  journal={Annals of Statistics},
  volume={39},
  number={1},
  pages={131--173},
  year={2011},
  publisher={NIH Public Access}
}

@article{bojinov2019time,
  title={Time series experiments and causal estimands: exact randomization tests and trading},
  author={Bojinov, Iavor and Shephard, Neil},
  journal={Journal of the American Statistical Association},
  volume={114},
  number={528},
  pages={1665--1682},
  year={2019},
  publisher={Taylor \& Francis}
}

@article{shi2022dynamic,
  title={Dynamic causal effects evaluation in a/b testing with a reinforcement learning framework},
  author={Shi, Chengchun and Wang, Xiaoyu and Luo, Shikai and Zhu, Hongtu and Ye, Jieping and Song, Rui},
  journal={Journal of the American Statistical Association},
  volume={118},
  number={543},
  pages={2059--2071},
  year={2022},
  publisher={Taylor \& Francis}
}

@article{jia2022policy,
  title={Policy evaluation and temporal-difference learning in continuous time and space: A martingale approach},
  author={Jia, Yanwei and Zhou, Xun Yu},
  journal={Journal of Machine Learning Research},
  volume={23},
  number={154},
  pages={1--55},
  year={2022}
}

@article{huang1998projection,
  title={Projection estimation in multiple regression with application to functional ANOVA models},
  author={Huang, Jianhua Z},
  journal={The annals of statistics},
  volume={26},
  number={1},
  pages={242--272},
  year={1998},
  publisher={Institute of Mathematical Statistics}
}

@article{brunel2008parameter,
  title={Parameter estimation of {ODE}’s via nonparametric estimators},
  author={Brunel, Nicolas J-B},
  journal={Electronic Journal of Statistics},
  volume={2},
  pages={1242--1267},
  year={2008},
  publisher={Institute of Mathematical Statistics}
}

@article{zhang2024individualized,
  title={Individualized dynamic latent factor model for multi-resolutional data with application to mobile health},
  author={Zhang, J and Xue, F and Xu, Q and Lee, J and Qu, A},
  journal={Biometrika},
  pages={asae015},
  year={2024},
  publisher={Oxford University Press}
}

@article{schomaker2024causal,
  author = {Schomaker, Michael and others},
  title = {Causal Inference for Continuous Multiple Time Point Interventions},
  journal = {Statistics in Medicine},
  year = {2024},
  volume = {43},
  issue = {5},
  pages = {1121--1134},
  doi = {10.1002/sim.10246}
}

@article{doya2000reinforcement,
  title={Reinforcement learning in continuous time and space},
  author={Doya, Kenji},
  journal={Neural Computation},
  volume={12},
  number={1},
  pages={219--245},
  year={2000},
  publisher={MIT Press}
}

@article{munos2006policy,
  title={Policy gradient in continuous time},
  author={Munos, R{\'e}mi},
  journal={Journal of Machine Learning Research},
  volume={7},
  pages={771--791},
  year={2006}
}

@article{kallus2022efficiently,
  title={Efficiently breaking the curse of horizon in off-policy evaluation with double reinforcement learning},
  author={Kallus, Nathan and Uehara, Masatoshi},
  journal={Operations Research},
  volume={70},
  number={6},
  pages={3282--3302},
  year={2022}
}

@inproceedings{lewis2020double,
  title={Double/debiased machine learning for dynamic treatment effects via $g$-estimation},
  author={Lewis, Greg and Syrgkanis, Vasilis},
  booktitle={Advances in Neural Information Processing Systems},
  volume={33},
  pages={22293--22304},
  year={2020}
}

@article{stone1985additive,
  title={Additive regression and other nonparametric models},
  author={Stone, Charles J},
  journal={Annals of Statistics},
  volume={13},
  number={2},
  pages={689--705},
  year={1985}
}

@article{claeskens2009asymptotic,
  title={Asymptotic properties of penalized spline estimators},
  author={Claeskens, Gerda and Krivobokova, Tatyana and Opsomer, Jean D},
  journal={Biometrika},
  volume={96},
  number={3},
  pages={529--544},
  year={2009}
}

@article{zhou1998local,
  title={Local asymptotics for regression splines and confidence regions},
  author={Zhou, Shanggang and Shen, Xiaotong and Wolfe, Douglas A},
  journal={Annals of Statistics},
  volume={26},
  number={5},
  pages={1760--1782},
  year={1998}
}

@article{davis1984piecewise,
  title={Piecewise-deterministic {M}arkov processes: a general class of non-diffusion stochastic models},
  author={Davis, Mark H. A.},
  journal={Journal of the Royal Statistical Society: Series B},
  volume={46},
  number={3},
  pages={353--388},
  year={1984}
}

@article{benaim2012quantitative,
  title={Quantitative ergodicity for some switched dynamical systems},
  author={Bena{\"i}m, Michel and Le Borgne, St{\'e}phane and Malrieu, Florent and Zitt, Pierre-Andr{\'e}},
  journal={Electronic Communications in Probability},
  volume={17},
  pages={1--14},
  year={2012}
}

@article{benaim2015qualitative,
  title={Qualitative properties of certain piecewise deterministic {M}arkov processes},
  author={Bena{\"i}m, Michel and Le Borgne, St{\'e}phane and Malrieu, Florent and Zitt, Pierre-Andr{\'e}},
  journal={Annales de l'Institut Henri Poincar{\'e}, Probabilit{\'e}s et Statistiques},
  volume={51},
  number={3},
  pages={1040--1075},
  year={2015}
}

@article{cloez2015exponential,
  title={Exponential ergodicity for {M}arkov processes with random switching},
  author={Cloez, Bertrand and Hairer, Martin},
  journal={Bernoulli},
  volume={21},
  number={1},
  pages={505--536},
  year={2015}
}

@article{davydov1973mixing,
  title={Mixing conditions for {M}arkov chains},
  author={Davydov, Yu. A.},
  journal={Theory of Probability and Its Applications},
  volume={18},
  number={2},
  pages={312--328},
  year={1973}
}

\end{document}

\def\spacingset#1{\renewcommand{\baselinestretch}{#1}\small\normalsize}
\spacingset{1}
\renewcommand{\thesection}{S}
\renewcommand{\thetable}{S\arabic{table}}
\renewcommand{\theequation}{S.\arabic{equation}}
\renewcommand{\theequation}{S.\arabic{equation}}
\renewcommand{\thefigure}{S\arabic{figure}}

\title{\bf Supplementary Material for ``Testing Cumulative Treatment Effects in Continuous Time with ODE Dynamics and Noisy Observations''}
\author{}
\date{}

\maketitle

\section*{Supplementary Materials}

This Supplementary Material is organized as follows. Sections~S.1 and~S.2 collect the assumptions: the causal conditions for identification (with the potential outcome notation, their Rubin language reading, and their plausibility in mobile health) and the conditions for asymptotic normality (with the block notation convention under which they cover general policy contrasts). Section~S.3 motivates each condition and explains how the two approximation families enter the asymptotics, Section~S.4 proves stationarity and mixing under the primitive behavior mechanism conditions, Section~S.5 establishes the first step spline rate condition from primitive conditions, and Section~S.6 shows that the sieve approximation error is second order. Sections~S.7--S.9 prove the identification, asymptotic normality, and discretization bias theorems. Sections~S.10--S.12 report additional simulation results (sample sizes and effect magnitudes, the regular grid comparison, and the SDE sensitivity analysis), Section~S.13 collects implementation details and design diagnostics for the OhioT1DM analyses, Section~S.14 reports the robustness grid for those analyses, and Section~S.15 reviews two step ODE drift estimation.

\subsection{Assumptions for identification}
\label{subsec:causal_assumptions}

This section collects the causal conditions underlying Theorem~\ref{thm:bellman}, together with the potential outcome notation in which they are stated and their discussion; they are gathered here to keep the main text compact.

The assumptions are stated in potential outcome notation, which we introduce here at its point of use. For participant $i$, let $\bar{a}_{it} = \{a_{is}: 0 \leq s < t,\ a_{is} \in \mathcal{A}\}$ denote a treatment history over the half open interval $[0,t)$, and write $S_{it}^{\star}(\bar{a}_{it})$ for the counterfactual state at time $t$ under this history and $Y_{it}^{\star}(\bar{a}_{it},a_{it})$ for the counterfactual outcome at time $t$ when the history is $\bar a_{it}$ and the current action is $a_{it}$; the state depends on the strictly earlier history only, while the instantaneous outcome may depend additionally on the current action, matching the reward $r(a,s)$ of Assumption~\ref{asp:CMIA}. The half open convention encodes the timing: at each instant the state is observed first and the action is set immediately after, so the state at time $t$ depends only on the strictly earlier history $\bar a_{it}$; in particular the initial state $S_{i0}$ precedes the first action $a_{i0}$, and under a regime $\pi$ the actions are generated as $a_{is}=\pi(S_{is}^\star(\bar a_{is}))$, which makes the counterfactual state well defined and non anticipating.

The first two conditions expand the consistency and sequential randomization assumptions in multistage decision making problems to continuous time.

\begin{assumption}
\label{asp:CA}
    (Consistency Assumption, CA) $S_{it}=S_{it}^*\left(\bar{A}_{it}\right)$ and $Y_{it}=$ $Y_{it}^*\left(\bar{A}_{it},A_{it}\right)$ for all $t \geq 0$ 
\end{assumption} 
\begin{assumption}
\label{asp:SR}
     (Continuous Time Sequential Randomization, CTSR) There exists a bounded function $\varepsilon(t, \delta)>0$ with $\int_0^{\infty} \varepsilon(t, \delta) d t \rightarrow 0$ as $\delta \rightarrow 0$, such that for any $\bar{a}_{it} \in \mathcal{A}, t \in[0, \infty)$, $\delta>0$
$$
\left\|\mathbb{E}\left\{Y_{it}^{\star} (\bar{a}_{it},a_{it}) \mid \bar{A}_{it}, S_{it}\right\}-\mathbb{E}\left\{Y_{it}^{\star} (\bar{a}_{it},a_{it}) \mid \bar{A}_{i,t-\delta}, S_{it}\right\}\right\|_1<\varepsilon(t, \delta)
$$
\end{assumption}

Assumption \ref{asp:SR} states that given $\bar{A}_{i,t-\delta}$ and $S_{it}$, the expectation of $Y_{it}^{\star} (\bar{a}_{it},a_{it})$ depends only on a short period of treatment assignment between $[t-\delta, t)$. This dependence is upper bounded by this bounded function $\varepsilon(t, \delta)$ whose integral over $t \in [0, \infty)$ tends to zero, as the gap $\delta$ goes to zero. The condition $\int_0^{\infty} \varepsilon(t, \delta) d t \rightarrow 0$ is needed to incorporate longitudinal studies in time range $[0, \infty)$. In discrete time longitudinal studies, this implies sequential randomization assumption that $Y_{i1}^{\star},\cdots,Y_{ik}^{\star} \perp A_{ik} \mid \bar{A}_{i,k-1}, S_{ik}$. Assumptions~\ref{asp:CA} and~\ref{asp:SR} enter the identification argument only through the following \emph{conditional mean sequential exchangeability} property: for each action $a$,
\[
\mathbb E\big\{Y_{it}^{\star}(\bar A_{it},a)\mid S_{it},\,A_{it}=a\big\}=\mathbb E\big\{Y_{it}^{\star}(\bar A_{it},a)\mid S_{it}\big\},
\]
so that conditioning on the realized current assignment does not distort the conditional mean of the corresponding potential outcome given the recorded state; CTSR delivers this in the limit of vanishing decision spacing, and a reader preferring a single primitive condition may take the display itself as the assumption.
\begin{assumption}
\label{asp:MA}
(Markov Assumption, MA) Under the ODE dynamics \eqref{eq:ode}, there exists a transition semigroup $(P)_{t \ge 0}$ such that for any $t^{\prime} \geq 0$ and $\mathcal{S} \subseteq \mathbb{R}^d$, we have
$\operatorname{Pr}\left\{S_{i,t+t^{\prime}}^*\left(\bar{a}_{it}\right) \in \mathcal{S} \mid S_{it}^*\left(\bar{a}_{it}\right)\right\}=P_{it^{\prime}}\left(\mathcal{S} ; a_{it}, S_{it}^*\left(\bar{a}_{it}\right)\right).$
\end{assumption}
\begin{assumption}
\label{asp:CMIA} 
(Conditional Mean Independence Assumption, CMIA) there exists a function $r$ such that for any $ t \ge 0$, we have 
$\mathbb{E}\left\{Y_{it}^*\left(\bar{a}_{it},a_{it}\right) \mid S_{it}^*\left(\bar{a}_{it}\right)\right\}=r\left(a_{it}, S_{it}^*\left(\bar{a}_{it}\right)\right).$
\end{assumption}

\begin{assumption}[Policy overlap/positivity]
\label{asp:positivity}
For each target policy $\pi$ under consideration and every state $s$ in the support of the discounted occupancy measure of $\pi$ initialized at $\mathbb{G}$ (the distribution $|\log\gamma|\int_0^\infty \gamma^u\,\mathrm{Law}\{S^{\star}_{u}(\pi)\mid S_0\sim\mathbb G\}\,du$ of states reachable from $\mathbb G$ under the target policy), the mechanism that assigned the observed treatments takes the action prescribed by $\pi$ with positive probability: $\Pr\{A_{it}=\pi(S_{it})\mid S_{it}=s\}\ge c_\pi>0$. This overlap requirement is a separate assumption: the behavior mechanism conditions of Assumption~\ref{asp:mixing} deliver it only at treatment decision times (Proposition~\ref{prop:behavior}).
\end{assumption}
Assumption~\ref{asp:positivity} is the continuous time analogue of the standard positivity/overlap condition, and it is the assumption that makes the policy contrast estimand~\eqref{eq:ate_policy} well posed. It requires the observed data to contain treatment decisions that agree with the target policy across the relevant state region, which is exactly what allows the drift $b(s,\pi(s))$ and the reward under $\pi$ to be recovered on the agreement subsample $\{A_{it}=\pi(S_{it})\}$. For a constant regime it reduces to $\Pr\{A_{it}=a\mid S_{it}=s\}>0$ for the relevant $a$; for a threshold policy it only requires treated decisions where the policy treats and untreated decisions where it does not. The latter is typically far better supported by observational data than the always treat extreme, whose overlap for the real data we quantify in Section~\ref{sec:realdata}. A practical diagnostic is to report, for each candidate policy, the empirical fraction of observation times on which $A_{it}=\pi(S_{it})$, stratified across the observed state range.

It is useful to restate the causal structure in Rubin's language. The confounder in this setting is the latent state process, which drives both the treatment assignment and the outcome; Assumption~\ref{asp:SR}, which requires the recorded state to capture the information used in assigning treatment, is the corresponding no unmeasured confounding condition. For the constant target regimes always treat and never treat there is no confounding of the regimes \emph{per se}, since these policies do not respond to the state; but overlap is still required: the observed data must contain both treated and untreated decisions across the relevant state range, which is exactly the positivity condition of Assumption~\ref{asp:positivity}. Together, Assumptions~\ref{asp:SR} and~\ref{asp:positivity} are the continuous time counterparts of the familiar no-unmeasured-confounding-plus-overlap pair.

Assumptions \ref{asp:MA} and \ref{asp:CMIA} are continuous time analogues of standard Markov decision process conditions used in sequential causal inference. If Assumption \ref{asp:MA} holds, Assumption \ref{asp:CMIA} automatically holds when $Y_{it}$ is a deterministic function of $\boldsymbol{S}_{it}$, $A_{it}$ and $\boldsymbol{S}_{i,t+t^{\prime}}$ that measures the system’s status at time $t + t^{\prime}$. The latter assumption is commonly made in the reinforcement learning literature, which is a stronger condition than Assumption \ref{asp:CMIA}.

In mobile health applications, these assumptions are scientifically plausible when treatment decisions are protocolized or are based on observable physiological signals: Assumption~\ref{asp:CA} (consistency) holds whenever the treatment assigned is the one actually delivered to the patient, and Assumption~\ref{asp:SR} (continuous time sequential randomization) is supported when the recorded state vector $S_{it}$ captures the information used to make the treatment decision (e.g., recent glucose and recent activity). The Markov assumption~\ref{asp:MA} is reasonable when the state vector summarizes the relevant short term history (e.g., minute resolution glucose and accelerometer signals carry the dynamics needed for next step prediction); when relevant history extends further back, the state can be augmented with lagged summaries. Modest violations, such as omission of an informative covariate, bias the estimating equation toward the conditional expectation given the recorded state. We recommend supplementing the analysis with sensitivity diagnostics: comparing test conclusions across plausible state augmentations and across patient subgroups, and stratifying when domain knowledge identifies confounders external to the recorded state.

The identification result in Theorem~\ref{thm:bellman} relies on the following regularity conditions on the generator, which we state explicitly here; the main text refers to them as the ``generator regularity conditions.''
\begin{assumption}[Generator regularity]
\label{asp:generator_reg}
For each stationary target policy $\pi$ under consideration---in particular each constant policy $\pi\equiv a$, $a\in\{0,1\}$---under the follow-$\pi$ dynamics the state process is Markov with a Feller semigroup $(P_h^\pi)_{h\ge0}$; the value function $V^\pi$ lies in the domain $\mathcal D(\mathcal L_\pi)$ of the corresponding generator; and the discounted reward $u\mapsto\gamma^u\,\mathbb E[Y_{t+u}^{(\pi)}\mid S_t^{(\pi)}=s]$ is integrable and locally continuous, so that the semigroup limit defining $\mathcal L_\pi$ exists. This covers the generality of Theorem~\ref{thm:bellman}, which identifies the value function of an arbitrary stationary policy; for the constant policies we write $\mathcal L_a$ and $V_a$ as elsewhere.
\end{assumption}
These conditions hold, for instance, when the drift $b(\cdot,a)$ is Lipschitz on the compact state space (so the ODE flow is well posed and Feller) and $V_a$ is continuously differentiable.

\subsection{Additional assumptions for asymptotic normality}
\label{subsec:assumptions_normality}

\noindent\emph{Block notation convention.} Fix the target policy pair $(\pi_1,\pi_2)$ of the main text. Throughout the remainder of this supplement the block label $a\in\{0,1\}$ indexes this pair, with $a=1$ corresponding to $\pi_1$ and $a=0$ to $\pi_2$: the indicator $\mathbb I\{A_{it}=a\}$ denotes the policy agreement indicator $\mathbb I\{A_{it}=\pi^{(a)}(S_{it})\}$ where $\pi^{(1)}=\pi_1$ and $\pi^{(0)}=\pi_2$, the drift $b(s,a)$ denotes the target drift $b(s,\pi^{(a)}(s))$, the value function $V_a$ denotes $V^{\pi^{(a)}}$, the density ratio $\omega_t(a;\cdot)$ denotes the follow-$\pi^{(a)}$ density ratio, and the generator $\mathcal L_a$ denotes $\mathcal L_{\pi^{(a)}}$. Under this convention every assumption, lemma, and proof below is stated for the general policy contrast $\tau(\pi_1,\pi_2)$; for the constant policies $\pi_1\equiv1$, $\pi_2\equiv0$ the notation coincides with its literal action indexed reading, which is the case evaluated in the first simulation studies of the main text.

We impose the following additional assumptions for the asymptotic analysis in Theorem~\ref{thm:normal}.

\begin{assumption}
\label{asp:sampling}
Each subject $i$ is followed over $[0,T_i]$ and carries two (possibly irregular) observation grids, which play distinct roles. (i)~\emph{State grid (infill).} The noisy state $X_{it}$ is recorded on $\mathbb T_i^X$ with $m_i^X:=|\mathbb T_i^X|$ points; the grids are quasi uniform (bounded mesh ratio) and infill, $\max_j \Delta t^X_{ij}\to0$, so $m_i^X/T_i\to\infty$. (ii)~\emph{Outcome grid (bounded spacing).} The outcome $Y_{it}$ is recorded, and the estimating equation is evaluated, on $\mathbb T_i^Y$ with spacing bounded below and above, $0<\underline\Delta\le \Delta t^Y_{ij}\le \bar\Delta<\infty$, so $m_i^Y:=|\mathbb T_i^Y|\asymp T_i$. The $I$ subjects are independent. The normalizing count is the total number of outcome observations, $N_Y:=\sum_{i=1}^I m_i^Y$, which by (ii) satisfies $N_Y\asymp\sum_{i=1}^I T_i$ and may therefore be interpreted equivalently as the total follow up time; no equivalence with the state observation count $\sum_i m_i^X$ is used anywhere.
\end{assumption}

\begin{assumption}[Behavior mechanism and dynamics]
\label{asp:mixing}
(i)~For each subject, treatment decision times form a Poisson process with rate $\lambda\in(0,\infty)$, independent across subjects and of everything else; at each decision time the treatment is redrawn from a continuous behavior policy $\pi_b$ with $c_b\le\pi_b(s)\le1-c_b$ for some $c_b>0$, and is held constant between decisions. (ii)~The drifts $b(\cdot,0)$ and $b(\cdot,1)$ are smooth and bounded, leave the compact state space $\mathcal S$ positively invariant, and are dissipative: $\langle s-s',\,b(s,a)-b(s',a)\rangle\le-\kappa\|s-s'\|^{2}$ for some $\kappa>0$, all $s,s'\in\mathcal S$ and each $a$. (iii)~Either the two drifts coincide, or the strong bracket condition of \citet[Definition~4.3]{benaim2015qualitative} holds at a point of the accessible set of the switching dynamics; in one dimension the latter reduces to the two drifts differing at an accessible point. (iv)~The initial pair $(S_{i0},A_{i0})$ is drawn from the invariant distribution $\mu^\star$ of Proposition~\ref{prop:behavior}.
\end{assumption}

Assumption~\ref{asp:mixing} no longer postulates any mixing property: it consists of primitive, verifiable conditions on the treatment assignment mechanism and the dynamics. Under (i)--(iv) the joint state--treatment process is strictly stationary and exponentially $\beta$-mixing; this is \emph{proved}, by reduction to ergodicity results for piecewise deterministic Markov processes, in Proposition~\ref{prop:behavior} of Section~\ref{subsec:behavior_mechanism}. Only the derived stationarity and mixing property is used in the proofs of this supplement, so any behavior mechanism for which that property can be established may replace conditions (i)--(iv) without change elsewhere.

Let $\omega_t(a;\bar A_t, s)$ denote the marginalized representer that transfers the reference distribution $\mathbb G$ to the behavior distribution through the value equation operator (the identity~\eqref{eq:representer} of Section~\ref{subsec:sieve_bias}, where its sign convention is fixed and a primitive existence condition is given; under that convention $\omega_t$ is a discounted occupation density ratio between the target policy $\pi=$ always-$a$ and the behavior mechanism) at time $t$, conditional on the treatment history $\bar A_t$ and the current state $s$; this representer appears in the long run covariance representation used in the proof of Theorem~\ref{thm:normal}.

\begin{assumption}
\label{asp:basisfunction}
There exist coefficient vectors $\boldsymbol{\beta}^* = ((\boldsymbol{\beta}^*_0)^\top, (\boldsymbol{\beta}^*_1)^\top)^\top$ and $\{\boldsymbol{\theta}^*_{a,t}\}_{a,t}$ such that
\[
\sup_{a \in \{0,1\},\, s \in \mathcal{S}} \Big\{\big| V_a(s) - \Psi(s)^\top \boldsymbol{\beta}^*_a \big| + \big\| \nabla V_a(s) - \nabla\Psi(s)^\top \boldsymbol{\beta}^*_a \big\|\Big\} = o\{(N_Y)^{-1/4}\},
\]
and
\[
\sup_{a \in \{0,1\},\, s \in \mathcal{S},\, t > 0} \big| \omega_t\big(a;\bar{A}_t, s\big) - \Psi(s)^\top \boldsymbol{\theta}^*_{a,t} \big| = o\{(N_Y)^{-1/4}\}.
\]
Here the first display is stated in the $C^1$ norm because the value equation operator involves the gradient of the value function; spline and polynomial sieves approximate a function and its first derivative at the same polynomial order, so the $C^1$ statement costs one degree of smoothness relative to the sup norm version. The second display presumes, as part of this assumption, that a bounded representer $\omega_t$ satisfying the identity~\eqref{eq:representer} of Section~\ref{subsec:sieve_bias} exists; a primitive sufficient condition is given there.

Two scoping remarks are part of this assumption. First, the sieve dimension $M=\dim(\Psi)$ is held \emph{fixed} throughout the asymptotic analysis: the proof of Theorem~\ref{thm:normal} treats $\boldsymbol\beta$ as a parameter of fixed dimension. With a fixed sieve and a fixed data generating law, a nonzero approximation error cannot be $o\{(N_Y)^{-1/4}\}$; the fixed dimensional theorem therefore requires either exact sieve representation ($V_a$, and under the block convention $V^{\pi_k}$, and the representer lie in the span of $\Psi$) or a local misspecification sequence along which the approximation errors satisfy the stated product rates. Section~\ref{subsec:sieve_bias} establishes the second order product bound conditional on these rates; a full growing sieve result is not claimed. The undersmoothing calculation of Section~\ref{subsec:assumption_roles}, in which the dimension grows as $J\asymp(N_Y)^{\kappa}$, indicates \emph{which} growing sieve regime would deliver these rates for H\"older classes, but a growing dimension version of Theorem~\ref{thm:normal} would additionally require explicit rates for $M_{N_Y}$, uniform eigenvalue control of $\Sigma_a$ along the sieve, growth conditions on the basis envelope $\sup_s\|\Psi(s)\|$, a central limit argument for the scalar functional $U^\top\widehat{\boldsymbol\beta}$ under growing dimension, and consistency of the sandwich variance estimator; we leave the growing sieve theory to future work. Second, for a hard threshold target policy the value function has a derivative discontinuity on the switching surface $\{g(s)=c\}$, and no basis of continuously differentiable functions can approximate it in the global $C^1$ norm. In that case the basis is required to contain hinge functions matched to the (analyst specified) surface---a knot of multiplicity one less than the spline order at the threshold in one dimension (a triple knot for a cubic spline, which retains continuity while freeing the first derivative)---and the two displays are understood with the supremum taken over $\mathcal S$ away from the switching surface: the fitted and true gradients then share the same jump locus and the approximation holds piecewise, while the surface itself is visited with probability zero under the margin condition (Assumption~\ref{asp:margin}) and its shrinking neighborhoods contribute negligibly by the counting argument of Lemma~\ref{lem:indicator}.
\end{assumption}

\begin{assumption}
\label{asp:varianceInvertable}
For some $c_1>0$, $\inf_{a,s}\mathrm{Var}\{\varepsilon_{it,a}\mid S_t=s\}\ge c_1$, where $\varepsilon_{it,a} := Y_{it} + \log(\gamma)\,V_a(S_{it}) + \langle \nabla V_a(S_{it}),\, D\rangle$ is the temporal difference error at the observation time of subject $i$, index $t$, for action $a$.
\end{assumption}

\begin{assumption}
\label{asp:moments}
$\Psi(s)$ and $\nabla\Psi(s)$ are uniformly bounded. In addition, there exists some $\delta>0$ such that
\[
\mathbb{E}[\|\Psi(S_t)\mathbb{I}\{A_t=a\}\varepsilon_{t,a}\|^{2+\delta}]<\infty.
\]
Define
\[
\Sigma_a(T) := -\frac{1}{T}\int_0^T \mathbb{E}\!\Big[\Psi(S_t)\,\mathbb{I}\{A_t=a\}\,\big\{\log\gamma\,\Psi(S_t)+\nabla\Psi(S_t)^\top b(S_t,a)\big\}^{\!\top}\Big]dt,
\]
and assume $\Sigma_a(T)\to \Sigma_a$ as $T\to\infty$, with each $\Sigma_a$ nonsingular. Let $\Sigma := \mathrm{diag}(\Sigma_0,\Sigma_1)$.
\end{assumption}

\medskip\noindent\emph{Policy specific conditions.} Under the block notation convention above, Assumptions~\ref{asp:basisfunction}--\ref{asp:moments} apply to the general policy contrast $\tau(\pi_1,\pi_2)$ exactly as stated. Four conditions nonetheless deserve policy specific comment. (i)~The sieve rate of Assumption~\ref{asp:basisfunction} must hold for $V^{\pi_k}$, which requires $V^{\pi_k}$ to possess the H\"older smoothness used in the undersmoothing calculation; for a hard threshold policy $V^{\pi}$ can have a derivative discontinuity at the switching surface $\{g(s)=c\}$; since the surface is specified by the analyst, the sieve is adapted to it (a knot of multiplicity one less than the spline order at the threshold in one dimension, a triple knot for a cubic spline, or, equivalently and more generally, hinge functions of $(g(s)-c)_+$), so piecewise H\"older smoothness on either side of the surface suffices, with the $C^1$ approximation of Assumption~\ref{asp:basisfunction} understood piecewise away from the surface as stated there, as described in Section~\ref{sec:methodology}.2 of the main text. (ii)~The density ratio approximation of Assumption~\ref{asp:basisfunction} must likewise hold for $\omega_t(\pi_k;\cdot)$. (iii)~Nonsingularity of $\Sigma_{\pi_k}$ is where the positivity condition (Assumption~\ref{asp:positivity} in the main text) does substantive work: the Jacobian averages over the agreement subsample $\{A_{it}=\pi_k(S_{it})\}$ and degenerates if the behavior mechanism rarely agrees with $\pi_k$ over part of the state range, which is exactly what the overlap diagnostic of Section~\ref{sec:realdata} in the main text detects. (iv)~The residual variance floor of Assumption~\ref{asp:varianceInvertable} must hold for the policy residual $\varepsilon_{it,\pi_k}$, as it does whenever the outcome carries conditional noise given the state. Nothing else changes. Assumptions~\ref{asp:sampling}--\ref{asp:mixing} and Lemma~\ref{lem:spline_rate} concern the sampling design, the behavior process, and the first stage spline recovery, none of which involves the target policy. On the agreement subsample the observed derivative satisfies $dS_{it}/dt=b(S_{it},\pi_k(S_{it}))$, so the estimating equation is computed by the same formula as in the constant regime case. Finally, the orthogonality and representer arguments of Section~\ref{subsec:sieve_bias} and the perturbation and central limit arguments in the proof of Theorem~\ref{thm:normal} use the block label only through the agreement weight in the score, so they hold for general policy contrasts under the block notation convention.

\medskip\noindent\emph{(v) Hard threshold policies: the estimated agreement indicator.} Conditions (i)--(iv) concern the population moment. For hard threshold policies one further first stage condition is needed, because Algorithm~1 of the main text evaluates the agreement indicator at the \emph{estimated} state, $\mathbb I\{A_{it}=\pi(\widehat S_{it})\}$: near the switching surface a small state estimation error can flip the indicator even when the value function itself is handled by the knot adapted sieve of condition~(i). A margin condition and the resulting negligibility lemma are stated at the end of Section~\ref{subsec:spline_rate} (Assumption~\ref{asp:margin} and Lemma~\ref{lem:indicator}), together with a remark on existence and uniqueness of the policy induced ODE at the discontinuous switching surface (Remark~\ref{rem:switching}).

\subsection{Roles of Assumptions~\ref{asp:sampling}--\ref{asp:moments} and the two approximation families}
\label{subsec:assumption_roles}

We motivate each condition, indicate a concrete sufficient condition or diagnostic, and make explicit how the two approximation families used by the procedure enter the asymptotics.

\emph{Assumption~\ref{asp:sampling} (sampling).} This specifies the two grid design and assigns each grid its role. The infill requirement applies to the \emph{state} grid only, and is what supports sup norm recovery of the latent derivative (Lemma~\ref{lem:spline_rate} and Remark~\ref{rem:regime}). The bounded below spacing of the \emph{outcome} grid serves two purposes: it makes the normalizing count $N_Y$ equivalent to total follow up time, and it guarantees that successive score summands are separated by at least $\underline\Delta$ units of real time, so that under Assumption~\ref{asp:mixing} their $\beta$-mixing coefficients decay geometrically in the summand index and the dependence between successive summands does \emph{not} deteriorate as the state grid refines; the central limit theorem is never applied along a densifying grid. In the application, the physiological state is recorded at one to five minute resolution, so the infill regime is the empirically relevant one. Diagnostic: report the per subject state and outcome sampling densities and confirm the state grid is substantially finer.

\emph{Assumption~\ref{asp:mixing} (behavior mechanism and dynamics).} This assumption consists of primitive, verifiable conditions rather than a mixing postulate: a refresh type assignment mechanism with a behavior policy bounded away from $0$ and $1$, dissipative drifts on a compact positively invariant state space, and an accessibility condition. What the proofs use is the property these conditions deliver: strict stationarity and exponential $\beta$-mixing of the joint state--treatment process, established in Proposition~\ref{prop:behavior} by reduction to ergodicity theory for piecewise deterministic Markov processes, which controls the serial dependence of the score and yields the long run covariance. Dissipativity supplies the contraction of the flows, the Poisson refresh supplies the exogenous mixing randomness, and the accessibility condition rules out decomposable dynamics. Any behavior mechanism for which the same stationarity and mixing property can be established may be substituted without change elsewhere. Diagnostic: inspect the decay of the empirical autocorrelation of the temporal difference residuals---a consistency check only: autocorrelation decay of the residuals neither verifies stationarity nor $\beta$-mixing of the underlying process, and the mechanism should be read as a sufficient theoretical example rather than as a description validated for any particular data set.

\emph{Assumption~\ref{asp:basisfunction} (sieve approximation).} This is the first of the two approximation families: a parametric sieve $\Psi^\top\beta$ for the value function $V_a$ and $\Psi^\top\theta_{a,t}$ for the marginalized density ratio $\omega_t$. A sufficient primitive condition is that $V_a$ and $\omega_t(a;\cdot)$ lie in a H\"older class of smoothness $p$ on the compact state space and $\Psi$ is a spline or polynomial sieve of dimension $J\asymp (N_Y)^{\kappa}$ with $\kappa$ chosen so that the $J^{-p/d}$ approximation error is $o\{(N_Y)^{-1/4}\}$; this is the standard undersmoothing calculation. As recorded in Assumption~\ref{asp:basisfunction}, the formal statement of Theorem~\ref{thm:normal} holds the sieve dimension fixed; the undersmoothing calculation identifies the growing sieve regime that would deliver the displayed rates for H\"older classes, not a case covered by the present proof. Diagnostic: monitor the second stage fit residuals and the stability of $\widehat\tau$ across sieve dimensions.

\emph{Assumption~\ref{asp:varianceInvertable} (non-degenerate residuals).} A strictly positive conditional variance of the temporal difference error rules out a degenerate score and guarantees a well defined limiting variance. It holds whenever the outcome retains conditional noise given the state (e.g. measurement error in $Y$), and can be checked by a lower bound on the estimated residual variance.

\emph{Assumption~\ref{asp:moments} (moments and Jacobian).} Bounded $\Psi,\nabla\Psi$ and a $2+\delta$ moment give the Lindeberg control for the mixing central limit theorem, and nonsingularity of $\Sigma_a$ makes the estimating equation locally invertible. Bounded state support with sub-Gaussian measurement noise is a sufficient primitive condition; the nonsingularity of $\Sigma_a$ is diagnosable through the condition number of its plug in estimate.

\emph{How the two approximation families enter the limit.} The procedure uses (a) the nonparametric spline recovery of the latent trajectory and its derivative, and (b) the parametric sieve for the value function and density ratio. Family (a) enters Theorem~\ref{thm:normal} only through the rate condition~\eqref{eq:spline_rate}, which Lemma~\ref{lem:spline_rate} shows is $o_p((N_Y)^{-1/2})$ under (P1)--(P4); the perturbation bounds~\eqref{eq:eta_perturbation_rewrite} and~\eqref{eq:sigma_perturbation_rewrite} in the proof below carry this error explicitly and show it is asymptotically negligible. Family (b) enters through Assumption~\ref{asp:basisfunction}. The inferential target is the value function contrast at the sieve pseudo true parameter $\beta^*$, and the $o\{(N_Y)^{-1/4}\}$ rates are calibrated so that the terms that actually reach the studentized statistic (the product of the value function and density ratio approximation errors, and the product of the sieve and spline errors) are each $o_p((N_Y)^{-1/2})$ and hence do not affect the limiting variance, while mild undersmoothing renders the remaining first order sieve bias $o_p((N_Y)^{-1/2})$ as well. This is why the approximation families appear in the conditions of Theorem~\ref{thm:normal} through~\eqref{eq:spline_rate} and Assumption~\ref{asp:basisfunction} rather than as additional terms in the limiting distribution.

\subsection{A concrete behavior mechanism: derived stationarity, mixing, and positivity}
\label{subsec:behavior_mechanism}

This subsection proves the stationarity, mixing, and positivity properties announced after Assumption~\ref{asp:mixing}: under the primitive conditions (i)--(iii) of that assumption (the Poisson refresh mechanism with a behavior policy bounded away from $0$ and $1$, smooth bounded dissipative drifts on a compact positively invariant state space, and the accessibility condition) the joint state--treatment process is stationary and exponentially $\beta$-mixing, so that the mixing used in the proofs is derived rather than assumed. Between decisions the state follows the ODE~\eqref{eq:ode}. In the strong bracket condition of \citet[Definition~4.3]{benaim2015qualitative}, the difference $b(\cdot,1)-b(\cdot,0)$, together with its iterated Lie brackets with the two drifts, spans $\mathbb R^{d}$ at a point of the accessible set; in one dimension it reduces to the two drifts differing at an accessible point, that is, to a nonzero instantaneous treatment effect on the drift.

\begin{proposition}[Stationarity, mixing, and positivity under the refresh mechanism]
\label{prop:behavior}
Under Assumption~\ref{asp:mixing}(i)--(iii): (i) the joint process $(S_t,A_t)_{t\ge0}$ is a piecewise deterministic Markov process with a unique invariant distribution $\mu^\star$, and initialized from $\mu^\star$ it is strictly stationary; (ii) initialized from $\mu^\star$, the joint state--treatment process is exponentially $\beta$-mixing; under the strong bracket branch of Assumption~\ref{asp:mixing}(iii) the stronger statement holds that its law converges to $\mu^\star$ exponentially fast in total variation, uniformly over initial conditions; (iii) the retained score process inherits exponential $\beta$-mixing at the same rate---for any stationary target policy, whose agreement indicator $\mathbb I\{A_t=\pi(S_t)\}$ is a fixed measurable function of $(S_t,A_t)$---because the measurement and outcome errors are i.i.d.\ and independent of $(S,A)$; (iv) at every decision time and every state, each action has conditional probability at least $c_b$, so the overlap condition of Assumption~\ref{asp:positivity} holds with $c_\pi=c_b$ along the decision time skeleton, for every stationary target policy $\pi$.
\end{proposition}

\medskip\noindent\emph{Verification of the hypotheses of \citet{benaim2015qualitative}.} The load bearing step of the proof is an application of Theorem~4.6 of \citet{benaim2015qualitative}, which concerns a switching process $(X_t,Y_t)$ on $M\times E$ with $E$ finite, and requires: (a) smooth, bounded vector fields $F^i$ admitting a compact set $M$ that is positively invariant under every flow; (b) state dependent jump rates $\lambda(x,i,j)$ that are continuous in $x$, nonnegative, and form an irreducible matrix at every $x\in M$; and (c) the strong bracket condition of their Definition~4.3 at a point of the accessible set. In our setting $E=\{0,1\}$ is the action set, $M=\mathcal S$, and $F^a=b(\cdot,a)$. Hypothesis (a) is Assumption~\ref{asp:mixing}(ii). Hypothesis (b) holds because the intensity of a switch from action $a$ to $1-a$ at state $s$ equals $\lambda\,\pi_b(1-a\mid s)\in[\lambda c_b,\lambda(1-c_b)]$, which is continuous in $s$ and strictly positive by Assumption~\ref{asp:mixing}(i), so the two state jump matrix is irreducible at every state. Hypothesis (c) is the bracket branch of Assumption~\ref{asp:mixing}(iii). The theorem therefore applies verbatim to $(S_t,A_t)$, and its conclusion is exponential convergence in total variation to the unique invariant law, with constants not depending on the initial condition.

\begin{proof}[Proof (by reduction to known results)]
(i) The pair $(S_t,A_t)$ is a piecewise deterministic Markov process in the sense of \citet{davis1984piecewise}: deterministic Lipschitz flow between jumps whose state dependent intensity is bounded in $[\lambda c_b,\lambda]$. The space $\mathcal S\times\{0,1\}$ is compact and the semigroup is Feller, so an invariant distribution exists by the Krylov--Bogolyubov argument; uniqueness follows from (ii) under the strong bracket branch, and is established directly within the coincident drift branch of (ii) below.
(ii) Under Assumption~\ref{asp:mixing}(ii) each flow is exponentially contracting, $\|\varphi^a_t(s)-\varphi^a_t(s')\|\le e^{-\kappa t}\|s-s'\|$, with a unique exponentially stable equilibrium in $\mathcal S$. Suppose first that the two drifts coincide (the first branch of Assumption~\ref{asp:mixing}(iii)), with common flow $\varphi_t$ and attractor $s^\star$. Exponential convergence in \emph{total variation} from an arbitrary initial condition fails in this branch, and is not claimed: started from $s\neq s^\star$ the state component is the point mass $\delta_{\varphi_t(s)}$, whose total variation distance from $\delta_{s^\star}$ equals one for every finite $t$ (the contraction gives Wasserstein, not total variation, convergence). What holds, and what the proofs of this supplement use, is $\beta$-mixing of the \emph{stationary} process. The invariant law is $\mu^\star=\delta_{s^\star}\otimes\nu$, where $\nu$ is the invariant action law of the refresh chain: the state component is invariant because $\varphi_t(s^\star)=s^\star$, and uniqueness follows since every trajectory converges to $s^\star$, forcing any invariant law to concentrate there. Initialized from $\mu^\star$ (Assumption~\ref{asp:mixing}(iv)), the state is almost surely constant at $s^\star$ while the action redraws from $\pi_b(s^\star)\in[c_b,1-c_b]$ at Poisson decision times; the action chain therefore regenerates at each decision time, giving $\|P_t\{(s^\star,a),\cdot\}-\mu^\star\|_{\mathrm{TV}}\le e^{-\lambda t}+Ce^{-\lambda c_b t}$ (no decision by time $t$, plus the geometric convergence of the two state refresh chain). By Davydov's identity for stationary Markov processes, $\beta(t)=\int\|P_t(z,\cdot)-\mu^\star\|_{\mathrm{TV}}\,\mu^\star(dz)$ \citep{davydov1973mixing}, and since $\mu^\star$ charges only $z=(s^\star,\cdot)$, this yields $\beta(t)\le C'e^{-\lambda c_b t}$: exponential $\beta$-mixing of the stationary process, which is conclusion (ii) for this branch. We note that in this branch the stationary state is degenerate at $s^\star$, so the Jacobian nonsingularity of Assumption~\ref{asp:moments} typically fails; the branch establishes mixing only, and the operative case for the inferential results is the strong bracket branch. Under the strong bracket branch, by the hypothesis verification preceding the proof, Theorem~4.6 of \citet{benaim2015qualitative} (see also \citealp{benaim2012quantitative,cloez2015exponential}) yields constants $C,\rho>0$, not depending on the initial condition, such that $\sup_{(s,a)}\|P_t\{(s,a),\cdot\}-\mu^\star\|_{\mathrm{TV}}\le Ce^{-\rho t}$. For a stationary Markov process the $\beta$-mixing coefficient equals the mean total variation distance \citep{davydov1973mixing}, whence $\beta(t)\le Ce^{-\rho t}$, proving both statements of (ii) in this branch.
(iii) The score at an observation time is a fixed measurable function of $(S_t,A_t)$ and of measurement and outcome errors that are i.i.d.\ and independent of the trajectory; this covers any stationary target policy, since the agreement indicator $\mathbb I\{A_t=\pi(S_t)\}$ is such a function. Enlarging the $\sigma$-fields of an exponentially $\beta$-mixing process by an independent i.i.d.\ sequence does not increase the mixing coefficients at positive lags, so the score process is exponentially $\beta$-mixing with the same rate.
(iv) At each decision time the action is drawn afresh with $\pi_b(S_{T_k})\in[c_b,1-c_b]$, so $\Pr\{A_{T_k}=\pi(S_{T_k})\mid S_{T_k}=s\}\ge c_b$ for every target policy $\pi$ and every state $s$; decisions recur at rate $\lambda$, so every observation time lies within an $\mathrm{Exp}(\lambda)$ interval of a decision governed by this bound.
\end{proof}

Two remarks are in order. First, claim (iv) gives the clean constant $c_b$ at the decision times, which is where treatment selection operates. At an arbitrary observation time the held action is a deterministic carryover of the most recent draw; there, absolute continuity of the invariant law \citep[Theorem~4.5]{benaim2015qualitative}, which the strong bracket condition implies, together with the bounded speed of the flows, gives $\Pr\{A_t=\pi(S_t)\mid S_t=s\}>0$ for $\mu^\star$-almost every $s$, with a lower bound that is uniform on compact subsets of the interior of the accessible set. Near the boundary of the accessible set the conditional probability of the action that drives the state away from the boundary can degenerate, so the uniform constant in Assumption~\ref{asp:positivity} should be read on the support of the reference distribution $\mathbb G$, taken inside the interior of the accessible set. Second, the block randomized behavior policy used in the policy contrast simulation of Section~5 in the main text (each action drawn with probability $1/2$ on blocks of one time unit) is the deterministic clock analogue of Assumption~\ref{asp:mixing}(i) with $c_b=1/2$, so the Type-I error and power results of that simulation double as an empirical check of the regime covered by Proposition~\ref{prop:behavior}. In applications the true mechanism is unknown; the residual autocorrelation diagnostic of Section~\ref{subsec:assumption_roles} remains the practical check that the mixing implied by mechanisms of this kind is consistent with the data.

\subsection{Primitive conditions for the first step rate condition~\eqref{eq:spline_rate}}
\label{subsec:spline_rate}

The rate condition~\eqref{eq:spline_rate} is the only channel through which the first step spline recovery enters Theorem~\ref{thm:normal}. We give explicit primitive conditions under which~\eqref{eq:spline_rate} provably holds, and we make transparent the role of the multi resolution sampling design. Two observation counts must be distinguished. Let $m_i^X:=|\mathbb T_i^X|$ be the number of state measurement times of subject $i$, which are used to fit the spline $\widehat S_i$ and its derivative $\widehat D_i$, and let $m_i^Y:=|\mathbb T_i^Y|$ be the number of outcome times entering the second step moment, so that $N_Y=\sum_{i=1}^I m_i^Y$. Set $\underline m^X:=\min_{1\le i\le I} m_i^X$. The key point is that the accuracy of $(\widehat S_i,\widehat D_i)$ is governed by the (dense) state grid $m_i^X$, whereas the normalization in~\eqref{eq:spline_rate} is $\sqrt{N_Y}$, driven by the (typically much coarser) outcome grid.

The conditions must respect the behavior mechanism of Assumption~\ref{asp:mixing}: treatment is held piecewise constant and refreshed at Poisson decision times, so whenever $b(s,0)\neq b(s,1)$ the state derivative $S_i'(t)=b\{S_i(t),A_i(t)\}$ \emph{jumps} at each treatment change. A globally H\"older trajectory assumption would therefore contradict the assumed treatment mechanism outside the null. Accordingly, the smoothness condition below is stated piecewise, with breakpoints at the treatment change times---which are \emph{observed}, since the treatment path is recorded---and the spline construction places repeated knots at those times so that the estimation space carries the matching derivative jumps. We use the following primitive conditions.
\begin{itemize}
\item[(P1)] Between consecutive treatment change times, each coordinate of the latent trajectory $S_{i,\cdot}$ lies in a H\"older class $\mathcal H^{r}$ of smoothness $r>3$ with H\"older norm bounded uniformly over segments and subjects; the trajectory itself is Lipschitz on the whole observation window, its derivative jumping only at the treatment change times. Let $N^{\mathrm{sw}}_i$ denote the number of treatment changes of subject $i$. In addition, the segments are uniformly observable: the minimum inter switch gap of subject $i$ is bounded below by a fixed multiple of the knot spacing $K_i^{-1}$ of (P3), so that every segment contains at least order $m_i^X/K_i$ state observations (a uniform local sample size condition; under the Poisson mechanism arbitrarily short segments occur with vanishing probability per switch and may alternatively be merged into a neighboring segment). The state grids are quasi uniform (bounded mesh ratio).
\item[(P2)] The measurement errors $\{\boldsymbol\epsilon_{it}\}$ are independent, mean zero, and sub-Gaussian with proxy variance at most $\sigma^2<\infty$.
\item[(P3)] $\widehat S_i$ is a spline estimator of order $q\ge r$ on $K_i\asymp (m_i^X)^{1/(2r+1)}$ quasi uniform knots (mild undersmoothing relative to generalized cross validation), augmented by a knot of multiplicity $q-1$ at each observed treatment change time, so that the spline space contains functions that are continuous with a derivative jump exactly there (an ordinary simple interior knot would leave the spline $C^{q-2}$ and could not carry the jump, while multiplicity $q$ would permit a spurious discontinuity of the function itself), and $\widehat D_i=\widehat S_i'$. The augmented knots do not alter the dimension order: $N^{\mathrm{sw}}_i=O_p(K_i)$. This is automatic in the bounded horizon infill regime relevant to the application, where each subject unit is a bounded window (a patient day) so $N^{\mathrm{sw}}_i=O_p(1)$ under the Poisson mechanism while $K_i\to\infty$; for growing horizons it amounts to the state grid densifying polynomially in the follow up time, $T_i=O\{(m_i^X)^{1/(2r+1)}\}$.
\item[(P4)] The state grids densify fast enough relative to the outcome sample size:
\[
N_Y\,(\underline m^X)^{-2(r-1)/(2r+1)}\,\log(\underline m^X)\,\log I \longrightarrow 0 .
\]
\end{itemize}

\begin{lemma}
\label{lem:spline_rate}
Under the infill state grid of Assumption~\ref{asp:sampling}(i) and conditions (P1)--(P4), the first step rate condition~\eqref{eq:spline_rate} holds; that is,
\[
\sqrt{N_Y}\Big(\sup_{i,t}\|\widehat S_{it}-S_{it}\|+\sup_{i,t}\|\widehat D_{it}-D_{it}\|\Big)=o_p(1).
\]
\end{lemma}

\begin{proof}
We first record that the piecewise setting of (P1) with the repeated knot construction of (P3) reduces to the classical smooth case segment by segment. Placing a knot of multiplicity $q-1$ at each treatment change time reduces the spline space over the whole window to the spline spaces on the inter switch segments coupled only through continuity at the change times---a constraint the (continuous) latent trajectory itself satisfies---so the approximation problem is the segmentwise approximation of functions that are H\"older $r$ on each segment and continuous across breakpoints; the approximation bound $K_i^{-r}$ (and $K_i^{-(r-1)}$ for the derivative) therefore applies segmentwise with constants uniform over segments by (P1), where $K_i$ is the quasi uniform knot count, whose spacing bounds the local spacing of the augmented knot sequence. The stochastic term is governed by the total spline dimension, which is $K_i+(q-1)N^{\mathrm{sw}}_i\asymp K_i$ by (P3). At a treatment change time the population derivative has left and right limits, and the fitted derivative converges to the corresponding one sided limits; the sup norm over $t$ below is understood with one sided derivatives at the (finitely many) switch times. Since the treatment decision times of Assumption~\ref{asp:mixing}(i) are continuously distributed and independent of the observation grids, they almost surely coincide with no outcome observation time, so the second stage moment never evaluates $\widehat D$ at a switch point.
By standard sup norm results for spline smoothing of noisy data \citep{stone1985additive,huang1998projection,zhou1998local,claeskens2009asymptotic}, applied on each inter switch segment as just described, under (P1)--(P3) the per subject errors satisfy, for a spline with $K_i$ knots,
\[
\|\widehat S_i-S_i\|_\infty=O_p\!\big(K_i^{-r}+\sqrt{K_i\log K_i/m_i^X}\big),\quad
\|\widehat D_i-D_i\|_\infty=O_p\!\big(K_i^{-(r-1)}+\sqrt{K_i^{3}\log K_i/m_i^X}\big),
\]
the extra factor $K_i^{2}$ in the derivative variance arising because differentiating an order $q$ spline scales its coefficients by the inverse knot spacing $\asymp K_i$. Choosing $K_i\asymp (m_i^X)^{1/(2r+1)}$ as in (P3) balances squared bias and variance for the derivative and dominates the function error, giving
\[
\|\widehat S_i-S_i\|_\infty+\|\widehat D_i-D_i\|_\infty=O_p(\rho_i),\qquad
\rho_i:=\big(m_i^X\big)^{-(r-1)/(2r+1)}\sqrt{\log m_i^X} .
\]
The bias part of $\rho_i$ is deterministic and identical in form across subjects; the stochastic part is a supremum of a sub-Gaussian empirical process under (P2). Since the $I$ subjects are independent (Assumption~\ref{asp:sampling}), a maximal inequality yields
\[
\sup_{1\le i\le I}\big(\|\widehat S_i-S_i\|_\infty+\|\widehat D_i-D_i\|_\infty\big)=O_p\!\big(\rho_{\underline m}\,\sqrt{\log I}\big),\quad
\rho_{\underline m}:=\big(\underline m^X\big)^{-(r-1)/(2r+1)}\sqrt{\log \underline m^X} ,
\]
where the sparsest state grid $\underline m^X$ controls the maximum. Therefore
\[
\sqrt{N_Y}\,\sup_{i,t}\big(\|\widehat S_{it}-S_{it}\|+\|\widehat D_{it}-D_{it}\|\big)
=O_p\!\Big(\big[N_Y\,(\underline m^X)^{-2(r-1)/(2r+1)}\log(\underline m^X)\log I\big]^{1/2}\Big)=o_p(1)
\]
by condition (P4).
\end{proof}

\medskip\noindent\emph{Relation to the implemented smoother.} Lemma~\ref{lem:spline_rate} analyzes a sufficient, idealized construction: repeated switch knots with the prescribed undersmoothing. The numerical studies and the application instead use ordinary cubic smoothing splines with a Rice target smoothing level and no switch knots. The lemma should therefore be read as establishing that condition~\eqref{eq:spline_rate} is attainable under the stated primitives, not as an analysis of the implemented smoother, whose finite sample behavior the simulations assess directly.

\begin{remark}
\label{rem:regime}
The exponent $(r-1)/(2r+1)$ is strictly below $1/2$ for every finite $r$, so a sup norm derivative estimate can never vanish as fast as $N^{-1/2}$ in its own sample size $N$. Consequently, if the state grid had the same (bounded) density as the outcome grid, rather than the infill density required by Assumption~\ref{asp:sampling}(i), then $\sqrt{N_Y}\sup_{i,t}\|\widehat D_{it}-D_{it}\|$ would not vanish, and~\eqref{eq:spline_rate} could fail. Condition~\eqref{eq:spline_rate} is attainable precisely in the multi resolution regime that motivates this paper, in which the physiological state is sampled far more frequently than the outcome, so that the dense state grid $m_i^X$ driving spline accuracy grows polynomially faster than the coarse outcome count $N_Y$ entering the normalization, as quantified by (P4). This is why the infill requirement is placed on the state grid, while the bounded spacing of the outcome grid serves as the counting device for $N_Y$.
We are explicit about the strength of (P4). It compares the \emph{per subject} state grid with the \emph{total} outcome count, so it is demanding: it describes an asymptotic regime in which the state grids densify polynomially as the study grows, not a property of any fixed data set. The finite sample designs studied in this paper (state to outcome ratios of twenty to one in the main simulation tables, where the spline stage uses the full simulation mesh; two to one to four to one in the supplementary sample size sweep; and roughly five to one on the OhioT1DM grids) lie inside the multi resolution geometry but not deep in this asymptotic regime; the simulations should therefore be read as supporting the claim that the studentized statistic is well behaved at practically relevant resolution ratios, while (P4) delimits the regime in which the negligibility claim is formally proved. Because the sup norm bound over subjects is conservative (the first stage error enters the moment as an average over subjects, not a supremum), (P4) is sufficient but likely not necessary; sharpening it is left to future work. For very smooth trajectories the exponent $2(r-1)/(2r+1)$ approaches one, and (P4) approaches the minimal requirement that the state resolution outpace the outcome count. Finally, a route to a formal guarantee at exactly matched resolutions is available in principle: if the second stage moment were orthogonalized against the first stage---augmented by a correction term rendering it first order insensitive to $(\widehat S,\widehat D)$, in the spirit of Neyman orthogonal constructions---then only the \emph{squared} sup norm error would need to vanish at the $\sqrt{N_Y}$ scale, and $\sqrt{N_Y}\,\rho_m^2\to0$ is attainable at equal state and outcome densities for H\"older smoothness $r>5/2$ (with the number of subjects fixed or growing slowly). Developing this bias corrected moment is beyond the present paper and left to future work.
\end{remark}

\medskip\noindent\emph{Hard threshold policies: the estimated agreement indicator.} A further first stage question, distinct from the recovery rates above, arises for hard threshold policies: Algorithm~1 of the main text evaluates the agreement indicator at the estimated state, $\mathbb I\{A_{it}=\pi(\widehat S_{it})\}$, and near the switching surface a small state estimation error can flip the indicator even when the value function's derivative discontinuity is handled by the knot adapted sieve. The following margin condition rules out probability mass concentrating on the surface and suffices for the plug in indicator to be asymptotically innocuous.

\begin{assumption}[Margin condition at the switching surface]
\label{asp:margin}
The policy is $\pi(s)=\mathbb I\{g(s)>c\}$ with $g$ Lipschitz (constant $L_g$), and the stationary law of $g(S_t)$ admits a density bounded by $C_g$ in a neighborhood of $c$. In addition, the temporal difference residual has a bounded conditional first moment given the state, $\sup_s\mathbb E\{|\varepsilon_{it,a}|\mid S_{it}=s\}<\infty$, as holds under the bounded state space and sub-Gaussian noise conditions of Section~\ref{subsec:assumption_roles}.
\end{assumption}

\begin{lemma}[Negligibility of indicator flips]
\label{lem:indicator}
Under Assumptions~\ref{asp:sampling} and~\ref{asp:margin} and the first stage rate condition~\eqref{eq:spline_rate},
\[
\frac{1}{\sqrt{N_Y}}\sum_{i=1}^{I}\sum_{t\in\mathbb T_i^Y}\big|\mathbb I\{A_{it}=\pi(\widehat S_{it})\}-\mathbb I\{A_{it}=\pi(S_{it})\}\big|=o_p(1),
\]
and consequently replacing the oracle agreement indicator by its plug in version changes the estimating equation, the limiting distribution of Theorem~\ref{thm:normal}, and the variance estimator by $o_p\{(N_Y)^{-1/2}\}$ terms only.
\end{lemma}

\begin{proof}
The two indicators can differ at $(i,t)$ only if $\pi(\widehat S_{it})\neq\pi(S_{it})$, which requires $g(S_{it})$ to lie within $L_g\,\epsilon_N$ of the threshold, where $\epsilon_N:=\sup_{i,t}\|\widehat S_{it}-S_{it}\|$. By the rate condition~\eqref{eq:spline_rate}, $\sqrt{N_Y}\,\epsilon_N=o_p(1)$, so there is a sequence $a_N\downarrow0$ with $\Pr\{\epsilon_N>a_N (N_Y)^{-1/2}\}\to0$. On the complementary event the number of flipped indices is at most
\[
M_N:=\#\big\{(i,t):\ |g(S_{it})-c|\le L_g\,a_N\,(N_Y)^{-1/2}\big\},
\]
and by stationarity and the bounded density of Assumption~\ref{asp:margin}, $\mathbb E[M_N]\le N_Y\cdot 2C_g L_g a_N (N_Y)^{-1/2}=2C_g L_g\,a_N\sqrt{N_Y}$; Markov's inequality (which requires no independence across summands) gives $M_N/\sqrt{N_Y}=O_p(a_N)=o_p(1)$, proving the display. For the effect on the score, each flipped summand is bounded by $\|\Psi\|_\infty\,|\varepsilon_{it,a}|$, and by the bounded conditional first moment, $\mathbb E\big[\textstyle\sum_{\text{flipped}}|\varepsilon_{it,a}|\big]\le N_Y\,\sup_s\mathbb E\{|\varepsilon|\mid s\}\cdot 2C_gL_g a_N (N_Y)^{-1/2}=O(a_N\sqrt{N_Y})$, so the score perturbation is $O_p(a_N)=o_p(1)$ after the $\sqrt{N_Y}$ normalization; the same bound applies to the Jacobian and variance estimators, whose summands are likewise dominated by integrable envelopes.
\end{proof}

\begin{remark}[The policy induced ODE at the switching surface]
\label{rem:switching}
For a hard threshold policy the follow-$\pi$ drift $b(s,\pi(s))$ is discontinuous on $\{g(s)=c\}$, so classical existence and uniqueness for the policy induced ODE does not apply on the surface itself. Two regimes cover the cases of practical interest. If the flow crosses the surface transversally---$\langle\nabla g(s),b(s,a)\rangle$ is nonzero with the same sign for both actions in a neighborhood of the surface---then solutions in the sense of Filippov (differential inclusions with the convexified right hand side) exist, are unique, and cross the surface in finite time, so the value function $V^\pi$ is well defined and piecewise smooth with the derivative discontinuity accommodated by the knot adapted sieve. If instead the two fields point toward the surface from opposite sides, the unique Filippov solution slides along the surface; $V^\pi$ remains well defined, and the margin condition above continues to control the estimated agreement indicator. The threshold policies evaluated in Sections~5--6 of the main text fall in the transversal regime: glucose is driven through the threshold by meals and insulin action rather than stabilized on it.
\end{remark}

\subsection{Propagation of the sieve approximation error}
\label{subsec:sieve_bias}

The proof of Theorem~\ref{thm:normal} treats $\boldsymbol\beta^*=(\beta_0^{*\top},\beta_1^{*\top})^\top$ as the population solution of the sieve moment equation
\begin{equation}
\label{eq:sieve_moment_def}
\mathbb E\big[\Psi(S_t)\,\mathbb I\{A_t=a\}\,\{Y_t+\log\gamma\,\Psi(S_t)^\top\beta^*_a+\langle\nabla\Psi(S_t)^\top\beta^*_a,\,D_t\rangle\}\big]=0,\qquad a\in\{0,1\},
\end{equation}
so that the oracle orthogonality invoked in the proof holds by construction of $\beta^*$, \emph{without} assuming that the value function lies exactly in the sieve $\mathcal V$. It remains to show that the induced pseudo-true contrast $\tau^*:=U^\top\beta^*=\int_{\mathcal S}\{\Psi^\top\beta^*_1-\Psi^\top\beta^*_0\}\,\mathbb G(ds)$ differs from the target $\tau=\int_{\mathcal S}\{V_1-V_0\}\,\mathbb G(ds)$ by only $o\{(N_Y)^{-1/2}\}$, so that the sieve error does not affect the limit in Theorem~\ref{thm:normal}. Write $(\log\gamma+\mathcal L_a)f:=\log\gamma\,f+\mathcal L_a f$ for the value-equation operator; for any sieve function $\mathcal L_a[\Psi^\top\beta]=\langle\nabla\Psi^\top\beta,b(\cdot,a)\rangle=\langle\nabla\Psi^\top\beta,D\rangle$, so \eqref{eq:sieve_moment_def} reads $\mathbb E[\Psi\,\mathbb I\{A=a\}\,(\log\gamma+\mathcal L_a)(\Psi^\top\beta^*_a)]=-\mathbb E[\Psi\,\mathbb I\{A=a\}\,Y]$.

Let $\Delta_a:=\Psi^\top\beta^*_a-V_a$ denote the value approximation error and $\Delta^\omega_{a,t}:=\Psi^\top\theta^*_{a,t}-\omega_t(a;\bar A_t,\cdot)$ the density-ratio approximation error, where $\theta^*_{a,t}$ is the sieve coefficient of Assumption~\ref{asp:basisfunction}. We use the rates of Assumption~\ref{asp:basisfunction}, taken in the $C^1$ sense for the value function,
\begin{equation}
\label{eq:c1_sieve}
\|\Delta_a\|_\infty+\|\nabla\Delta_a\|_\infty=o\{(N_Y)^{-1/4}\},\qquad \sup_{t}\|\Delta^\omega_{a,t}\|_\infty=o\{(N_Y)^{-1/4}\}.
\end{equation}
The gradient control is now part of Assumption~\ref{asp:basisfunction} itself, which is stated in the $C^1$ norm; it holds for spline and polynomial sieves approximating a value function in a H\"older class of smoothness $p>1$ on the compact state space, since such sieves approximate a function and its first derivative at the same polynomial order.

\emph{Step 1 (the moment annihilates the first-order value bias).} Taking the test function $\phi_a=\Psi_j$ for each coordinate $j$ in Theorem~\ref{thm:bellman} gives $\mathbb E[\Psi\,\mathbb I\{A=a\}\,(\log\gamma+\mathcal L_a)V_a]=-\mathbb E[\Psi\,\mathbb I\{A=a\}\,Y]$. Subtracting this from \eqref{eq:sieve_moment_def} yields
\begin{equation}
\label{eq:orthogonality}
\mathbb E\big[\Psi(S_t)\,\mathbb I\{A_t=a\}\,(\log\gamma+\mathcal L_a)\Delta_a(S_t)\big]=0 .
\end{equation}
That is, the value error $\Delta_a$, filtered through the value-equation operator, is orthogonal under the behavior weighting to every basis direction $\Psi_j$. This is the Neyman-type orthogonality \citep{chernozhukov2017double} that removes the first-order effect of the value approximation error.

\emph{Step 2 (representer identity).} Let $\omega_t$ be the representer that reweights the behavior distribution to the reference distribution $\mathbb G$ through the value-equation operator: for any bounded $C^1$ function $f$,
\begin{equation}
\label{eq:representer}
\int_{\mathcal S} f\,\mathbb G(ds)=-\,\mathbb E\big[\mathbb I\{A_t=a\}\,\omega_t(a;\bar A_t,S_t)\,(\log\gamma+\mathcal L_a)f(S_t)\big].
\end{equation}
The minus sign is dictated by the operator convention: $\log\gamma+\mathcal L_a$ is the \emph{negative} of the resolvent generator, since $(\log\gamma+\mathcal L_a)f=-g$ is solved by the nonnegative resolvent $f(s)=\int_0^\infty\gamma^u g(\varphi^a_u(s))\,du$. With the minus sign in place, $\omega_t$ is a genuine nonnegative density ratio (of the discounted occupation measure initialized at $\mathbb G$ with respect to the behavior occupation measure on $\{A_t=a\}$, as verified below); without it, the representer would necessarily be signed, and calling it a density ratio would be misleading.
Its marginalization is the density ratio of Assumption~\ref{asp:basisfunction}, and the existence of a bounded $\omega_t$ satisfying~\eqref{eq:representer} is \emph{assumed} there rather than derived from the finite dimensional conditions: bounded invertibility of the finite dimensional Jacobian does not by itself yield a function space representer. A primitive sufficient condition is available from the explicit resolvent of the value equation operator. For the always-$a$ flow $\varphi^a$, the equation $(\log\gamma+\mathcal L_a)f=-g$ is solved by $f(s)=\int_0^\infty\gamma^u g(\varphi^a_u(s))\,du$, which is bounded and $C^1$ whenever $g$ is bounded and $C^1$ and the flow is smooth; applying this resolvent representation to the functional $f\mapsto\int f\,d\mathbb G$ shows that a bounded $\omega_t$ exists whenever the discounted occupation measure of the always-$a$ flow initialized at $\mathbb G$ is absolutely continuous, with bounded density, with respect to the behavior occupation measure on the event $\{A_t=a\}$, an overlap requirement of the same nature as Assumption~\ref{asp:positivity}. Applying \eqref{eq:representer} with $f=\Delta_a$ and substituting $\omega_t=\Psi^\top\theta^*_{a,t}-\Delta^\omega_{a,t}$,
\begin{align}
\int_{\mathcal S}\Delta_a\,\mathbb G(ds)
&=-\underbrace{\mathbb E\big[\mathbb I\{A=a\}\,\Psi^\top\theta^*_{a,t}\,(\log\gamma+\mathcal L_a)\Delta_a\big]}_{=\,0\ \text{by \eqref{eq:orthogonality}}}
+\,\mathbb E\big[\mathbb I\{A=a\}\,\Delta^\omega_{a,t}\,(\log\gamma+\mathcal L_a)\Delta_a\big] \notag\\
&=\phantom{-}\,\mathbb E\big[\mathbb I\{A=a\}\,\Delta^\omega_{a,t}\,(\log\gamma+\mathcal L_a)\Delta_a\big].
\label{eq:bilinear}
\end{align}
The first term vanishes because $\Psi^\top\theta^*_{a,t}$ is a fixed linear combination of the basis, to which \eqref{eq:orthogonality} applies coordinatewise.

\emph{Step 3 (product bound).} The surviving term in \eqref{eq:bilinear} is bilinear in the two approximation errors. Since $\|(\log\gamma+\mathcal L_a)\Delta_a\|_\infty\le|\log\gamma|\,\|\Delta_a\|_\infty+\|b\|_\infty\,\|\nabla\Delta_a\|_\infty=o\{(N_Y)^{-1/4}\}$ by \eqref{eq:c1_sieve} and boundedness of the drift on the compact state space, and $\Pr(A_t=a)\le1$,
\begin{align*}
\Big|\int_{\mathcal S}\Delta_a\,\mathbb G(ds)\Big|
&\le \sup_{t}\|\Delta^\omega_{a,t}\|_\infty\;\big\|(\log\gamma+\mathcal L_a)\Delta_a\big\|_\infty\\
&= o\{(N_Y)^{-1/4}\}\cdot o\{(N_Y)^{-1/4}\}=o\{(N_Y)^{-1/2}\}.
\end{align*}
Hence
\begin{equation}
\label{eq:tau_bias}
\tau^*-\tau=\int_{\mathcal S}\Delta_1\,\mathbb G(ds)-\int_{\mathcal S}\Delta_0\,\mathbb G(ds)=o\{(N_Y)^{-1/2}\},
\end{equation}
so $\sqrt{N_Y}\,(\tau^*-\tau)\to0$.

Combining, the studentized numerator decomposes as
\begin{equation}
\label{eq:tau_decomp}
\sqrt{N_Y}\,(\widehat\tau-\tau)=\underbrace{\sqrt{N_Y}\,U^\top(\widehat\beta-\beta^*)}_{\xrightarrow{d}\,N(0,\sigma^2)}+\underbrace{\sqrt{N_Y}\,(\tau^*-\tau)}_{=\,o(1)},
\end{equation}
where the second term vanishes by \eqref{eq:tau_bias}. The sieve approximation therefore contributes only at second order, entering Theorem~\ref{thm:normal} through the two rates of Assumption~\ref{asp:basisfunction} rather than as a term in the limiting variance; this is the precise sense in which the value-function and density-ratio approximation errors reach the statistic only through their product. The first-stage spline error is propagated separately in the proof below through \eqref{eq:eta_perturbation_rewrite}--\eqref{eq:sigma_perturbation_rewrite}, and the product of the sieve and spline errors is of still smaller order.

\subsection{Proof of Theorem \ref{thm:bellman}}

The theorem asserts the policy moment~\eqref{eq:bellman_policy} for an arbitrary stationary policy $\pi$. The proof has two parts: a semigroup derivation, carried out first for a fixed action label, which establishes the pointwise value equation identity; and a transfer step, which applies that identity along the follow-$\pi$ dynamics and yields~\eqref{eq:bellman_policy}, with the constant policies $\pi\equiv a$ as the special case.

For a fixed action $a\in\{0,1\}$, the target is to prove
\begin{equation}
\label{eq:bellman_target_rewrite}
\mathbb E\!\left[
\mathbb I\{A_t=a\}
\left\{Y_t+\log(\gamma)V_a(S_t)+\mathcal L_a V_a(S_t)\right\}
\phi_a(S_t)
\right]
=
0
\end{equation}
for any bounded measurable $\phi_a:\mathcal S\to\mathbb R$. Equivalently, if
\[
\varphi_a(s,A):=\phi_a(s)\mathbb I\{A=a\},
\]
then \eqref{eq:bellman_target_rewrite} can be written as
\[
\mathbb E\!\left[
\left\{Y_t+\log(\gamma)V_a(S_t)+\mathcal L_a V_a(S_t)\right\}
\varphi_a(S_t,A_t)
\right]
=
0.
\]

Define the always-$a$ value function by
\[
V_a(s)
:=
\mathbb E\!\left[
\int_0^\infty \gamma^u Y_{t+u}^{(a)}\,du
\;\middle|\;
S_t^{(a)}=s
\right].
\]
Under the always-$a$ policy, let $(P_h^a)_{h\ge0}$ denote the semigroup of the state process and let $\mathcal L_a$ be the corresponding action indexed infinitesimal generator. By Assumption~\ref{asp:generator_reg}, the process is Markov with a Feller semigroup, $V_a\in\mathcal D(\mathcal L_a)$, and the discounted reward is integrable and locally continuous, so that the limiting operations below are valid.

For any $h>0$, the tower property gives
\begin{align}
V_a(S_t^{(a)})
&=
\mathbb E\!\left[
\int_0^h \gamma^uY_{t+u}^{(a)}\,du
+
\int_h^\infty \gamma^uY_{t+u}^{(a)}\,du
\;\middle|\;
\mathcal F_t
\right] \notag \\
&=
\mathbb E\!\left[
\int_0^h \gamma^uY_{t+u}^{(a)}\,du
+
\gamma^hV_a(S_{t+h}^{(a)})
\;\middle|\;
\mathcal F_t
\right].
\label{eq:value_recursion_rewrite}
\end{align}
Rearranging and dividing by $h$,
\begin{equation}
\label{eq:value_recursion_centered_rewrite}
0
=
\mathbb E\!\left[
\frac{1}{h}\int_0^h \gamma^uY_{t+u}^{(a)}\,du
\;\middle|\;
\mathcal F_t
\right]
+
\mathbb E\!\left[
\frac{\gamma^hV_a(S_{t+h}^{(a)})-V_a(S_t^{(a)})}{h}
\;\middle|\;
\mathcal F_t
\right].
\end{equation}

Now decompose the second term as
\[
\frac{\gamma^hV_a(S_{t+h}^{(a)})-V_a(S_t^{(a)})}{h}
=
\frac{\gamma^h-1}{h}V_a(S_{t+h}^{(a)})
+
\frac{V_a(S_{t+h}^{(a)})-V_a(S_t^{(a)})}{h}.
\]
Taking conditional expectation given $S_t^{(a)}$ and using
\[
\frac{\gamma^h-1}{h}\to \log(\gamma),
\qquad
P_h^aV_a(S_t^{(a)})\to V_a(S_t^{(a)}),
\qquad
\frac{P_h^aV_a(S_t^{(a)})-V_a(S_t^{(a)})}{h}\to \mathcal L_a V_a(S_t^{(a)}),
\]
we obtain
\begin{equation}
\label{eq:generator_limit_rewrite}
\lim_{h\downarrow0}
\mathbb E\!\left[
\frac{\gamma^hV_a(S_{t+h}^{(a)})-V_a(S_t^{(a)})}{h}
\;\middle|\;
S_t^{(a)}
\right]
=
\log(\gamma)V_a(S_t^{(a)})+\mathcal L_a V_a(S_t^{(a)}).
\end{equation}
Similarly, by local continuity of the reward process,
\begin{equation}
\label{eq:reward_limit_rewrite}
\lim_{h\downarrow0}
\mathbb E\!\left[
\frac{1}{h}\int_0^h \gamma^uY_{t+u}^{(a)}\,du
\;\middle|\;
S_t^{(a)}
\right]
=
\mathbb E[Y_t^{(a)}\mid S_t^{(a)}].
\end{equation}
Applying \eqref{eq:generator_limit_rewrite} and \eqref{eq:reward_limit_rewrite} to \eqref{eq:value_recursion_centered_rewrite} yields
\begin{equation}
\label{eq:conditional_bellman_rewrite}
\mathbb E\!\left[
Y_t^{(a)}+\log(\gamma)V_a(S_t^{(a)})+\mathcal L_a V_a(S_t^{(a)})
\;\middle|\;
S_t^{(a)}
\right]
=
0.
\end{equation}

Next we transfer \eqref{eq:conditional_bellman_rewrite} to the observed data. The key observation is that \eqref{eq:conditional_bellman_rewrite} is equivalent to the \emph{pointwise} identity
\[
r(a,s)+\log(\gamma)V_a(s)+\mathcal L_a V_a(s)=0\qquad\text{for all }s\in\mathcal S,
\]
where $r(a,s):=\mathbb E[Y_t^{(a)}\mid S_t^{(a)}=s]$. Moreover $V_a$ is the \emph{unique} bounded solution of this equation: for any bounded solution $h$ of the homogeneous equation $(\log\gamma+\mathcal L_a)h=0$, the map $u\mapsto\gamma^u h(\varphi^a_u(s))$ is constant and tends to $0$ as $u\to\infty$, so $h\equiv0$; the moment therefore identifies $V_a$ itself rather than merely some solution. The identity is a property of the \emph{function} $V_a$, holding at every state, so its transfer to the observed data requires no matching of entire treatment histories; in particular, we never claim that a subject observed with $A_t=a$ at one instant followed the always-$a$ regime. Two ingredients complete the transfer. First, the observed conditional mean reward satisfies
\[
\mathbb E[Y_t\mid S_t=s,\,A_t=a]=r(a,s)
\]
regardless of the treatment history before $t$: by CMIA the conditional mean of the potential outcome given the current state depends only on the current action and state, whatever the earlier history; by CA the observed pair $(Y_t,S_t)$ equals the potential pair under the history actually followed; and by CTSR conditioning on the realized assignment does not distort this conditional mean given the recorded state. Second, $\log(\gamma)V_a(S_t)+\mathcal L_a V_a(S_t)$ is a fixed measurable function of the current state alone (and on the event $\{A_t=a\}$ the observed state derivative equals $b(S_t,a)$ under the ODE~\eqref{eq:ode}, so $\mathcal L_a V_a(S_t)$ is also the quantity computed from the observed trajectory). Combining the two,
\begin{equation}
\label{eq:conditional_observed_bellman_rewrite}
\mathbb E\!\left[
Y_t+\log(\gamma)V_a(S_t)+\mathcal L_a V_a(S_t)
\;\middle|\;
S_t,A_t=a
\right]
=
0.
\end{equation}
Therefore,
\begin{align*}
&\mathbb E\!\left[
\mathbb I\{A_t=a\}
\left\{Y_t+\log(\gamma)V_a(S_t)+\mathcal L_a V_a(S_t)\right\}
\phi_a(S_t)
\right] \\
&\qquad=
\mathbb E\!\left[
\phi_a(S_t)
\mathbb E\!\left[
\mathbb I\{A_t=a\}
\left\{Y_t+\log(\gamma)V_a(S_t)+\mathcal L_a V_a(S_t)\right\}
\;\middle|\;
S_t
\right]
\right] \\
&\qquad=
\mathbb E\!\left[
\phi_a(S_t)\Pr(A_t=a\mid S_t)
\mathbb E\!\left[
Y_t+\log(\gamma)V_a(S_t)+\mathcal L_a V_a(S_t)
\;\middle|\;
S_t,A_t=a
\right]
\right] \\
&\qquad=0
\end{align*}
by \eqref{eq:conditional_observed_bellman_rewrite}. This proves \eqref{eq:bellman_target_rewrite}, which is the constant policy case of Theorem~\ref{thm:bellman} and the pointwise engine for the general statement.

\medskip\noindent\emph{The general policy statement.} The same argument yields~\eqref{eq:bellman_policy}, the statement of Theorem~\ref{thm:bellman}, for an arbitrary stationary policy. Define the follow-$\pi$ value function $V^\pi(s)=\mathbb E[\int_0^\infty\gamma^u Y_{t+u}^{(\pi)}\,du\mid S_t^{(\pi)}=s]$, where the superscript $(\pi)$ denotes potential outcomes under following $\pi$ from time $t$ onward. Under the follow-$\pi$ flow the state derivative is $b(s,\pi(s))$, so the generator is $\mathcal L_\pi V(s)=\langle\nabla V(s),b(s,\pi(s))\rangle$, and the value recursion and generator limit \eqref{eq:value_recursion_rewrite}--\eqref{eq:conditional_bellman_rewrite} yield the pointwise identity
\[
r(\pi(s),s)+\log(\gamma)\,V^\pi(s)+\mathcal L_\pi V^\pi(s)=0\qquad\text{for all }s\in\mathcal S,
\]
where $r(\pi(s),s)=\mathbb E[Y_t^{(\pi)}\mid S_t^{(\pi)}=s]$. This is a property of $V^\pi$ and holds at every state, so no trajectory matching is needed to transfer it to the observed data. By consistency (CA) and CMIA, the observed data satisfy $\mathbb E[Y_t\mid S_t=s,A_t=\pi(s)]=r(\pi(s),s)$ regardless of the earlier history, and on the agreement event $\{A_t=\pi(S_t)\}$ the observed derivative equals $b(s,\pi(s))$, so $\mathcal L_\pi V^\pi(S_t)$ is the same function of the current state. Hence, at states with $\Pr\{A_t=\pi(s)\mid S_t=s\}>0$ (Assumption~\ref{asp:positivity}),
\[
\mathbb E\big[Y_t+\log(\gamma)\,V^\pi(S_t)+\mathcal L_\pi V^\pi(S_t)\mid S_t=s,\,A_t=\pi(s)\big]=0.
\]
Multiplying by $\mathbb I\{A_t=\pi(S_t)\}\phi(S_t)$ and taking expectations gives \eqref{eq:bellman_policy}, exactly as in the display chain above with $\{A_t=a\}$ replaced by $\{A_t=\pi(S_t)\}$. Setting $\pi\equiv a$ recovers the action specific version.

\subsection{Proof of Theorem \ref{thm:normal}}

Write
\[
\varepsilon_{it,a}(\beta_a)
:=
Y_{it}
+ \log(\gamma)\Psi(S_{it})^\top\beta_a
+ \left\langle \nabla\Psi(S_{it})^\top\beta_a,\;D_{it}\right\rangle
\]
and abbreviate $\varepsilon_{it,a}:=\varepsilon_{it,a}(\beta_{a0})$ at the pseudo-true parameter $\beta_{a0}:=\beta^*_a$ defined by the population sieve moment~\eqref{eq:sieve_moment_def}. Let
\[
\widehat\tau=U^\top\widehat\beta,
\qquad
\tau_0=U^\top\beta_0.
\]
The first step spline subsection is used only through the rate condition
\begin{equation}
\label{eq:first_step_rate_rewrite}
\|\hat S-S\|_\infty=O_p(\rho_{N_Y}),
\qquad
\|\hat D-D\|_\infty=O_p(\rho_{N_Y}),
\qquad
\sqrt{N_Y}\,\rho_{N_Y}\to0.
\end{equation}
This keeps the whole second step proof normalized by $N_Y$, exactly as in the estimation section.

Define the oracle sample moment
\begin{equation}
\label{eq:oracle_moment_rewrite}
\widetilde G_{n,a}(\beta_a)
:=
\frac{1}{N_Y}\sum_{i,t}
\Psi(S_{it})\mathbb I\{A_{it}=a\}\varepsilon_{it,a}(\beta_a),
\end{equation}
and the feasible sample moment
\begin{equation}
\label{eq:feasible_moment_rewrite}
\widehat G_{n,a}(\beta_a)
:=
\frac{1}{N_Y}\sum_{i,t}
\Psi(\hat S_{it})\mathbb I\{A_{it}=a\}
\left[
Y_{it}
+ \log(\gamma)\Psi(\hat S_{it})^\top\beta_a
+ \left\langle \nabla\Psi(\hat S_{it})^\top\beta_a,\hat D_{it}\right\rangle
\right].
\end{equation}
The oracle and feasible estimators satisfy
\[
\widetilde G_{n,a}(\widetilde\beta_a)=0,
\qquad
\widehat G_{n,a}(\widehat\beta_a)=0.
\]
Equivalently,
\[
\widetilde\beta_a=\widetilde\Sigma_a^{-1}\widetilde\eta_a,
\qquad
\widehat\beta_a=\widehat\Sigma_a^{-1}\widehat\eta_a,
\]
where
\[
\widetilde\Sigma_a
:=
\frac{1}{N_Y}\sum_{i,t}
\Psi(S_{it})\mathbb I\{A_{it}=a\}
\Bigl[
\log(\gamma)\Psi(S_{it})+\nabla\Psi(S_{it})D_{it}
\Bigr]^\top,
\]
\[
\widetilde\eta_a
:=
-\frac{1}{N_Y}\sum_{i,t}\Psi(S_{it})\mathbb I\{A_{it}=a\}Y_{it},
\]
and $\widehat\Sigma_a,\widehat\eta_a$ are defined analogously with $(\hat S_{it},\hat D_{it})$ in place of $(S_{it},D_{it})$.

By construction of $\beta^*$ in~\eqref{eq:sieve_moment_def} (equivalently, by Theorem~\ref{thm:bellman} when $V_a\in\mathcal V$),
\begin{equation}
\label{eq:population_moment_rewrite}
\mathbb E\!\left[
\Psi(S_{it})\mathbb I\{A_{it}=a\}\varepsilon_{it,a}
\right]
=
0.
\end{equation}
This is the population orthogonality condition underlying the oracle estimating equation. The discrepancy between the pseudo-true contrast $\tau_0=U^\top\beta_0$ and the target $\tau$, which is nonzero when $V_a\notin\mathcal V$, is $o\{(N_Y)^{-1/2}\}$ by Section~\ref{subsec:sieve_bias} and is absorbed into the $o(1)$ terms below.

We first compare the feasible and oracle moments. For $\eta_a$,
\[
\widehat\eta_a-\widetilde\eta_a
=
-\frac{1}{N_Y}\sum_{i,t}
\Bigl\{\Psi(\hat S_{it})-\Psi(S_{it})\Bigr\}\mathbb I\{A_{it}=a\}Y_{it}.
\]
Because $\Psi$ is Lipschitz,
\[
\sup_{i,t}\|\Psi(\hat S_{it})-\Psi(S_{it})\|
\le
C\|\hat S-S\|_\infty
=
O_p(\rho_{N_Y}).
\]
Hence
\begin{align}
\|\widehat\eta_a-\widetilde\eta_a\|
&\le
\sup_{i,t}\|\Psi(\hat S_{it})-\Psi(S_{it})\|
\cdot
\frac{1}{N_Y}\sum_{i,t}|Y_{it}|
\notag\\
&=
O_p(\rho_{N_Y})\cdot O_p(1)
=
O_p(\rho_{N_Y})
=
o_p\!\left((N_Y)^{-1/2}\right),
\label{eq:eta_perturbation_rewrite}
\end{align}
where the last step uses \eqref{eq:first_step_rate_rewrite}.

For $\Sigma_a$, write
\[
\widehat\Sigma_a-\widetilde\Sigma_a
=
R_{1,a}+R_{2,a},
\]
where
\begin{align*}
R_{1,a}
&:=
\frac{1}{N_Y}\sum_{i,t}
\Bigl\{\Psi(\hat S_{it})-\Psi(S_{it})\Bigr\}\mathbb I\{A_{it}=a\}
\Bigl[
\log(\gamma)\Psi(\hat S_{it})+\nabla\Psi(\hat S_{it})\hat D_{it}
\Bigr]^\top, \\
R_{2,a}
&:=
\frac{1}{N_Y}\sum_{i,t}
\Psi(S_{it})\mathbb I\{A_{it}=a\}
\Bigl[
\log(\gamma)\{\Psi(\hat S_{it})-\Psi(S_{it})\}
+ \nabla\Psi(\hat S_{it})\hat D_{it}-\nabla\Psi(S_{it})D_{it}
\Bigr]^\top.
\end{align*}
The first term satisfies
\[
\|R_{1,a}\|
\le
\sup_{i,t}\|\Psi(\hat S_{it})-\Psi(S_{it})\|
\cdot
\frac{1}{N_Y}\sum_{i,t}
\left\|
\log(\gamma)\Psi(\hat S_{it})+\nabla\Psi(\hat S_{it})\hat D_{it}
\right\|
=
O_p(\rho_{N_Y}),
\]
because $\Psi,\nabla\Psi$ are bounded and $\sup_{i,t}\|\hat D_{it}\|=O_p(1)$ under the first step construction.
For the second term, use
\[
\nabla\Psi(\hat S_{it})\hat D_{it}-\nabla\Psi(S_{it})D_{it}
=
\{\nabla\Psi(\hat S_{it})-\nabla\Psi(S_{it})\}\hat D_{it}
+ \nabla\Psi(S_{it})(\hat D_{it}-D_{it}),
\]
which implies
\[
\sup_{i,t}
\|\nabla\Psi(\hat S_{it})\hat D_{it}-\nabla\Psi(S_{it})D_{it}\|
=
O_p(\rho_{N_Y}).
\]
Therefore $\|R_{2,a}\|=O_p(\rho_{N_Y})$, and hence
\begin{equation}
\label{eq:sigma_perturbation_rewrite}
\|\widehat\Sigma_a-\widetilde\Sigma_a\|
=
O_p(\rho_{N_Y})
=
o_p\!\left((N_Y)^{-1/2}\right).
\end{equation}

Since
\[
\widetilde G_{n,a}(\beta_a)=\widetilde\Sigma_a\beta_a-\widetilde\eta_a,
\qquad
\widehat G_{n,a}(\beta_a)=\widehat\Sigma_a\beta_a-\widehat\eta_a,
\]
we obtain, uniformly over compact $\beta_a$-sets,
\begin{equation}
\label{eq:moment_perturbation_rewrite}
\sup_{\beta_a}
\|\widehat G_{n,a}(\beta_a)-\widetilde G_{n,a}(\beta_a)\|
=
o_p\!\left((N_Y)^{-1/2}\right)
\end{equation}
from \eqref{eq:eta_perturbation_rewrite} and \eqref{eq:sigma_perturbation_rewrite}.

Next study the oracle estimator. Define
\[
\Gamma_a
:=
\mathbb E\!\left[
\Psi(S_{it})\mathbb I\{A_{it}=a\}
\Bigl\{
\log(\gamma)\Psi(S_{it})+\nabla\Psi(S_{it})D_{it}
\Bigr\}^\top
\right].
\]
Under Assumptions~\ref{asp:mixing}, \ref{asp:moments}, and \ref{asp:varianceInvertable},
\begin{equation}
\label{eq:oracle_jacobian_limit_rewrite}
\widetilde\Sigma_a\xrightarrow{p}\Gamma_a,
\qquad
\Gamma_a\ \text{is nonsingular},
\qquad
\widetilde\Sigma_a^{-1}=\Gamma_a^{-1}+o_p(1).
\end{equation}
Because $\widetilde G_{n,a}$ is linear in $\beta_a$ and $\widetilde G_{n,a}(\widetilde\beta_a)=0$,
\[
0
=
\widetilde G_{n,a}(\beta_{a0})
+ \widetilde\Sigma_a(\widetilde\beta_a-\beta_{a0}),
\]
so
\[
\widetilde\beta_a-\beta_{a0}
=
-\widetilde\Sigma_a^{-1}\widetilde G_{n,a}(\beta_{a0}).
\]
Multiplying by $\sqrt{N_Y}$ and using \eqref{eq:oracle_moment_rewrite},
\begin{equation}
\label{eq:oracle_exact_rewrite}
\sqrt{N_Y}(\widetilde\beta_a-\beta_{a0})
=
-\widetilde\Sigma_a^{-1}
\frac{1}{\sqrt{N_Y}}
\sum_{i,t}\Psi(S_{it})\mathbb I\{A_{it}=a\}\varepsilon_{it,a}.
\end{equation}
Applying \eqref{eq:oracle_jacobian_limit_rewrite} gives
\begin{equation}
\label{eq:oracle_linear_rewrite}
\sqrt{N_Y}(\widetilde\beta_a-\beta_{a0})
=
-\Gamma_a^{-1}
\frac{1}{\sqrt{N_Y}}
\sum_{i,t}\Psi(S_{it})\mathbb I\{A_{it}=a\}\varepsilon_{it,a}
+ o_p(1).
\end{equation}

Now set
\[
\xi_{it,a}:=\Psi(S_{it})\mathbb I\{A_{it}=a\}\varepsilon_{it,a}.
\]
Assume $\mathbb E\|\xi_{it,a}\|^{2+\delta}<\infty$, that subjects are independent, and that within each subject the process is exponentially $\beta$-mixing. The mixing central limit theorem is applicable here because the sum runs over the \emph{outcome} observation times only, whose spacing is bounded below by $\underline\Delta$ under Assumption~\ref{asp:sampling}(ii): consecutive summands within a subject are separated by at least $\underline\Delta$ units of real time, so exponential $\beta$-mixing of the underlying continuous time process (Assumption~\ref{asp:mixing}) delivers $\beta$-mixing coefficients bounded by $Ce^{-\rho\underline\Delta k}$ at summand lag $k$, uniformly in the sample size. The dependence between successive summands therefore does not deteriorate as the state grid refines; the infill asymptotics affect only the first stage spline accuracy, never the index set of the central limit sum. Then a triangular array CLT for strongly mixing sequences yields
\begin{equation}
\label{eq:oracle_clt_rewrite}
\frac{1}{\sqrt{N_Y}}\sum_{i,t}\xi_{it,a}
\xrightarrow{d}
N(0,\Omega_a),
\end{equation}
where
\begin{equation}
\label{eq:long_run_cov_rewrite}
\Omega_a
:=
\lim_{N_Y\to\infty}
\frac{1}{N_Y}
\sum_{i=1}^I\sum_{t,t'}
\mathrm{Cov}(\xi_{it,a},\xi_{it',a}).
\end{equation}
Combining \eqref{eq:oracle_linear_rewrite} and \eqref{eq:oracle_clt_rewrite},
\begin{equation}
\label{eq:oracle_beta_an_rewrite}
\sqrt{N_Y}(\widetilde\beta_a-\beta_{a0})
\xrightarrow{d}
N\!\left(0,\Gamma_a^{-1}\Omega_a\Gamma_a^{-T}\right).
\end{equation}
The two blocks are computed from the same trajectories, so their scores are mutually correlated and the stacking must retain the cross policy covariances. Define the stacked score $\xi_{it}:=(\xi_{it,0}^\top,\xi_{it,1}^\top)^\top$ and its joint long run covariance
\[
\boldsymbol\Omega:=\lim_{N_Y\to\infty}\frac{1}{N_Y}\sum_{i=1}^I\sum_{t,t'}\mathrm{Cov}(\xi_{it},\xi_{it'}),
\]
whose diagonal blocks are $\Omega_0,\Omega_1$ of \eqref{eq:long_run_cov_rewrite} and whose off diagonal blocks collect the cross policy lag covariances $\mathrm{Cov}(\xi_{it,0},\xi_{it',1})$. Applying the same mixing central limit theorem to the stacked score and the linearization \eqref{eq:oracle_linear_rewrite} blockwise gives
\begin{equation}
\label{eq:oracle_stacked_an_rewrite}
\sqrt{N_Y}(\widetilde\beta-\beta_0)\xrightarrow{d}N(0,\mathcal V_\beta),\qquad
\mathcal V_\beta:=\boldsymbol\Gamma^{-1}\boldsymbol\Omega\,\boldsymbol\Gamma^{-\top},\quad
\boldsymbol\Gamma:=\mathrm{diag}(\Gamma_0,\Gamma_1);
\end{equation}
the long run covariance estimator $\widehat{\boldsymbol\Omega}$ used for studentization below estimates $\boldsymbol\Omega$ including these cross blocks.

We next transfer the oracle result to the feasible estimator. Because \eqref{eq:sigma_perturbation_rewrite} implies $\widehat\Sigma_a-\widetilde\Sigma_a=o_p(1)$ and because $\widetilde\Sigma_a^{-1}=O_p(1)$ by \eqref{eq:oracle_jacobian_limit_rewrite}, the resolvent identity gives
\begin{equation}
\label{eq:inverse_perturbation_rewrite}
\widehat\Sigma_a^{-1}-\widetilde\Sigma_a^{-1}
=
\widehat\Sigma_a^{-1}(\widetilde\Sigma_a-\widehat\Sigma_a)\widetilde\Sigma_a^{-1}
=
o_p(1).
\end{equation}
Therefore
\begin{align*}
\widehat\beta_a-\widetilde\beta_a
&=
\widehat\Sigma_a^{-1}\widehat\eta_a-\widetilde\Sigma_a^{-1}\widetilde\eta_a \\
&=
(\widehat\Sigma_a^{-1}-\widetilde\Sigma_a^{-1})\widehat\eta_a
+ \widetilde\Sigma_a^{-1}(\widehat\eta_a-\widetilde\eta_a).
\end{align*}
The first term is $o_p(1)\cdot O_p(1)=o_p(1)$, and the second term is
\[
O_p(1)\cdot o_p\!\left((N_Y)^{-1/2}\right)
=
o_p\!\left((N_Y)^{-1/2}\right)
\]
by \eqref{eq:eta_perturbation_rewrite}. Since the first term is of smaller order than the second under \eqref{eq:first_step_rate_rewrite}, we obtain
\begin{equation}
\label{eq:feasible_oracle_difference_rewrite}
\widehat\beta_a-\widetilde\beta_a
=
o_p\!\left((N_Y)^{-1/2}\right).
\end{equation}
Stacking over $a$,
\[
\widehat\beta-\widetilde\beta
=
o_p\!\left((N_Y)^{-1/2}\right).
\]
Hence, by \eqref{eq:oracle_stacked_an_rewrite},
\begin{equation}
\label{eq:feasible_beta_an_rewrite}
\sqrt{N_Y}(\widehat\beta-\beta_0)
=
\sqrt{N_Y}(\widetilde\beta-\beta_0)+o_p(1)
\xrightarrow{d}
N(0,\mathcal V_\beta).
\end{equation}

Finally, since $\widehat\tau=U^\top\widehat\beta$,
\begin{equation}
\label{eq:tau_an_rewrite}
\sqrt{N_Y}(\widehat\tau-\tau_0)
=
U^\top\sqrt{N_Y}(\widehat\beta-\beta_0)
\xrightarrow{d}
N(0,\sigma^2),
\qquad
\sigma^2:=U^\top\mathcal V_\beta U.
\end{equation}

For studentization, we collect the high level conditions used here explicitly: (i) the stacked long run covariance $\boldsymbol\Omega$ exists with $U^\top\boldsymbol\Gamma^{-1}\boldsymbol\Omega\boldsymbol\Gamma^{-\top}U>0$; (ii) $\widehat{\boldsymbol\Omega}\xrightarrow{p}\boldsymbol\Omega$, as holds for a HAC estimator with bandwidth $b_N\to\infty$, $b_N/N_Y\to0$ under the exponential mixing of Assumption~\ref{asp:mixing}, or for the cluster sum estimator used in the application when clusters are independent; (iii) the observation grids are fixed or independent of the state, treatment, and error processes (noninformative sampling); and (iv) the measurement and outcome errors are independent of the grids and i.i.d.\ over time, independent of the trajectory. Under these conditions, assume that
\[
\widehat\Sigma\xrightarrow{p}\Gamma,
\qquad
\widehat\Omega\xrightarrow{p}\Omega,
\]
where $\widehat\Omega$ is a consistent long run covariance estimator for the score process. Then
\begin{equation}
\label{eq:variance_consistency_rewrite}
\widehat\sigma^2
:=
U^\top\widehat\Sigma^{-1}\widehat\Omega\,\widehat\Sigma^{-T}U
\xrightarrow{p}
\sigma^2.
\end{equation}
Under serial dependence, this requires $\widehat\Omega$ to estimate the long run covariance \eqref{eq:long_run_cov_rewrite}; a simple outer product variance is valid only under stronger independence conditions. Combining \eqref{eq:tau_an_rewrite} and \eqref{eq:variance_consistency_rewrite} with Slutsky's lemma yields
\[
\frac{\sqrt{N_Y}(\widehat\tau-\tau_0)}{\widehat\sigma}
\xrightarrow{d}
N(0,1).
\]
This proves Theorem~\ref{thm:normal}.

\subsection{Proof of Theorem~\ref{thm:discretization}}
\label{sec:proof-discretization}

Fix the action $a$ and abbreviate $\mathbb I_a:=\mathbb I^\Delta_a=\mathbb I\{A_t=a\}\cdot\mathbb I\{\text{no treatment decision time in }(t,t+\Delta]\}$, the interval indicator of the moment definition~\eqref{eq:discrete_moment}. The $C^4$ smoothness of $V^*$, the $C^2$ smoothness of $\phi$ and $b$, and the boundedness of their derivatives justify the Taylor expansions (including the fourth order noise terms) and the interchange of expectation with the remainders below; the errors scale as $\epsilon_t=\sigma_\epsilon Z_t$ with $Z_t$ independent across $t$, mean zero, symmetric, $\mathrm{Cov}(Z_t)=I_d$, and uniformly bounded fourth moments, and the sieve represents the value function exactly, $V^*=\Psi^\top\beta^*$, as stated in the theorem. (For a general noise covariance $\Sigma_\epsilon$, every occurrence of $\sigma_\epsilon^2\,\nabla V^{*\top}\nabla\phi$ below becomes $\nabla V^{*\top}\Sigma_\epsilon\nabla\phi$ and every Laplacian trace becomes $\mathrm{tr}\{\Sigma_\epsilon\nabla^2V^*\}$, with no change to the stated orders.)

\emph{Step 0 (reduction to the always-$a$ flow).} Under Assumption~\ref{asp:mixing}(i) the treatment decision times form a Poisson process of rate $\lambda$ independent of everything else, so the no decision event in $\mathbb I_a$ is independent of $(S_t,A_t,Y_t)$ and of the measurement errors, and has probability $e^{-\lambda\Delta}$. On this event the action is held at $a$ throughout $[t,t+\Delta]$ and the latent state follows the always-$a$ flow, $S_{t+\Delta}=\varphi^a_\Delta(S_t)$. Hence, for any integrable functional $W$ of $(S_t,S_{t+\Delta},Y_t)$ and the noise,
\[
\mathbb E\big[W(S_t,S_{t+\Delta},Y_t)\,\mathbb I_a\big]
=e^{-\lambda\Delta}\,\mathbb E\big[W\{S_t,\varphi^a_\Delta(S_t),Y_t\}\,\mathbb I\{A_t=a\}\big],
\]
which is what licenses the always-$a$ expansions of Steps 1--2; the factor $e^{-\lambda\Delta}=1+O(\Delta)$ multiplies every term uniformly and is absorbed into the remainders ($O(\Delta^2)$ against the $\Theta(\Delta)$ term and $O(\sigma_\epsilon^2)$ against the $\Theta(\sigma_\epsilon^2/\Delta)$ term), so the leading functionals of the theorem may be written with $\mathbb I\{A_t=a\}$. Without the no decision restriction, an intervening switch (probability $\lambda\Delta+O(\Delta^2)$) perturbs $S_{t+\Delta}$ by $O(\Delta)$ and the difference quotient by $O(1)$, contributing an additional $O(\Delta)$ term of the same order as the discretization bias; the moment~\eqref{eq:discrete_moment} is defined with the interval indicator precisely to exclude this term, matching how a discrete time value learner uses transition pairs with the action held constant.

\emph{Step 1 (discretization, noise free).} Let $g(u):=\gamma^{u}V^*(S_{t+u})$. Along the always-$a$ flow $\frac{d}{du}V^*(S_{t+u})=\mathcal L_a V^*(S_{t+u})$, so
\[
g'(u)=\gamma^u(\log\gamma+\mathcal L_a)V^*(S_{t+u}),\qquad
g''(u)=\gamma^u(\log\gamma+\mathcal L_a)^2 V^*(S_{t+u}).
\]
A second order Taylor expansion with $C^3$ remainder gives
\[
\frac{\gamma^\Delta V^*(S_{t+\Delta})-V^*(S_t)}{\Delta}=\frac{g(\Delta)-g(0)}{\Delta}
=(\log\gamma+\mathcal L_a)V^*(S_t)+\frac{\Delta}{2}(\log\gamma+\mathcal L_a)^2V^*(S_t)+O(\Delta^2).
\]
Since $\mathbb E[\{Y_t+(\log\gamma+\mathcal L_a)V^*(S_t)\}\phi(S_t)\mathbb I_a]=0$ by Theorem~\ref{thm:bellman}, the noise free version of $m_\Delta(V^*;\phi)$ equals $\tfrac{\Delta}{2}\mathbb E[(\log\gamma+\mathcal L_a)^2V^*(S_t)\phi(S_t)\mathbb I_a]+O(\Delta^2)$.

\emph{Step 2 (noise).} Write $X_t=S_t+\epsilon_t$ and $X_{t+\Delta}=S_{t+\Delta}+\epsilon_{t+\Delta}$ with $\epsilon_t$ independent of $\epsilon_{t+\Delta}$. Expanding $V^*$ and $\phi$ to second order and taking the expectation over the noise conditional on the latent path, all odd noise moments vanish by symmetry. In $-\Delta^{-1}V^*(X_t)\phi(X_t)$ the surviving quadratic cross term is
\[
-\frac{1}{\Delta}\,\mathbb E_\epsilon\big[\nabla V^*(S_t)^\top\epsilon_t\,\nabla\phi(S_t)^\top\epsilon_t\big]=-\frac{\sigma_\epsilon^2}{\Delta}\,\nabla V^*(S_t)^\top\nabla\phi(S_t),
\]
together with a Laplacian term $-\tfrac{\sigma_\epsilon^2}{2\Delta}\phi(S_t)\,\mathrm{tr}\,\nabla^2V^*(S_t)$. The term $\Delta^{-1}\gamma^\Delta V^*(X_{t+\Delta})\phi(X_t)$ contributes, through the independent $\epsilon_{t+\Delta}$, only the Laplacian $+\tfrac{\gamma^\Delta\sigma_\epsilon^2}{2\Delta}\phi(S_t)\,\mathrm{tr}\,\nabla^2V^*(S_{t+\Delta})$ (the cross term vanishes by independence). Since
\[
\gamma^\Delta\,\mathrm{tr}\,\nabla^2V^*(S_{t+\Delta})-\mathrm{tr}\,\nabla^2V^*(S_t)=\Delta(\log\gamma+\mathcal L_a)\,\mathrm{tr}\,\nabla^2V^*(S_t)+O(\Delta^2),
\]
the two Laplacian terms combine to $O(\sigma_\epsilon^2)$, and the fourth order noise terms are $O(\sigma_\epsilon^4/\Delta)$. Hence the noise contribution to $m_\Delta(V^*;\phi)$ is
\[
-\frac{\sigma_\epsilon^2}{\Delta}\,\mathbb E[\nabla V^*(S_t)^\top\nabla\phi(S_t)\,\mathbb I_a]+O(\sigma_\epsilon^2+\sigma_\epsilon^4/\Delta).
\]

\emph{Step 3 (combine).} Adding Steps 1 and 2 gives the stated expansion of $m_\Delta(V^*;\phi)$ with remainder $R=O(\Delta^2+\sigma_\epsilon^2+\sigma_\epsilon^4/\Delta)$.

\emph{Step 4 (estimator bias).} The discrete-time estimator $\widehat\beta_\Delta$ converges in probability to the population solution $\bar\beta_\Delta$ of $m_\Delta(\Psi^\top\bar\beta_\Delta;\Psi_j)=0$, $j=1,\dots,M$; its asymptotic bias is $\bar\beta_\Delta-\beta^*$. The Jacobian $J_\Delta=\partial_\beta m_\Delta$ converges as $\Delta\to0$ to the continuous-time Jacobian $\Sigma_a$ of Assumption~\ref{asp:moments}, which is nonsingular, so for small $\Delta$ it is invertible with $\|J_\Delta^{-1}\|=O(1)$. A first order expansion around $\beta^*$, using $\Psi^\top\beta^*=V^*$, gives
\[
\bar\beta_\Delta-\beta^*=-J_\Delta^{-1}\,m_\Delta(V^*;\Psi)\,(1+o(1)),\qquad
m_\Delta(V^*;\Psi)=\big(m_\Delta(V^*;\Psi_1),\dots,m_\Delta(V^*;\Psi_M)\big)^\top.
\]
By Steps 1--3 the $j$th component is $\tfrac{\Delta}{2}A_j-\tfrac{\sigma_\epsilon^2}{\Delta}B_j+R_j$ with $A_j=\mathbb E[(\log\gamma+\mathcal L_a)^2V^*\Psi_j\mathbb I_a]$ and $B_j=\mathbb E[\nabla V^{*\top}\nabla\Psi_j\mathbb I_a]$. If the vectors $A=(A_j)_j$ and $B=(B_j)_j$ are not collinear, then $\tfrac{\Delta}{2}A-\tfrac{\sigma_\epsilon^2}{\Delta}B$ is nonzero for every $\Delta>0$, so $\|m_\Delta(V^*;\Psi)\|=\Theta(\Delta+\sigma_\epsilon^2/\Delta)$ and hence $\|\bar\beta_\Delta-\beta^*\|=\Theta(\Delta+\sigma_\epsilon^2/\Delta)$. Minimizing $\Delta+\sigma_\epsilon^2/\Delta$ over $\Delta>0$ yields $\Delta^\star\asymp\sigma_\epsilon$ with value $\Theta(\sigma_\epsilon)$. Finally, changing the reward-scaling convention (immediate reward entering as $Y_t$ versus $Y_t\Delta$) rescales $m_\Delta$ by $1+O(\Delta)$ and leaves these orders unchanged. This proves Theorem~\ref{thm:discretization}.

\subsection{Additional simulations}
\label{sec:repro}

\noindent\emph{Implementation details of the compared methods.} For reproducibility, the implementations are as follows. DML uses the partially linear regression with random forest nuisance learners ($100$ trees, maximum depth $5$, minimum leaf size $2$, one candidate feature per split), the package default five fold cross fitting applied to the pooled observation level data (folds are not stratified by subject), and the package asymptotic variance. SAVE solves the discrete time Bellman regression on the common grid with the fixed per step discount $\gamma=0.8$, the same polynomial value function basis as the proposed method, action blocked least squares, and a plug in sandwich variance for its contrast. The proposed method uses cubic smoothing splines with the Rice target smoothing level $\lambda=n_s\widehat\sigma^2_\varepsilon$, polynomial value bases of degree $3$ (univariate) and $2$ (multivariate), discount $\gamma=0.8$, and a standard normal reference distribution $\mathbb G$; the supplementary sweeps select the discount and basis degree per cell from a small grid ($\gamma\in\{0.5,0.7,0.8\}$, degree $\in\{2,3\}$) in a separate tuning run documented with the sweep outputs.

Table \ref{tb:sim2} summarizes the power of the competing methods across different sample sizes under the two dimensional setting Sim1 at treatment effect $\delta=0.05$, where the smaller treatment effect avoids saturation and reveals method differences. Unlike the main text multivariate tables, in which the proposed method's spline stage uses the full $0.01$ simulation mesh (a twenty to one state to outcome ratio) with fixed tuning, this sweep \emph{controls} the state grid at $n_s\in\{50,100,200\}$ state observations per subject with per cell tuned discount and basis degree; SAVE always uses only the outcome grid, which is why its entries here match the main tables while the proposed method's power at $(n_s,n_y)=(100,50)$ is lower than the $0.994$ of Table~2 in the main text. The t-test and DML-test reject at or below their nominal levels across all sample sizes; because these methods test contemporaneous null hypotheses that are nearly true under this data generating process, their rows calibrate what a contemporaneous analysis detects rather than measure a power deficiency. The proposed method's edge over the SequentiAl Value Evaluation (SAVE) strengthens with sample size: at the largest configuration $(n_s = 200, n_y = 50)$, the proposed method achieves a power of $0.926$ versus SAVE's $0.216$, a four fold advantage. At the moderate configuration $(n_s = 100, n_y = 50)$, the proposed method attains $0.474$ versus SAVE's $0.224$. At the smallest tested configuration $(n_s = 50, n_y = 25)$, the two methods are within Monte Carlo noise of each other ($0.284$ for the proposed method versus $0.353$ for SAVE), reflecting the difficulty of stably fitting splines and the continuous time estimating equation with very few state observations. This pattern indicates that the proposed method's multi resolution and continuous time formulation pays off most when the state observation grid is sufficiently dense for spline smoothing to recover the latent trajectory.

\begin{table}[!h]
\centering

\begin{tabular}{c|cccc}
\hline
\hline
Sample size  & t-test & DML-test & SAVE  & Proposed  \\
\hline
$n_s = 50, n_y = 25$  & 0.000 (0.001) & 0.044 (0.017)  &   \textbf{0.353 (0.034)}  &  0.284 (0.032) \\
$n_s = 100, n_y = 50$ & 0.001 (0.002) & 0.040 (0.012)  &   0.224 (0.027)            &  \textbf{0.474 (0.039)} \\
$n_s = 200, n_y = 50$ & 0.001 (0.002) & 0.035 (0.012)  &   0.216 (0.029)            &  \textbf{0.926 (0.018)} \\
\hline
\hline
\end{tabular}
 \caption{Power of competing methods on Sim1 at $\delta=0.05$ under different state/outcome sample sizes. Entries are rejection rates over $20$ Monte Carlo runs of $200$ replications each.}
 \label{tb:sim2}
\end{table}

Additionally, we compare the performance of the four methods under varying levels of treatment effect $\delta$ in Table \ref{tb:sim3}, under the same controlled $(n_s,n_y)=(100,50)$ design and per cell tuning as Table~\ref{tb:sim2}. The proposed method experiences far less power loss than the SequentiAl Value Evaluation (SAVE) at small treatment effects. At the smallest nonzero effect $\delta = 0.025$, the proposed method attains a power of $0.231$ while SAVE manages only $0.039$, a roughly six fold advantage at the regime where carryover effects are most difficult to detect. At $\delta = 0.05$ the proposed method achieves $0.559$ versus SAVE's $0.212$, and at $\delta = 0.075$ the proposed method achieves $0.844$ versus SAVE's $0.563$. Both methods approach saturation as $\delta \to 0.1$, where the proposed method reaches $0.975$ and SAVE reaches $0.846$. The t-test and DML-test stay within Monte Carlo noise of the nominal $0.05$ level across all $\delta$, confirming that neither captures the cumulative continuous time signal. These findings are particularly relevant in healthcare applications where detecting small treatment effects is critical, as the proposed method exhibits improved robustness and rejection probability over the discrete time value equation benchmark (SAVE), which targets the same cumulative alternative, in the small effect regime.

\begin{table}[!h]
\centering

\begin{tabular}{c|cccc}
\hline
\hline
$\delta$   & t-test & DML-test & SAVE  & Proposed  \\
\hline
0     & 0.001 (0.002) & 0.042 (0.014) & 0.004 (0.005) & 0.059 (0.018) \\
0.025 & 0.000 (0.000) & 0.042 (0.017) & 0.039 (0.013) & \textbf{0.231 (0.028)} \\
0.050 & 0.000 (0.001) & 0.038 (0.014) & 0.212 (0.033) & \textbf{0.559 (0.035)} \\
0.075 & 0.000 (0.000) & 0.041 (0.012) & 0.563 (0.039) & \textbf{0.844 (0.024)} \\
0.100 & 0.000 (0.001) & 0.043 (0.016) & 0.846 (0.022) & \textbf{0.975 (0.014)} \\
\hline
\hline
\end{tabular}
 \caption{Power of competing methods on Sim1 across treatment effects $\delta$. Entries are rejection rates over $20$ Monte Carlo runs of $200$ replications each.}
 \label{tb:sim3}
\end{table}

\subsection{Regular grid comparison under observation noise}
\label{sec:regular-grid}

To address how the proposed method compares to SAVE and DML when $Y_t$ is observed on a regular grid (so that the multi resolution advantage is switched off), we run the following targeted experiment. Under the 1D simulation setup of Section~5 with deterministic drift $-0.2\,S_t+\delta\,A_t$, we generate trajectories on a fine time mesh with $\Delta=0.01$ and $T=10$ and then record noisy observations $X_{it}=S_{it}+\varepsilon_{it}^X$ and $Y_{it}=S_{it}+\varepsilon_{it}^Y$, with $\varepsilon_{it}^X,\varepsilon_{it}^Y\overset{\text{iid}}{\sim}\mathcal{N}(0,\epsilon_{obs}^2)$, on a regular grid of spacing $h\in\{0.1,0.2\}$. All methods see the same grid points of $X$ and $Y$. The proposed method performs spline smoothing on $X$ at the fine observation frequency; SAVE fits a discrete time Bellman equation on the grid; DML applies the partial linear regression of \citet{chernozhukov2017double} to grid point observations. Treatment follows the long period specification. Table~\ref{tb:regular_grid} reports power and Type-I error across two grid spacings and two observation noise levels over $30$ Monte Carlo runs of $100$ repetitions each with $I=50$ subjects.

The results show that the proposed method matches SAVE in the clean regime and dominates it under realistic observation noise. When observations are nearly clean ($\epsilon_{obs}=0.1$), the proposed method and SAVE attain identical near perfect power ($1.000$) at both grid spacings, confirming that the continuous time formulation incurs no efficiency cost when the discrete Bellman discretization is essentially unbiased. When observation noise rises to $\epsilon_{obs}=0.3$, the proposed method retains power $0.989$ at $h=0.2$ and $0.993$ at $h=0.1$, while SAVE's power drops to $0.328$ and $0.166$, respectively. The ratio of proposed to SAVE power thus widens from about three fold at $h=0.2$ to roughly six fold at $h=0.1$: as the grid refines, SAVE's discrete Bellman residuals are increasingly dominated by observation noise, whereas the proposed method's spline smoothing recovers the latent trajectory and benefits from the denser sampling. DML and the t-test stay within Monte Carlo noise of the nominal $0.05$ level across all eight cells, confirming that neither captures the cumulative continuous time signal. All methods control Type-I error at or below the nominal $0.05$ level; the proposed method's Type-I error ranges from $0.028$ to $0.044$ across the four Type-I rows.

\begin{table}[!h]
\centering

\begin{tabular}{c|c|c|cccc}
\hline\hline
Grid $h$ & $\epsilon_{obs}$ & $\delta$ & t-test & DML & SAVE & Proposed \\
\hline
  0.2 & 0.1 & 0.3 & 0.001 (0.002) & 0.051 (0.036) & \textbf{1.000 (0.000)} & \textbf{1.000 (0.000)} \\
  0.2 & 0.1 & 0.0 & 0.000 (0.000) & 0.045 (0.027) & 0.002 (0.004)          & 0.034 (0.018) \\
  0.2 & 0.3 & 0.3 & 0.024 (0.017) & 0.034 (0.027) & 0.328 (0.039)          & \textbf{0.989 (0.010)} \\
  0.2 & 0.3 & 0.0 & 0.016 (0.012) & 0.037 (0.026) & 0.013 (0.012)          & 0.028 (0.017) \\
  0.1 & 0.1 & 0.3 & 0.000 (0.002) & 0.055 (0.027) & \textbf{1.000 (0.000)} & \textbf{1.000 (0.000)} \\
  0.1 & 0.1 & 0.0 & 0.000 (0.002) & 0.054 (0.028) & 0.000 (0.000)          & 0.044 (0.020) \\
  0.1 & 0.3 & 0.3 & 0.019 (0.014) & 0.047 (0.030) & 0.166 (0.026)          & \textbf{0.993 (0.010)} \\
  0.1 & 0.3 & 0.0 & 0.018 (0.015) & 0.036 (0.030) & 0.009 (0.009)          & 0.035 (0.020) \\
\hline\hline
\end{tabular}
\caption{Regular grid comparison under deterministic ODE simulation with additive observation noise $X_t=S_t+\varepsilon^X_{obs}$ and $Y_t=S_t+\varepsilon^Y_{obs}$. The proposed method uses fine frequency noisy observations for spline smoothing; SAVE and DML use grid point noisy observations directly. Columns report Monte Carlo power / Type-I error with standard deviations across the $30$ Monte Carlo runs. Best power per row in bold.}
\label{tb:regular_grid}
\end{table}

\subsection{Sensitivity to latent process noise}
\label{sec:sensitivity-sde}

To assess the robustness of the proposed test to violations of the deterministic ODE assumption, we run an additional simulation in which the latent state evolves according to a stochastic differential equation (SDE) with nontrivial diffusion:
\begin{equation}
    \begin{aligned}
    dS_t &= (-0.2\,S_t + \delta\,A_t)\,dt + \sigma\,dW_t,\\
    Y_t  &= S_t,
\end{aligned}
\label{sim_sde}
\end{equation}
where $\sigma\,dW_t$ is a Brownian motion latent process noise term distinct from the additive observation noise in Section~5. We consider $\sigma \in \{0.1, 0.3\}$, treatment effects $\delta \in \{0, 0.3\}$, and the same short period (Treatment~1) and long period (Treatment~2) treatment schedules used in the main 1D simulation. The proposed ODE method is fit as is, without any modification to account for the diffusion component. Initial value $S_0 = 0$, time step $0.01$, $T = 10$, $I = 50$ subjects, and $50$ outer Monte Carlo replicates with $200$ inner repetitions each.

Table~\ref{tb:sim_sde} reports power and Type-I error of the four methods. The proposed method retains the highest power across all alternative cells and controls Type-I error at or near the nominal $0.05$ level. Even at $\sigma=0.3$ in the most challenging short period setting, the proposed method retains power $0.561$, exceeding SAVE's $0.191$ by a wide margin; for the long period setting at the same noise level, the proposed method attains $0.867$ versus SAVE's $0.758$. These results demonstrate finite sample robustness in the simulated settings: spline smoothing recovers the drift driven component of the latent trajectory even in the presence of Brownian process noise, supporting the ODE specification as a working model under modest process noise rather than a strictly deterministic commitment. The $\delta=0$ rows of Table~\ref{tb:sim_sde} coincide across the two diffusion levels because the null replicates share common random paths and the standardized, studentized statistic is insensitive to the scale of the common diffusion component, so identical rejection decisions are expected rather than anomalous. A formal extension to SDE dynamics is discussed as future work in Section~\ref{sec:discussion} of the main manuscript.

\begin{table}[!h]
\centering
\begin{tabular}{c|c|c|cccc}
\hline\hline
               & $\sigma$ & $\delta$
               & t-test        & DML-test        & SAVE            & Proposed \\
\hline
\multirow{4}{*}{Treatment 1}
  & \multirow{2}{*}{0.1} & 0.3
      & 0.031 (0.012) & 0.027 (0.021) & 0.221 (0.028) & \textbf{0.701 (0.031)} \\
  &                      & 0.0
      & 0.009 (0.007) & 0.031 (0.027) & 0.049 (0.013) & 0.018 (0.008) \\ \cline{2-7}
  & \multirow{2}{*}{0.3} & 0.3
      & 0.016 (0.009) & 0.025 (0.020) & 0.191 (0.029) & \textbf{0.561 (0.035)} \\
  &                      & 0.0
      & 0.009 (0.007) & 0.031 (0.027) & 0.049 (0.013) & 0.018 (0.008) \\
\hline
\multirow{4}{*}{Treatment 2}
  & \multirow{2}{*}{0.1} & 0.3
      & 0.002 (0.003) & 0.024 (0.019) & 0.960 (0.014) & \textbf{0.988 (0.006)} \\
  &                      & 0.0
      & 0.038 (0.011) & 0.045 (0.030) & 0.048 (0.014) & 0.033 (0.013) \\ \cline{2-7}
  & \multirow{2}{*}{0.3} & 0.3
      & 0.038 (0.013) & 0.048 (0.033) & 0.758 (0.025) & \textbf{0.867 (0.029)} \\
  &                      & 0.0
      & 0.038 (0.011) & 0.045 (0.030) & 0.048 (0.014) & 0.033 (0.013) \\
\hline\hline
\end{tabular}
\caption{Sensitivity to latent process noise: data generated from the SDE in equation~\eqref{sim_sde} with diffusion $\sigma$ representing latent Brownian noise (distinct from the observation noise used in Section~5). The proposed ODE method is applied without modification. Columns report Monte Carlo power / Type-I error with standard deviations across $50$ replications. Best power per row in bold.}
\label{tb:sim_sde}
\end{table}

\subsection{Implementation details and design diagnostics for the OhioT1DM analyses}
\label{sec:realdata-implementation}

\emph{Implementation choices,} fixed across all analyses of Section~6 in the main text: time measured in hours, discount $\gamma=0.8$ (effective horizon $1/|\log\gamma|\approx4.5$ hours); a degree $2$ polynomial sieve in the state coordinates of dimension $9$---an intercept, the four coordinates, and their four squares; no cross products and \emph{no threshold hinge terms}, so the basal threshold analyses of Section~6 are exploratory relative to the hinge adapted sieve required by the formal threshold policy guarantee; cubic smoothing splines per channel and day, with the smoothing level set by a Rice estimator of the noise variance; each channel standardized by its global standard deviation, so $\widehat\tau$ is a discounted cumulative quantity on the standardized glucose $\times$ hours scale; the reference distribution $\mathbb G$ taken as the \emph{empirical} distribution of the day initial states, a finite sample plug in choice: it is supported on the observed state range where the overlap condition is diagnosed, though as a discrete measure it sits outside the bounded density sufficient condition under which the theory is stated, while a Gaussian fit to the day initial states instead has unbounded support extending outside the observed range (the two choices change every reported estimate only in the second or third decimal: the bolus contrast is unchanged at $+0.168$ to three decimals, and the basal threshold interval endpoints move by at most $0.06$; analytic inference conditions on the realized reference distribution, and in the leave one patient out and bootstrap refits $\mathbb G$ is recomputed, day weighted, from the day initial states of the included days); days excluded when they carry fewer than $120$ CGM readings or a physiological channel is observed on less than $10\%$ of its grid, leaving $188$ patient days from the six patients of the 2020 cohort ($18$--$41$ days per patient). Patient clustered inference sums scores within patient and uses the $t_{5}$ reference; hierarchical bootstrap intervals resample the six patients with replacement; leave one patient out (LOPO) ranges recompute each estimate excluding one patient at a time. For the thin overlap basal contrasts the Jacobian condition numbers are of order $10^{7}$, quantifying how little information the treat stratum carries.

\emph{Hierarchical bootstrap results.} The $500$ draw patient bootstrap percentile intervals are $[-2.76,+0.48]$ for the bolus observed regime versus never contrast, and $[-6.98,+6.97]$, $[-4.38,+5.01]$, and $[-8.38,+5.88]$ for the three basal contrasts of Table~5 in the main text ($\pi_{150}$ vs.\ $\pi_{180}$, $\pi_{150}$ vs.\ $\pi\equiv0$, $\pi_{180}$ vs.\ $\pi\equiv0$). All four intervals cover zero; their widths vary substantially relative to the $t_5$ intervals (wider for the bolus and $\pi_{180}$ vs.\ $\pi\equiv0$ contrasts, narrower for the two $\pi_{150}$ contrasts). Two caveats attach: with six patients, bootstrap draws that replicate a single patient several times produce ill conditioned Jacobians and occasional extreme replicates, inflating the tails, and the percentile interval is itself only first order accurate at this cluster count. We therefore report the $t_5$ clustered intervals as primary and read the bootstrap as confirming that patient level uncertainty, however quantified, dominates the analysis.

\emph{Overlap diagnostics for the basal threshold policies.} Table~\ref{tb:realdata_overlap} collects the per policy overlap diagnostics summarized in Section~6 of the main text: the marginal agreement rate and the treat stratum agreement $\widehat{\Pr}\{A=1\mid\mathrm{glucose}>c\}$ with the count of agreeing treated epochs; the stratum sizes differ because each policy defines its stratum by its own threshold, and the no elevation policy has no treat stratum.

\begin{table}[!h]
\centering
\begin{tabular}{l|c|c}
\hline\hline
Policy & marginal agree & treat stratum ($n$) \\
\hline
$\pi_{150}$              & $0.48$ & $0.15$ ($3251$) \\
$\pi_{180}$              & $0.62$ & $0.15$ ($2027$) \\
$\pi\equiv0$ (no elevation) & $0.85$ &  \\
\hline\hline
\end{tabular}
\caption{Per policy overlap diagnostics for the basal threshold policies of Table~5 in the main text: marginal agreement rate and treat stratum agreement $\widehat{\Pr}\{A=1\mid\mathrm{glucose}>c\}$, with the count of agreeing treated epochs in parentheses.}
\label{tb:realdata_overlap}
\end{table}

\emph{Conditional agreement propensity.} To refine the stratified overlap diagnostic of Table~\ref{tb:realdata_overlap}, we estimate the agreement propensity by a logistic regression of the treatment indicator on the four standardized state channels, evaluated at every epoch: $\widehat q(s)=\widehat e(s)$ where the policy prescribes treatment and $1-\widehat e(s)$ elsewhere, with $\widehat e(s)=\widehat{\Pr}(A_{it}=1\mid S_{it}=s)$. On the treat stratum, $\widehat q$ has median $0.14$, fifth percentile $0.12$, and first percentile $0.004$ for both thresholds; the Kish effective sample size of the agreement subsample, weighting each agreeing epoch by $1/\widehat q$, is $6{,}057$ of $20{,}499$ agreeing epochs for $\pi_{150}$ and $7{,}065$ of $26{,}205$ for $\pi_{180}$ (and $36{,}170$ of $36{,}202$ for $\pi\equiv0$, whose support is excellent). The near zero first percentile shows that conditional positivity comes close to failing in part of the treat region, sharpening the stratified diagnostic and supporting the inconclusive reading of the basal contrasts.

\emph{Why the bolus contrast is reported descriptively.} Meals, the principal bolus trigger, are self reported and substantially under reported: the data contain $2.2$ boluses per reported meal, only about a third of boluses fall within an hour of a reported meal, and boluses without a nearby meal report have median glucose $182$ mg/dL, splitting between corrections and doses that evidently accompany unlogged meals or snacks. The observed dosing regime is therefore a rule of triggers the recorded state only partially captures, and unrecorded meals can affect both dosing and glucose. A causal reading of the regime versus never dose contrast would require conditional support for never dosing in the meal related and hyperglycemic states where boluses occur, together with sequential randomization given the recorded state; the marginal untreated fraction of $0.98$ does not establish that conditional support. Nor can the regime be replaced by an implementable rule of the recorded variables: a policy that doses at a reported meal or when glucose exceeds $180$ mg/dL covers only $62\%$ of the observed boluses, and its treat stratum agreement on the CGM grid is $0.07$---thinner than the basal diagnostics---because dosing is a point event while its triggers persist across many five minute epochs. These diagnostics are why the main text reports the bolus contrast as a descriptive cumulative timing association.

\subsection{Robustness of the OhioT1DM findings across time of day windows and thresholds}
\label{sec:realdata-robustness}

Because glucose dynamics have strong circadian structure, we verify that the real data conclusions of Section~6 in the main text are not an artifact of pooling day and night. We repeat the always treat versus never treat contrast and the threshold versus never contrasts ($c\in\{140,150,160,180\}$ mg/dL) on three time of day windows: the full day, daytime (06:00--24:00), and overnight (00:00--06:00), the clinically standard fasting window on which basal titration is judged. All analyses use the same estimating equation structure as the main text---the four dimensional augmented state, the generator based moment, and two sided tests---but with alternative uncertainty and reference distribution choices: the \emph{day} clustered long run variance and a Gaussian reference distribution (the main text tables use the patient clustered $t_5$ inference and the empirical reference of Section~\ref{sec:realdata-implementation}, the latter shifting estimates only in the second decimal); days without adequate coverage in a window are excluded. This grid is therefore a descriptive sensitivity analysis across windows and thresholds, not a rerun of the primary pipeline; the primary inference of Section~6 in the main text is patient clustered ($t_5$ reference), which widens the uncertainty substantially (for the observed regime bolus contrast of Table~4 in the main text, $z$ falls from $6.4$ under day clustering to $2.0$ under patient clustering), so the $z$ and $p$ entries below should be read as window to window comparisons, not as confirmatory tests. Table~\ref{tb:realdata_robust} reports the full grid.

\begin{table}[!h]
\centering
\begin{tabular}{ll|cc|c|cc|c}
\hline\hline
 & Window & days & treated & always vs.\ never $\widehat\tau$ ($z$; $p$) & prop.\ rr & t test rr & thr.\ $p$ range \\
\hline
Basal & full  & 188 & 0.14 & $+1.5$ ($0.2$; $0.87$)   & $0.032$ & $0.822$ & $0.84$--$0.98$ \\
Basal & day   & 185 & 0.12 & $+1.0$ ($0.1$; $0.96$)   & $0.044$ & $0.794$ & $0.41$--$0.99$ \\
Basal & night & 129 & 0.22 & $-3.1$ ($-0.4$; $0.69$)  & $0.032$ & $0.796$ & $0.65$--$0.77$ \\
Bolus & full  & 188 & 0.02 & $+6.6$ ($7.1$; $<0.001$) & $0.134$ & $0.488$ & $0.19$--$0.36$ \\
Bolus & day   & 185 & 0.02 & $+5.2$ ($4.4$; $<0.001$) & $0.104$ & $0.194$ & $0.61$--$0.98$ \\
Bolus & night & 129 & 0.01 & $-0.2$ ($0.0$; $0.99$)   & ---     & $0.579$ & $0.64$--$0.77$ \\
\hline\hline
\end{tabular}
\caption{Robustness grid for the OhioT1DM analyses. ``treated'' is the treated fraction of observation times in the window; the always versus never column gives the full data estimate with two sided $p$ from the day clustered variance; ``prop.\ rr'' and ``t test rr'' are resampled rejection rates ($20$ day draws) for the proposed test and the t test; ``thr.\ $p$ range'' is the range of full data $p$ values across the four threshold versus never contrasts. The overnight bolus proposed rate is omitted as degenerate: with $0.6\%$ treated observations, a $20$ night draw contains one or two bolus events and the resampled statistic is unstable. All $\widehat\tau$ estimates are on the standardized glucose scale used throughout Section~6 of the main text; note that the bolus rows evaluate the always treat versus never treat contrast, not the observed dosing regime versus never treat contrast of Table~4 in the main text, which is why the two sets of estimates differ.}
\label{tb:realdata_robust}
\end{table}

This grid also carries the always treat versus never treat basal comparison, which the main text no longer features: under the $20$ day resampling protocol used for the rejection rate columns of this grid (Table~\ref{tb:realdata} in the main text reports full data patient clustered analyses instead), the full resolution DML rejects at $0.358$ ($0.053$) for basal and $0.000$ for bolus, alongside the t test and proposed rates shown here. Three observations are qualitatively stable across the grid. First, no statistically detectable basal differences were found in any window or at any threshold, and the contrasts remain inconclusive; notably, the overnight window has the \emph{best} overlap in the grid (treated fraction $0.22$) and yields a negative point estimate (above average basal associated with lower cumulative fasting glucose, directionally consistent with pharmacology) but at $z=-0.4$ the evidence is far from significance, so the inconclusive reading of the main text persists in the most favorable window. Second, the always dose versus never dose bolus association in this grid is concentrated in the full day and daytime windows and attenuates under threshold regimes throughout, as the overlap diagnostics predict; because these rows evaluate a different contrast from the observed regime contrast of Table~4 in the main text, they are a descriptive companion rather than a confirmation of it. Third, the t test rejects at rates near $0.8$ for basal in \emph{every} window, including the overnight fasting window where no contemporaneous contrast is plausible; this illustrates the pseudoreplication incurred by treating serially dependent five minute readings as independent---day clustering removes the within day part of that dependence, while the patient clustered primary analysis of the main text also accounts for dependence across days within a patient.

\subsection{Two-step estimator of drift coefficient with noisy data}
\citet{brunel2008parameter} proposes a two step approach for parameter estimation in ordinary differential equations (ODEs). In the first step, the solution \( S_t \) of the ODE and its derivative \( {dS_t}/{dt} \) are estimated nonparametrically using smoothing splines, kernel regression, or local polynomial regression. The resulting estimators $\hat{S}_t$ and $d\hat{S}_t/dt$ are consistent for the true trajectory and its derivative under standard smoothness and bandwidth conditions, without requiring a parametric form for the ODE solution. This first step provides a foundation for the subsequent parameter estimation by ensuring that the estimated trajectories and derivatives closely approximate the true dynamics of the system.

In the second step, the drift parameter \( \vartheta \) is estimated by minimizing the discrepancy between the estimated derivative \( {d\hat{S}_t}/{dt} \) and the ODE system function \( b(t; \vartheta) \). (We use $\vartheta$ here to distinguish this drift parameter from the density ratio coefficients $\theta^*_{a,t}$ in Assumption~\ref{asp:basisfunction}.) Specifically, the criterion for optimization is defined as
\[
R_{q,n,w}(\vartheta) = \left\| \frac{d\hat{S}_t}{dt} - b(t; \vartheta) \right\|_{q,w},
\]
where \( \|\cdot\|_{q,w} \) represents the weighted \( L_q \)-norm. The parameter estimator \( \hat{\vartheta}_n \) is then obtained by solving the optimization problem
\[
\hat{\vartheta}_n = \arg \min_{\vartheta} R_{q,n,w}(\vartheta).
\]
When smoothing splines are employed in the first step, the solution \( S_t \) is represented as \( S_{it} = \chi^{\top} w_i \), where \( \chi \) denotes the spline basis functions and \( w_i \) are the coefficients. The estimated derivative entering the second step criterion is then the fitted spline derivative \( \nabla\chi^\top\hat w_i \); the drift parameter \( \hat{\vartheta}_n \) is obtained by the subsequent minimization and is a functional of, not identical to, the spline coefficients \( \hat{w}_i \).

\bibliographystyle{agsm}
\bibliography{reflist}